\documentclass[prd,%
preprint,%
tightenlines,superscriptaddress,showpacs,%
nofootinbib,eqsecnum,amsfonts,amsmath,amssymb]{revtex4}
\usepackage{bm}
\usepackage{graphicx}

\begin{document}

\title{Dimensional regularization of the third post-Newtonian
gravitational wave generation from two point masses}

\author{Luc Blanchet} \email{blanchet@iap.fr}
\affiliation{${\mathcal{G}}{\mathbb{R}}
\varepsilon{\mathbb{C}}{\mathcal{O}}$, Institut d'Astrophysique de
Paris, UMR 7095-CNRS, 98$^{\text{bis}}$ boulevard Arago, F-75014
Paris, France}

\author{Thibault Damour} \email{damour@ihes.fr} \affiliation{Institut
des Hautes \'Etudes Scientifiques, 35 route de Chartres, F-91440
Bures-sur-Yvette, France}

\author{Gilles \surname{Esposito-Far\`ese}} \email{gef@iap.fr}
\affiliation{${\mathcal{G}}{\mathbb{R}}
\varepsilon{\mathbb{C}}{\mathcal{O}}$, Institut d'Astrophysique de
Paris, UMR 7095-CNRS, 98$^{\text{bis}}$ boulevard Arago, F-75014
Paris, France}

\author{Bala R. Iyer}
\email{bri@rri.res.in}
\affiliation{Raman Research Institute, Bangalore 560 080, India}

\begin{abstract}
Dimensional regularization is applied to the computation of the
gravitational wave field generated by compact binaries at the third
post-Newtonian (3PN) approximation. We generalize the wave
generation formalism from isolated post-Newtonian matter systems to
$d$ spatial dimensions, and apply it to point masses (without
spins), modelled by delta-function singularities. We find that the
quadrupole moment of point-particle binaries in harmonic coordinates
contains a pole when $\varepsilon\equiv d-3\rightarrow 0$ at the 3PN
order. It is proved that the pole can be renormalized away by means
of the same shifts of the particle world-lines as in our recent
derivation of the 3PN equations of motion. The resulting
renormalized (finite when $\varepsilon\rightarrow 0$) quadrupole
moment leads to unique values for the ambiguity parameters $\xi$,
$\kappa$ and $\zeta$, which were introduced in previous computations
using Hadamard's regularization. Several checks of these values are
presented. These results complete the derivation of the
gravitational waves emitted by inspiralling compact binaries up to
the 3.5PN level of accuracy which is needed for detection and
analysis of the signals in the gravitational-wave antennas
LIGO/VIRGO and LISA.
\end{abstract}

\date{\today}

\maketitle

\section{Introduction}\label{secI}

A compelling motivation for accurate computations of the
gravitational radiation field generated by compact binary systems
(\textit{i.e.}, made of neutron stars and/or black holes) is the
need for accurate \textit{templates} to be used in the data analysis
of the current and future generations of laser interferometric
gravitational wave detectors. It is indeed recognized that the
\textit{inspiral} phase of the coalescence of two compact objects
represents an extremely important source for the ground-based
detectors LIGO/VIRGO, provided that their total mass does not exceed
say 10 or 20 $M_\odot$ (this includes the interesting case of double
neutron-star systems), and for space-based detectors like LISA, in
the case of the coalescence of two galactic black holes, if the
masses are within the range between say $10^5$ and $10^8\,M_\odot$.

For these sources the \textit{post-Newtonian} (PN) approximation
scheme has proved to be the appropriate theoretical tool in order to
construct the necessary templates. A program started long ago with
the goal of obtaining these templates with 3PN and even 3.5PN
accuracy.\footnote{Following the standard custom we use the
qualifier $n$PN for a term in the wave form or (for instance) the
energy flux which is of the order of $1/c^{2n}$ relatively to the
lowest-order Newtonian quadrupolar radiation.} Several studies
\cite{3mn,CF94,TNaka94,P95,DIS98,DIS00,BCV03a,DIJS03,AIRS05,AISS05}
have shown that such a high PN precision is probably sufficient, not
only for detecting the signals in LIGO/VIRGO, but also for analyzing
them and accurately measuring the parameters of the binary (such
high-accuracy templates will also be of great value for detecting
massive black-hole mergers in LISA). The templates have been first
completed through 2.5PN order, for both the phase
\cite{BDIWW95,BDI95,WWi96,B96} and wave amplitude
\cite{BIWW96,ABIQ04}. The 3.5PN accuracy for the templates (in the
case where the compact objects have negligible intrinsic spins) has
been achieved more recently, in essentially two steps.
\begin{enumerate}
\item[(1)]The first step has been to compute all the terms, in both
the 3PN equations of motion, either in Hamiltonian form
\cite{JaraS98,JaraS99,DJSpoinc,DJSequiv}
or using harmonic coordinates \cite{BF00,BFeom,ABF01,BI03CM}, and
the 3.5PN gravitational radiation field, using a multipolar wave
generation formalism \cite{B98tail,BIJ02,BFIJ02,BI04mult}, by means
of the Hadamard self-field regularization
\cite{Hadamard,Schwartz,Sellier,BFreg}, in short HR. (The 3.5PN
terms in the equations of motion have been added in
Refs.~\cite{PW00,KFS03,NB05}.) However, a few terms were left
undetermined by Hadamard's regularization, which corresponds to some
incompleteness of this regularization occurring at the 3PN order.
These terms could be parametrized by some unknown numerical
coefficients called \textit{ambiguity parameters}.
\item[(2)]The second step has been to fix the values of the
ambiguity parameters by means of dimensional regularization
\cite{tHooft,Bollini,Breitenlohner}, henceforth abbreviated as DR.
Technically, DR is based on analytic continuation in the dimension
of space $d=3+\varepsilon$. The ambiguity parameter $\lambda$
entering the 3PN equations of motion has been computed in
Refs.~\cite{DJSdim,BDE04}, with result $\lambda = -1987/3080$. (This
result has also been obtained with an alternative approach in
Refs.~\cite{IFA01,itoh1,itoh2}.) The three ambiguity parameters
appearing in the 3PN gravitational radiation field will be shown in
the present paper to have the following unique values
\begin{equation}\label{xikappazeta}
\xi = -\frac{9871}{9240}\,, \qquad \kappa = 0\,, \qquad \zeta =
-\frac{7}{33}\,,
\end{equation}
as already announced in Ref.~\cite{BDEI04}. The method we use for
applying DR essentially consists in computing the
\textit{difference} between DR and some appropriately defined
Hadamard-type regularization called below the pure-Hadamard-Schwartz
(pHS) regularization.
\end{enumerate}
Those results complete the determination of the 3.5PN-accurate phase
evolution as it suffices to insert into the formulas of
Ref.~\cite{BFIJ02} the value for $\lambda$, together with the values
given by (\ref{xikappazeta}). Actually, this phase evolution depends
only on $\lambda$ and on the following particular combination of
parameters,
\begin{equation}\label{theta}
\theta\equiv\xi+2\kappa+\zeta = -\frac{11831}{9240}\,.
\end{equation}
The present paper is devoted to the details of our DR computation
of the ambiguity parameters, \textit{item} (2) above, which has led
to the values (\ref{xikappazeta})--(\ref{theta}). We refer to
\cite{BDEI04} for a summary of our method and a general discussion.

Let us emphasize that the values (\ref{xikappazeta}), which
constitute the end result of the application of DR, have all been
confirmed by alternative methods. Our first independent check has
been the confirmation of one particular combination of the ambiguity
parameters, namely $\xi+\kappa$, which was shown to follow from the
requirement that the 3PN mass dipole moment of the binary, computed
in \cite{BI04mult} from the multipolar wave generation formalism,
should agree with the 3PN center-of-mass position, known from the
conservative part of the 3PN equations of motion in harmonic
coordinates \cite{ABF01}. Secondly, we have also obtained the value
of $\zeta$ by considering the limiting physical situation of a
\textit{boosted Schwarzschild solution}, corresponding to the case
where the mass of one of the particles is exactly zero, and the
other particle moves with uniform velocity \cite{BDI04zeta}. It can
be argued from this calculation that the value of $\zeta$ in
Eq.~(\ref{xikappazeta}) is a consequence of the global Poincar\'e
invariance of the multipolar wave generation formalism. Thirdly, in
Sec.~\ref{secVII} below, we shall be able to show that the value of
$\kappa$ is zero by a diagrammatic approach (where the ``diagrams''
are taken in the sense of \cite{Dgef96}), showing that no
dangerously divergent diagrams contributing to $\kappa$ appear at
this order. Those checks altogether provide a confirmation,
\textit{independent from DR}, for all the parameters
(\ref{xikappazeta}).

The plan of this paper is as follows. In Section~\ref{secII} we
investigate the symmetric-trace-free (STF) multipole decomposition
in $d$ dimensions for a scalar field with compact-support source. In
Section~\ref{secIII} we generalize to $d$ dimensions the known
results for the multipole expansion of the gravitational field and
the definition of the source-type multipole moments.
Section~\ref{secIV} is devoted to the explicit expressions of the
source terms in the latter source multipole moments at the 3PN order
in terms of a convenient set of retarded-like elementary potentials.
Then, in Section~\ref{secV}, we obtain a general formula for the
difference between DR and HR (in the pHS variant of it). This
difference is non-zero at the 3PN order because of the occurrence of
poles in $d$ dimensions (\textit{i.e.}, $\propto 1/\varepsilon$). In
Section~\ref{secVI} we deduce the ambiguity parameters from the DR
regularization of the 3PN mass quadrupole moment, and we check that
the 3PN mass dipole is in agreement with the known center-of-mass
position deduced from the equations of motion. Section~\ref{secVII}
deals with a direct computation of the pole part of the moments
using diagrams, their renormalization using shifts of the
world-lines, and the check that $\kappa=0$. In Section~\ref{secVIII}
we present an alternative derivation of the value of $\zeta$ based
on considering the physical situation of a single \textit{boosted}
point particle in $d$ dimensions (the result agrees with the recent
computation of the boosted Schwarzschild solution in
\cite{BDI04zeta}).

\section{Multipole expansion of a scalar field in $d$
dimensions}\label{secII}

A crucial input for the derivations we are going to perform in the
present article is the multipolar expansion of solutions of flat
space-time wave equations in $D=d+1$ dimensions. We denote by
$\Box=\eta^{\mu\nu}\partial_\mu\partial_\nu$ the flat d'Alembertian
operator, using the signature ``mostly plus'', \textit{i.e.},
$\Box=\Delta-c^{-2}\partial^2_t$, where
$\partial_t\equiv\partial/\partial t$ and $\Delta$ is the Laplace
operator. We first consider the case of a scalar wave equation, say
\begin{equation}\label{boxphi}
\Box \varphi(\mathbf{x},t) = S(\mathbf{x},t)\,,
\end{equation}
and shall postpone to Sec.~\ref{secIII} the case of tensorial wave
equations. Note that, in the present work, we shall not introduce
any numerical factor in the ``source'' $S$ on the right hand side
(RHS) of the inhomogeneous scalar wave equation (\ref{boxphi}).
Similarly, we define the scalar Green functions as the solutions of
\begin{equation}\label{green1}
\Box G(\mathbf{x},t) = \delta(t)\,\delta^{(d)}(\mathbf{x})\,,
\end{equation}
where $\delta^{(d)}(\mathbf{x})$ is a $d$-dimensional Dirac
distribution, such that $\int
d^d\mathbf{x}\,\delta^{(d)}(\mathbf{x})\,f(\mathbf{x})=f(\mathbf{0})
$. When $d=3$, the \textit{retarded} Green function takes the simple
form
\begin{equation}\label{green2}
G^{(3+1)}_\mathrm{Ret}(\mathbf{x},t) =
-\frac{\delta\left(t-\vert\mathbf{x}\vert/
c\right)}{4\pi\,\vert\mathbf{x}\vert}\,.
\end{equation}
Because of the presence of the factor $-1/4\pi$ in (\ref{green2}),
it was convenient, when working in $3+1$ dimensions, to introduce a
factor $-4\pi$ in front of the RHS's of (\ref{boxphi}) and
(\ref{green1}). However, there is no analogous, universally
simplifying factor in $D$ dimensions, so it is finally simpler to
introduce no factors at all in (\ref{boxphi})--(\ref{green1}).

The $D$-dimensional retarded Green
function has no simple expression in $(t,\mathbf{x})$ space. However,
starting from its well-known Fourier-space expression, one can write
the following simple integral expression (see \textit{e.g.}
\cite{Cardoso}),
\begin{equation}\label{green3}
G_\mathrm{Ret}(\mathbf{x},t)=-\frac{\theta(t)}{(2\pi)^{d/2}}
\int_0^{+\infty}dk \left(\frac{k}{r}\right)^{\frac{d}{ 2}-1}\sin
(c\,k\,t)\,J_{\frac{d}{2}-1}(k\,r)\,.
\end{equation}
Notice that this is in fact a function of $t$ and
$r\equiv\vert\mathbf{x}\vert$ only: say
$G_\mathrm{Ret}(\mathbf{x},t) = G_\mathrm{Ret}(r,t)$. Here
$\theta(t)$ is the Heaviside step function, and
$J_{\frac{d}{2}-1}(k\,r)$ the usual Bessel function. Actually, we
shall never need to use the explicit form (\ref{green3}) of the
Green function in $D$ dimensions. Indeed, we shall obtain the
$d$-dimensional generalizations of the $3$-dimensional
\textit{relativistic multipole moments}, obtained in
Refs.~\cite{B95,B98mult,PB02}, by working directly with the source
$S$ of the wave equation (\ref{boxphi}), or of its tensor
generalizations. To do this, we note first that the retarded
solution of (\ref{boxphi}) reads
\begin{equation}\label{sol1}
\varphi(\mathbf{x},t) = \int
d^d\mathbf{y}\,ds\,G_\mathrm{Ret}(\mathbf{x}-\mathbf{y},t-s)
\,S(\mathbf{y},s)\,.
\end{equation}

In this section, we shall consider sources $S(\mathbf{x},t)$ having
a \textit{spatially compact} support in $d$ space dimensions: say
$S(\mathbf{x},t)=0$ when $\vert\mathbf{x}\vert > a$, where $a$ is
the source's radius. We are interested in the \textit{multipolar
expansion} of the field $\varphi(\mathbf{x},t)$, \textit{i.e.}, its
decomposition (when considered in the external domain
$\vert\mathbf{x}\vert > a$) in $d$-dimensional spherical harmonics.
Traditionally, the multipolar expansion of $\varphi(\mathbf{x},t)$,
Eq.~(\ref{sol1}), is obtained by expanding the spatial kernel
$G_\mathrm{Ret}(\mathbf{x}-\mathbf{y})$ in powers of
$\vert\mathbf{y}\vert\rightarrow 0$. This introduces the (reducible)
multipole moments of the source, say $\int
d^d\mathbf{y}\,y^{i_1}\cdots y^{i_\ell}\,S(\mathbf{y})$. A simpler,
formally equivalent way of proceeding is to replace the continuous
source $S(\mathbf{x})$ by its ``\textit{distributional skeleton}'',
\textit{i.e.}, an expansion in increasing derivatives of the
$d$-dimensional Dirac distribution $\delta(\mathbf{x})$. [For
notational simplicity, we henceforth suppress the superscript $(d)$
on $\delta(\mathbf{x})$.] This skeletonized version of the source
$S$ is equivalent to a continuous function $S(\mathbf{x})$ with
compact support when (and only when) it is integrated by a regular
kernel $K(\mathbf{x},\mathbf{y})$, as in (\ref{sol1}). It reads
\begin{equation}\label{Sskel}
S^\mathrm{Skel}(\mathbf{x},t)=\sum_{\ell=0}^{+\infty}
\frac{(-)^\ell}{\ell!}\,S_L(t)\,\partial_L\delta(\mathbf{x})\,,
\end{equation}
where the coefficients are the reducible multipole moments
\begin{equation}\label{SL}
S_L(t)=\int d^d\mathbf{y}\,y_L \,S(\mathbf{y},t)\,.
\end{equation}
We recall our simplified notation: $L$ denotes a multi-index
$i_1\cdots i_\ell$ and we use the shorthands
$\partial_L\equiv\partial_{i_1}\cdots \partial_{i_\ell}$, where
$\partial_i\equiv\partial/\partial x^i$, and $y_L\equiv
y_{i_1}\cdots y_{i_\ell}$, where $y_i\equiv y^i$.

The skeleton expansion (\ref{Sskel}) does not yet give rise to a
multipole expansion because the various terms on the RHS of
(\ref{Sskel}) do not correspond to \textit{irreducible}
representations of the $d$-dimensional rotation group $O(d)$.
However, it is relatively simple to transform the expansion
(\ref{Sskel}) into irreducible components. To do this, it is enough
to decompose the symmetric tensors $S_L$ into irreducible symmetric
and trace-free (STF) pieces, which is easily done by using the STF
decomposition of $y_L$ in $d$ dimensions, obtained by recursively
separating the traces, like in $y_{ij}\equiv
\widehat{y}_{ij}+\frac{1}{d}\,\delta_{ij}\,\vert\mathbf{y}\vert^2$.
Here we denote the STF projection by means of a hat:
$\widehat{y}_L\equiv\mathrm{STF}[y_{i_1}\cdots y_{i_\ell}]$, or
sometimes by means of brackets surrounding the indices:
$\widehat{y}_L\equiv y_{\langle L\rangle}$. The general formula
defined by this recursion has already been given in
Ref.~\cite{BDE04}\footnote{We refer to the Appendix B of
\cite{BDE04} for a compendium of formulae for working in a space
with $d$ dimensions.} and reads
\begin{subequations}\label{yL}
\begin{eqnarray}
y_L&=&\sum_{k=0}^{[\frac{\ell}{2}]}\,a_\ell^k\,\delta_{\{i_1i_2}
\cdots \delta_{i_{2k-1}i_{2k}}\,\widehat{y}_{L -2K\}}\,
\vert\mathbf{y}\vert^{2k},\\
\text{with}\quad
a_\ell^k&\equiv&\frac{1}{2^k}\frac{\Gamma\left(\frac{d}{2}+
\ell-2k\right)}{\Gamma\left(\frac{d}{2}+\ell-k\right)}\,.
\end{eqnarray}
\end{subequations}
Here, $\delta_{ij}$ is the Kronecker symbol, $[\frac{\ell}{2}]$
denotes the integer part of $\frac{\ell}{2}$, $L-2K$ is a
multi-index with $\ell-2k$ indices, and $\Gamma$ is the usual
Eulerian function. The curly brackets surrounding the indices refer
to the (unnormalized, minimal) sum of the permutations of the
indices which keep the object fully symmetric in $L$, for instance
$\delta_{\{ij}V_{k\}}\equiv\delta_{ij}V_k
+\delta_{ik}V_j+\delta_{jk}V_i$ (for convenience we do not normalize
the latter sum).

We replace the STF decomposition (\ref{yL}) into (\ref{SL}) and
insert the resulting moments back into Eq.~(\ref{Sskel}). After some
simple manipulations we arrive at
\begin{subequations}\label{Smult1}\begin{eqnarray}
S^\mathrm{Skel}(\mathbf{x},t)&=&
\sum_{\ell=0}^{+\infty}\frac{(-)^\ell}{\ell!}
\sum_{k=0}^{+\infty}\alpha_\ell^k
\,\Delta^k\partial_L\left[\delta(\mathbf{x})\int
d^d\mathbf{y}\,\widehat{y}_L\,
\vert\mathbf{y}\vert^{2k}\,S(\mathbf{y},t)
\right],\\
\text{where}\quad
\alpha_\ell^k&\equiv&\frac{1}{2^{2k}k!}\frac{\Gamma\left(\frac{d}{2}+
\ell\right)}{\Gamma\left(\frac{d}{2}+\ell+k\right)}\,.
\label{Smult1b}
\end{eqnarray}\end{subequations}
At this point let us notice that any term in the skeletonized source
$S^\mathrm{Skel}(\mathbf{x},t)$ which is in the form of a d'Alembert
operator $\Box$ acting on spatial gradients or time derivatives of
the delta function, say
$\Box\left[\partial\,\delta(\mathbf{x})\right]$,\footnote{Here the
notation $\partial$ symbolizes any product of space or time
derivatives (so that, for instance, $\partial$ can involve any power
of the box operator $\Box$ itself)} will give no contribution to the
multipole expansion of $\varphi(\mathbf{x},t)$. Indeed, a term in
the source of the form $\Box^{i+1}
\left[f(t)\,\partial_L\delta(\mathbf{x})\right]$, with $i\geq 0$,
$\ell\geq 0$, will yield a contribution to the solution of the form
$\Box^{-1}_\mathrm{Ret}\left(\Box^{i+1}
\left[f(t)\,\partial_L\delta(\mathbf{x})\right]\right) = \Box^{i}
\left[f(t)\,\partial_L\delta(\mathbf{x})\right]$. Such a
contribution is localized at the spatial origin
$\mathbf{x}=\mathbf{0}$ and thus vanishes outside of the world tube
$r\leq a$ containing the source.

We now transform the Laplacians in
(\ref{Smult1}) into d'Alembertians using
\begin{equation}
\Delta^k=\left(\Box +
\frac{1}{c^2}\partial_t^2\right)^k=
\sum_{j=0}^k\frac{k!}{j!(k- j)!}\,\Box^j
\left(\frac{1}{c^2}\partial_t^2\right)^{k-j}.
\end{equation}
We then arrive at an \textit{irreducible} (STF) decomposition of
the skeletonized source $S$, which is of the type
\begin{equation}\label{Smult}
S^\mathrm{Skel}(\mathbf{x},t)=
\sum_{\ell=0}^{+\infty}\frac{(-)^\ell}{\ell!}
\,\widehat{S}_L(t)\,\partial_L\delta(\mathbf{x})
+\mathcal{O}\left(\Box\,\partial\,\delta\right).
\end{equation}
Here the last term, symbolically denoted
$\mathcal{O}\left(\Box\,\partial\,\delta\right)$, is an (infinite)
sum of terms of the form
$\Box^{i+1}[f(t)\partial_L\delta(\mathbf{x})]$ with $i\geq 0$,
$\ell\geq 0$. As we just said, these terms will not contribute to
the multipole expansion of the field $\varphi(\mathbf{x},t)$,
\textit{i.e.}, considered in the external domain $r >a$.

The most useful result for our purpose is the explicit expression of
the STF moments in Eq.~(\ref{Smult}) which we find to be
\begin{equation}\label{ShatL}
\widehat{S}_L(t)=\int
d^d\mathbf{y}\,\widehat{y}_L\,\overline{S}_\ell(\mathbf{y},t)\,,
\end{equation}
where we have introduced a convenient $\ell$-dependent weighted time
average given by the formal infinite PN series
\begin{equation}\label{Sell}
\overline{S}_\ell(\mathbf{y},t)=\sum_{k=0}^{+\infty}\alpha_\ell^k
\left(\frac{\vert\mathbf{y}\vert}{c}\frac{\partial}{\partial
t}\right)^{2k}S(\mathbf{y},t)\,.
\end{equation}
The coefficients $\alpha_\ell^k$ are those which have been
introduced in Eq.~(\ref{Smult1b}). When written out explicitly, the
``effective'' source $\overline{S}_\ell(\mathbf{y},t)$ reads,
\begin{eqnarray}\label{Sellexpl}
\overline{S}_\ell(\mathbf{y},t)&=&S(\mathbf{y},t)
+\frac{1}{2(2\ell+d)}
\left(\frac{\vert\mathbf{y}\vert}{c}\frac{\partial} {\partial
t}\right)^2S(\mathbf{y},t)+\cdots\\
&+&\frac{1}{(2k)!!(2\ell+d)(2\ell+d+2)\cdots(2\ell+d+2k-2)}
\left(\frac{\vert\mathbf{y}\vert}{c}\frac{\partial}{\partial
t}\right)^{2k}S(\mathbf{y},t)+\cdots\,,\nonumber
\end{eqnarray}
where $(2k)!!\equiv(2k)(2k-2)\cdots(2)$.

Note that the result (\ref{ShatL})--(\ref{Sellexpl}) for the scalar
relativistic multipoles in $d$ dimensions is a remarkably simple
generalization of the $3$-dimensional result obtained in
\cite{BD89}: It is enough to replace the explicit $3$'s, $5$'s
\textit{etc.} appearing in Eq.~(B.14b) of \cite{BD89} by $d$, $d+2$,
\textit{etc.}, without changing anything else. In \cite{BD89} it was
also shown that the expansion (\ref{Sellexpl}) was in $3$ dimensions
the PN expansion of the \textit{exact} result
\begin{subequations}\label{Sellcompact3}\begin{eqnarray}
\overline{S}_\ell^{\,(d=3)}(\mathbf{y},t)&=&\int_{-1}^1 dz
\,\delta_\ell^{(0)} (z)
\,S(\mathbf{y},t+z\vert\mathbf{y}\vert/c)\,,\label{expldelta3}\\
\text{with}\qquad \delta_\ell^{(0)} (z) &\equiv&
\frac{\Gamma\left(\ell+\frac{3}{2}\right)}{
\Gamma\left(\frac{1}{2}\right)\Gamma
\left(\ell+1\right)} \,(1-z^2)^\ell\,, \qquad
\int_{-1}^{1} dz\,\delta_\ell^{(0)}(z) = 1\,.\label{deltal3}
\end{eqnarray}\end{subequations}
The ratio of Gamma functions appearing in Eq.~(\ref{deltal3}) is
equal to $(2\ell+1)!!/(2^{\ell+1} \ell!)$. Note that since the
expansion is purely ``even'' (\textit{i.e.}, with only even powers
of $c^{-1}$), the time argument $t+z\vert\mathbf{y}\vert/c$ in
(\ref{expldelta3}) can be equivalently changed into
$t-z\vert\mathbf{y}\vert/c$.

Correspondingly, one can check that the $d$-dimensional result
(\ref{Sell})--(\ref{Sellexpl}) is the PN expansion of the following
simple generalization of the $3$-dimensional case:
\begin{equation}\label{Sellcompact}
\overline{S}_\ell^{(\varepsilon)}(\mathbf{y},t)=\int_{-1}^1 dz
\,\delta_\ell^{(\varepsilon)} (z)
\,S(\mathbf{y},t+z\vert\mathbf{y}\vert/c)\,,
\end{equation}
where we introduced $\varepsilon\equiv d-3$, and
\begin{equation}\label{deltal}
\delta_\ell^{(\varepsilon)} (z) \equiv
\frac{\Gamma\left(\ell+\frac{3}{2}+\frac{\varepsilon}{2}\right)}{
\Gamma\left(\frac{1}{2}\right)\Gamma
\left(\ell+1+\frac{\varepsilon}{2}\right)}
\,(1-z^2)^{\ell+\frac{\varepsilon}{2}},
\qquad\int_{-1}^{1}
dz\,\delta_\ell^{(\varepsilon)}(z) = 1\,.
\end{equation}
Consistently with what happened in Eq.~(\ref{Sellexpl}), the kernel
$\delta_\ell^{(\varepsilon)} (z)$ is simply obtained from its
$3$-dimensional limit by replacing everywhere $\ell$ by
$\ell+\frac{\varepsilon}{2}$ (\textit{i.e.}, $2\ell$ by $2\ell+d-3$):
\begin{equation}\label{deltaeps0}
\delta_\ell^{(\varepsilon)} (z) =
\delta_{\ell+\frac{\varepsilon}{2}}^{(0)} (z)\,.
\end{equation}
Let us mention in passing that the ``exact'' re-summed expression
(\ref{Sellcompact}) can also be directly derived from the
Fourier-space expression of the $d$-dimensional Green's function.

Finally, having obtained the STF decomposition of the source term
$S^\mathrm{Skel}$ in the form (\ref{Smult}), we obtain the
corresponding expression of the scalar field
$\varphi(\mathbf{x},t)$. As we pointed out above, the remainder term
in Eq.~(\ref{Smult}) does not contribute to the \textit{multipolar
expansion} of the field. Henceforth we shall denote by
$\mathcal{M}(\varphi)$ the multipolar expansion of $\varphi$, which
is therefore given by
\begin{equation}\label{phimult}
\mathcal{M}(\varphi)(\mathbf{x},t)=
\sum_{\ell=0}^{+\infty}\frac{(-)^\ell }{\ell!}
\,\Box_\mathrm{Ret}^{-1}\left(\widehat{S}_L(t)
\,\partial_L\delta(\mathbf{x})\right),
\end{equation}
since the terms $\Box_\mathrm{Ret}^{-1}
\mathcal{O}\left(\Box\,\partial\,\delta\right)$ give zero when
considered outside the compact support of the source. In terms of
the retarded Green's function the latter formula becomes
\begin{equation}\label{phimultG}
\mathcal{M}(\varphi)(\mathbf{x},t)=
\sum_{\ell=0}^{+\infty}\frac{(-)^\ell }{\ell!}
\,\partial_L\left[\int_{-\infty}^{+\infty}ds\,\widehat{S}_L(s)
\,G_\mathrm{Ret}(\mathbf{x},t-s)\right].
\end{equation}
Note that, in view of the retarded nature of the Green function
$G_\mathrm{Ret}(\mathbf{x},t-s)$, the integral is limited to $s<t$,
and even to $s<t-r/c$ with $r\equiv\vert\mathbf{x}\vert$. Equation
(\ref{phimultG}) generalizes what was the basic result for the
multipolar expansion of a $3$-dimensional inhomogeneous wave
equation $\Box^{\,(d=3)}\varphi = S$, namely
\begin{equation}\label{phimultG3}
\mathcal{M}(\varphi)^{\,(d=3)}(\mathbf{x},t)=-\frac{1}{4\pi}
\sum_{\ell=0}^{+\infty}\frac{(-)^\ell}{\ell!}
\,\partial_L\left(\frac{\widehat{S}_L^{\,(d=3)}(t-r/c)}{r}\right).
\end{equation}
A common feature of the result (\ref{phimultG3}) and its
$d$-dimensional generalization (\ref{phimultG}) is that each
``multipolar wave'' of degree $\ell$ is obtained by an $\ell$-tuple
differentiation, with respect to the spatial coordinates, of an
elementary \textit{spherically symmetric} (\textit{i.e.}, monopolar)
retarded solution; indeed, as mentioned above
$G_\mathrm{Ret}(\mathbf{x},t-s)$ depends only on $r$ and $t-s$. In
$3$ dimensions the elementary spherically symmetric retarded
solutions admit a simple expression in terms of the multipole
moments, namely $\widehat{S}_L^{\,(d=3)}(t-r/c)/r$. By contrast, the
$d$-dimensional analogue of each elementary spherically symmetric
solution is a more complicated non-local \textit{functional} of
$\widehat{S}_L(s)$, which involves an integral over its time
argument: $\int_{-\infty}^{t-r/c}ds\,\widehat{S}_L(s)
\,G_\mathrm{Ret}(r,t-s)$. This non-locality in time in the
expression of $\varphi$ in terms of $\widehat{S}_L$ comes in
addition to the non-locality in time entering the exact definition
(\ref{Sellcompact}) of the effective source term
$\overline{S}_\ell^{(\varepsilon)}(\mathbf{y},t)$. The former
non-locality is evidently related to the fact that the ``Huygens
principle'' holds only in $d=3$, $5$, $7$, $\cdots$ dimensions. In
these special dimensions, the support of the retarded Green function
$G_\mathrm{Ret}(r,t-s)$ is concentrated on the past light cone
$s=t-r/c$. On the other hand, in other dimensions (and notably in
dimensionally-continued complex ones) the support of the retarded
Green function $G_\mathrm{Ret}(r,t-s)$ extends over the interior of
the past light cone: $s\leq t-r/c$.

\section{Multipole decomposition of the gravitational
field}\label{secIII}

\subsection{$d$-dimensional generalization of the multipolar
post-Minkowskian formalism}\label{secIIIA}

The calculations of the 3.5PN templates,
Refs.~\cite{B98tail,BIJ02,BFIJ02,BI04mult}, applied the general
expressions of the relativistic multipole moments of
Refs.~\cite{B95,B98mult,PB02}, which are themselves to be inserted
into the (3-dimensional) multipolar post-Minkowskian (MPM) formalism
of Ref.~\cite{BD86}. Let us sketch how one can, in principle,
generalize this MPM formalism to arbitrary dimensions $d$. The basic
building blocks of the MPM formalism are:
\begin{enumerate}
\item[(i)] the parametrization of a general solution of the
linearized vacuum Einstein equations in harmonic coordinates, say
$h^{\mu\nu}$, by means of several sequences of \textit{irreducible}
multipole moments;
\item[(ii)] the definition of an integral operator, called
$\mathop{\mathrm{FP}}\Box^{-1}_\mathrm{Ret}$, which produces, when
it is applied to the non-linear effective MPM source $N_n^{\mu\nu} =
N_n^{\mu\nu}(h_1, h_2, \ldots, h_{n-1})$ appearing at the
$n^\text{th}$ non-linear iteration, a particular non-linear
solution, $p_n^{\mu\nu}$, of the inhomogeneous wave equation $\Box
p_n^{\mu\nu} = N_n^{\mu\nu}$;
\item[(iii)] the definition of a complementary homogeneous solution
$q_n^{\mu\nu}$ ($\Box q_n^{\mu\nu} = 0$) such that
$h_n^{\mu\nu}\equiv p_n^{\mu\nu} + q_n^{\mu\nu}$ satisfies the
harmonicity condition $\partial_\nu h_n^{\mu\nu} = 0$.
\end{enumerate}
Given these building blocks, the MPM formalism generates, by
iteration, a general solution of the non-linear vacuum Einstein
equations as a formal power series, $\sqrt{-g} g^{\mu\nu} =
\eta^{\mu\nu} + G h_1^{\mu\nu} + \cdots + G^n h_n^{\mu\nu} +
\cdots$, this solution being parametrized by the arbitrary ``seed''
multipole moments entering the definition of the first approximation
$h_1^{\mu\nu}$. We briefly indicate how the various building blocks
can be generalized to arbitrary dimensions $d$. We have in mind here
an extension to generic integer dimensions $d>3$, before defining a
formal continuation to complex dimensions. [We consider mainly
larger dimensions $d>3$ because they exhibit generic $d$-dependent
features, while lower integer dimensions, $d=1,2$, exhibit special
phenomena.]

In the previous section we have discussed the multipole expansion of
scalar fields, $\Box\varphi = S$, in arbitrary $d$. We have seen
that the general (retarded) solution outside the source $S$ could be
parametrized, in any $d$, by a set of symmetric trace-free (STF)
time-dependent tensors $\widehat{S}_L(t)$. The situation is somewhat
more complicated for other fields, notably the spin-2 field
$h^{\mu\nu}$ relevant for gravity in any $d$. As we shall discuss in
the next subsection, the multipole moments needed in a generic $d>3$
to parametrize a general gravitational field are more complicated
than what can be used in $d=3$. In $d=3$, one can use two
independent sets of STF tensors, say $M_L$ (the ``mass multipole
moments'') and $S_L$ (the ``spin multipole moments''). In a generic
$d>3$, one has still the analogue of the mass multipole moments,
\textit{i.e.}, STF tensors $M_L$ corresponding to a Young tableau
made of $\ell$ horizontal boxes
(\raisebox{-0.8mm}{\includegraphics[scale=0.5]{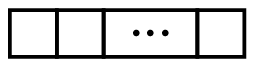}}).
However, the spin multipole moments must be described by a mixed
Young tableau having one vertical column of two boxes and $\ell-1$
complementary horizontal ones --- so that there are $\ell$ boxes on
the upper horizontal row
(\raisebox{-1.8mm}{\includegraphics[scale=0.5]{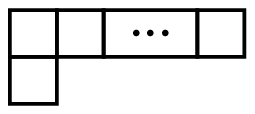}}). In
addition, one must introduce a third type of irreducible
representation of the $d$-dimensional rotation group $O(d)$, namely
a mixed Young tableau having two vertical columns of two boxes and
$\ell-2$ complementary horizontal ones
(\raisebox{-1.8mm}{\includegraphics[scale=0.5]{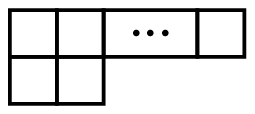}}). For
instance, when $\ell = 2$, this new irreducible representation has
the symmetry of a Weyl tensor in $d$ dimensions:
\raisebox{-1.8mm}{\includegraphics[scale=0.5]{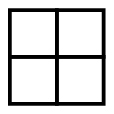}}. As is
well known, this representation does not occur in $d\leq 3$.
However, all these technical complications will have little impact
on what we will need to calculate here. Indeed, as discussed below,
it will be enough for our purpose of unambiguously computing the
3PN-level gravitational radiation emission to deal with the simpler
\textit{mass} multipole moments $M_L$, which admit a uniform
treatment in any dimension $d$ (actually we shall use a specific
definition for what we call the source-type mass multipole moments
and denote them by $I_L$ instead of $M_L$).

Let us turn to the generalization of the integral operator
$\mathop{\mathrm{FP}}\Box^{-1}_\mathrm{Ret}$. In $d=3$, the precise
definition of this operator was the following. Consider a typical
non-linear source generated by the MPM iteration, \textit{e.g.} $N_2
= N_2(h_1,h_1) \sim (\partial h_1)^2 + h_1\partial^2 h_1$, in which
$h_1$ is represented by its multipole expansion.
One formally assumes that the multipole expansion of $h_1 \sim \sum
\partial \left[M_L(t-r/c)/r\right]$ contains a \textit{finite}
number of multipoles. This ensures that the non-linear source
$N_2(h_1)$ is a finite sum of terms of the form $\widehat{n}_K
F(t-r/c)/r^q$, with angular factor
$\widehat{n}_K=\mathrm{STF}\left(n^{i_1}\cdots n^{i_k}\right)$,
$n^i=x^i/r$. We can further expand $F(t-r/c)$ in powers of $r/c$ and
get $N_2(h_1)$ as a sum of terms $\sim \widehat{n}_K F(t)/r^p$.
Though this multipole expansion of $N_2(h_1)$ is only physically
relevant in the region outside the source, say $r>a$, in the MPM
formalism we always mathematically extend its definition (by real,
analytic continuation in $r$) down to $r=0$. Then this formal
construction, $h_\mathrm{MPM}=G\,h_1+ G^2\,h_2+\cdots$, valid by
real analytic continuation for any $r>0$, is \textit{identified}
with the multipolar expansion, say $\mathcal{M}(h)$, of the physical
field $h$. While the physical $h$ takes different expressions inside
($r<a$) and outside ($r>a$) the source, the object
$\mathcal{M}(h)\equiv h_\mathrm{MPM}$ is mathematically defined
everywhere (except at $r=0$) by the same formal expression, but is
physically correct only when $r>a$ (see \cite{B98mult} for the
notation and further discussion).

To deal with the singular behavior near $r=0$ of the non-linear MPM
source terms, \textit{e.g.} $N_2(h_1)\sim \widehat{n}_K F(t)/r^p$,
one introduces a complex number $B$ and considers the action of the
retarded Green operator onto the product of the source by a
``regularization'' factor $(r/r_0)^B$, say
\begin{equation}\label{B1}
F_2^{(d=3)}(B) \equiv \Box_\mathrm{Ret}^{-1}\left[
\left(\frac{r}{r_0}\right)^B N_2(h_1) \right].
\end{equation}
The length scale $r_0$ represents an arbitrary dimensionful
parameter serving the purpose of adimensionalizing the above
regularization factor. It was shown in Ref.~\cite{BD86} that the
integral $F_2(B)$, Eq.~(\ref{B1}), is convergent when the real part
of $B$ is large enough, and that $F_2(B)$, considered as a function
of the complex number $B$, is a \textit{meromorphic} function of
$B$, which has in general (simple) poles at
$B=0$,\footnote{Actually, it was shown in \cite{BD88,B98quad} that
$F_2(B)$ happens to have no pole when $B\rightarrow 0$, due to the
particular structure of the quadratic-order interaction.} coming
{}from the singular behavior of the integrand $N_2(h_1)$ near $r=0$.
[One formally assumes that the multipole moments are
time-independent before some instant $-\mathcal{T}$, and at the end
of the calculation the limit $\mathcal{T} \rightarrow +\infty$ is
taken.] Therefore, the Laurent expansion of $F_2(B)$, near $B=0$, is
of the form
\begin{equation}\label{B2}
F_2^{(d=3)}(B) = \frac{C_{-1}(\mathbf{x},t)}{B} + C_0(\mathbf{x},t)
+ C_1(\mathbf{x},t) B + \mathcal{O}\left(B^2\right).
\end{equation}
One then defines, when $d=3$, the \textit{finite part} at $B=0$ of
$\Box^{-1}_\mathrm{Ret}N_2(h_1)$, denoted
$\mathrm{FP}\,\Box^{-1}_\mathrm{Ret}N_2(h_1)$, as the term
$C_0(\mathbf{x},t)$ in the Laurent expansion of $F_2(B)$. One proves
that $C_0(\mathbf{x},t)$ satisfies the equation $\Box C_0= N_2(h_1)$
and uses it as the ``particular'' second-order contribution
$p_2^{\mu\nu}$ to the second-order metric $h_2^{\mu\nu}$. Let us not
spend time on the construction of the additional homogeneous
contribution $q_2^{\mu\nu}$ necessary to satisfy the harmonicity
condition $\partial_\nu\left(p_2^{\mu\nu}+q_2^{\mu\nu}\right) = 0$
[an example of construction of such contribution will be given in
(\ref{q1munu}) below]. Having so constructed (in $d=3$) the
second-order term in the MPM expansion of the external metric,
$h_2^{\mu\nu} = p_2^{\mu\nu} + q_2^{\mu\nu}$, one continues the
iteration by considering the next order inhomogeneous equation $\Box
h_3 = N_3\left(h_1,h_2\right)$ and introducing $F_3(B) \equiv
\Box_\mathrm{Ret}^{-1}\left[ (r/r_0)^B N_3(h_1,h_2) \right]$. The
singular behavior near $r=0$ of $N_3$ is more complicated (it
contains logarithms of $r$), and, as a consequence, one finds that
$F_3(B)$ is still meromorphic in the complex $B$ plane, but will
contain \textit{double} poles at $B=0$. Again, one defines
$p_3=\mathop{\mathrm{FP}}\Box^{-1}_\mathrm{Ret}N_3$ as the
coefficient of the zero-th power of $B$ in the Laurent expansion of
$F_3(B)$ when $B\rightarrow 0$.

Having recalled the definition and properties of the operation
$\mathop{\mathrm{FP}}\Box^{-1}_\mathrm{Ret}$ in the 3-dimensional
MPM formalism, let us sketch what changes when working in $d$
dimensions. Let us start with the ``seed'' linearized metric
$h_1^{\mu\nu}$. As we see in Eq.~(\ref{phimult}), and will see below
with more details for the tensorial analogue of the scalar multipole
expansion, the multipole expansion $h_1$ is of the form $h_1 \sim
\sum \partial \,\Box_\mathrm{Ret}^{-1}\left[ M_L(t)
\delta(\mathbf{x})\right]$. Though one cannot write, in arbitrary
$d$, a simple, closed-form expression for the object
$\Box_\mathrm{Ret}^{-1}\left[ M_L(t) \delta(\mathbf{x})\right]$, it
is enough to write down its expansion when $r\rightarrow 0$ (which
is in fact the same as its PN expansion). Modulo regular terms near
the origin, this expansion is obtained as
\begin{eqnarray}\label{B3}
\Box_\mathrm{Ret}^{-1}\Bigl[M_L(t) \delta(\mathbf{x})\Bigr] &=&
\left(\Delta^{-1} +\frac{1}{c^2} \,\partial_t^2\Delta^{-2}
+\frac{1}{c^4} \,\partial_t^4\Delta^{-3} +\cdots\right) \Bigl[ M_L(t)
\delta(\mathbf{x})\Bigr] \nonumber\\
&&+~\text{regular terms}.
\end{eqnarray}
Using $\Delta^{-1}\delta(\mathbf{x}) \propto r^{2-d}$,
$\Delta^{-2}\delta(\mathbf{x}) \propto r^{4-d}$ \textit{etc.} we see
that the 3-dimensional form of the expansion of $h_1$ near $r=0$
(after taking into account the expansion of the retardation $r/c$),
takes in $d$ dimensions the form
\begin{equation}\label{B4}
h_1 \sim \sum \frac{\widehat{n}_K F(t)}{r^{p+\varepsilon}}\,,
\end{equation}
where $\widehat{n}_K\equiv \mathrm{STF}[n_{i_1}\cdots n_{i_k}]$,
$p$ is a (relative) integer, and $\varepsilon \equiv d-3$.
Inserting this expansion in the second-order source $N_2(h_1)
\sim \partial h_1 \partial h_1 + h_1 \partial^2 h_1$ yields
\begin{equation}\label{B5}
N_2\left(h_1\right) \sim \sum \frac{\widehat{n}_K
F(t)}{r^{p+2\varepsilon}}\,.
\end{equation}
At this stage, one could consider $\Box_\mathrm{Ret}^{-1} N_2$,
without inserting a factor $(r/r_0)^B$, by using the analytic
continuation in $d$. However, to ensure continuity with what was
done in $3$ dimensions, it is better to insert this factor and to
consider
\begin{equation}\label{B6}
F_2^{(d)}(B) \equiv \Box_\mathrm{Ret}^{-1}\left[
\left(\frac{r}{r_0}\right)^B N_2\left(h_1\right) \right].
\end{equation}
The main difference between (\ref{B6}) and its 3-dimensional
analogue (\ref{B1}) concerns the meromorphic structure of $F_2(B)$.
Indeed, in view of the shift by $+2\varepsilon$ of the integer
exponent $p$ in (\ref{B5}), and of the presence of $r^\varepsilon$
in the $d$-dimensional volume element $d^d \mathbf{x} =
r^{2+\varepsilon} dr \,d\Omega_{2+\varepsilon}$, one easily sees
that the (simple) \textit{poles} in $F_2(B)$ that were located at
$B=0$ when $d=3$ are no longer located at $B=0$ when $d \neq 3$, but
are \textit{shifted} at $B= 2\varepsilon - \varepsilon =
\varepsilon$. Alternatively, this can be explicitly verified by
using the expansion $\Box^{-1}_\mathrm{Ret} = \Delta^{-1} +c^{-2}
\partial_t^2\Delta^{-2} +\cdots$ (plus a regular kernel), and the
formula $\Delta^{-1} r^\alpha = r^{\alpha+2}/[(\alpha+2)(\alpha+d)]$
where the pole at $\alpha = -d$ is the only one which comes from the
ultraviolet (UV) behavior $r\rightarrow 0$. As a consequence, the
expansion (\ref{B2}) is now modified to
\begin{equation}\label{B7}
F_2^{(d)}(B) = \frac{C_{-1}^{(d)}(\mathbf{x},t)}{B-\varepsilon} +
C_0^{(d)}(\mathbf{x},t) + C_1^{(d)}(\mathbf{x},t) B +
\mathcal{O}\left(B^2\right).
\end{equation}
This expansion, and its analogues considered below, is considered
for $\varepsilon$ and $B$ both small (so that the expansion in
powers of $B$ makes sense), but without assuming any relative
ordering between the smallness of $B$ and that of $\varepsilon$. One
should neither re-expand $(B-\varepsilon)^{-1}$ in powers of
$B/\varepsilon$ nor in powers of $\varepsilon/B$.

Having in hand the above structure, one then \textit{defines} the
$d$-dimensional generalization of the finite part of $N_2(h_1)$ as
the coefficient of $B^0$ in Eq.~(\ref{B7}), namely $C_0^{(d)}$. We
denote such a finite part by
\begin{subequations}\label{B8}\begin{eqnarray}
\mathop{\mathrm{FP}}_B\Box^{-1}_\mathrm{Ret} \Bigl[\widetilde{r}^B
N_2\left(h_1\right)\Bigr] &\equiv&
C_0^{(d)}(\mathbf{x},t)\,,\\
\text{where}~~\widetilde{r}&\equiv&\frac{r}{r_0}\,,
\label{rtilde}
\end{eqnarray}\end{subequations}
or, more simply, by
\begin{equation}\label{B8prime}
\mathrm{FP}\,\Box^{-1}_\mathrm{Ret} N_2\left(h_1\right) \equiv
\mathrm{FP}\,F_2^{(d)} \equiv C_0^{(d)}(\mathbf{x},t)\,.
\end{equation}
Note the subtlety that the expansion (\ref{B7}) is neither a Laurent
expansion in powers of $B-\varepsilon$, nor a Laurent expansion in
powers of $B$. After subtracting the \textit{shifted} pole terms
$\propto (B-\varepsilon)^{-1}$, one expands the remainder in a
regular Taylor series in powers of $B$. The interest of this
specific definition is the fact that it ensures that
$C_0^{(d)}(\mathbf{x},t)$ is an exact solution of the equation we
initially wanted to solve, namely
\begin{equation}
\label{B9}
\Box^{(d)} C_0^{(d)} = N_2\left(h_1\right).
\end{equation}
Indeed, by its mere definition (\ref{B6}), one has $\Box
F_2^{(d)}(B) = (r/r_0)^B N_2(h_1)$. Comparing this result (which has
no pole) to the application of $\Box$ to (\ref{B7}), we first see
that the pole part must be a homogeneous solution, $\Box
C_{-1}^{(d)} = 0$. Then, identifying the successive powers of $B$
(using ${\widetilde{r}}^B = e^{B \ln\widetilde{r}}= 1 + B
\ln\widetilde{r} + \cdots$), yields $\Box C_0^{(d)} = N_2$, $\Box
C_1^{(d)} = \ln(r/r_0) N_2$, and so on. Another useful property of
the $d$-modified definition (\ref{B8}) is that it automatically
ensures the continuity between $d\rightarrow 3$ and $d=3$. Indeed,
the shift in the location of the pole in (\ref{B7}) was made to
``follow'' the pole that existed at $B=0$ when $d=3$. Therefore we
have $\lim_{d\rightarrow 3} C_{-1}^{(d)} = C_{-1}$, and similarly
$\lim_{d\rightarrow 3} C_0^{(d)} = C_0$, \textit{etc.}, where the
RHSs are those defined in Eq.~(\ref{B2}) when $d=3$.

The extension of the iteration to higher non-linear orders
introduces a new subtlety. Indeed, let us look more precisely at the
structure of the second-order contribution to the metric, $h_2 =
p_2+q_2$ where, as we said, the \textit{particular} solution $p_2$
is defined by the modified FP process: $p_2
\equiv\mathrm{FP}\,\Box^{-1}_\mathrm{Ret} N_2(h_1)$, and where $q_2$
is a complementary homogeneous solution. Most of the terms in the
integrand $N_2$ introduce no poles, and, for them, we simply find a
structure of the type $p_2^\mathrm{(no \, pole)} \sim \sum
r^{-p-2\varepsilon}$ (for simplicity, we henceforth suppress angular
factors). Let us now consider the terms in $N_2$ that generate
poles $\propto (B-\varepsilon)^{-1}$ in $F_2(B)$.\footnote{As said
above, these terms actually cancel among themselves because of the
particular structure of $N_2$. However, similar terms appear at
higher iteration orders, and their general structure is simpler to
describe if we start our induction reasoning at the quadratically
non-linear level.} We know that such terms introduce, when $d=3$,
some logarithms of the radial variable $r$. When $d\neq 3$, they no
longer introduce logarithms but they introduce a further technical
complication. Indeed, let us look at a typical example, namely a
dangerous term in $F_2(B)$ of the form $F_2^\mathrm{(pole)}(B) =
\Delta^{-1}(r^{B-3-2\varepsilon})$. Suppressing for simplicity a
factor $(B-1-2\varepsilon)^{-1}$ which is jointly analytic in $B$
and $\varepsilon$ near $B = 0$ and $\varepsilon = 0$ respectively
and therefore creates no problem, we have essentially
$F_2^\mathrm{(pole)}(B) = (B-\varepsilon)^{-1} r^{B-1-2\varepsilon}$.
According to Eq.~(\ref{B7}) the pole part of $F_2^\mathrm{(pole)}$
that we must subtract is, for instance, obtained by multiplying by
$B-\varepsilon$ and then taking the limit $B\rightarrow \varepsilon$
(and \textit{not} $B\rightarrow 0$). This pole part is therefore
given by $(B-\varepsilon)^{-1} r^{-1-\varepsilon}$. The finite part
of $F_2^\mathrm{(pole)}(B)$ is then obtained by subtracting the pole
part and taking the limit $B\rightarrow 0$; this yields
\begin{equation}\label{B10}
p_2^\mathrm{(pole)} = \mathrm{FP}\,F_2^\mathrm{(pole)} =
\frac{1}{\varepsilon}\left[ r^{-1-\varepsilon}-r^{-1-2\varepsilon}
\right].
\end{equation}
The subtlety is that there seems to appear poles in
$\varepsilon^{-1}$. However, the residue of the pole vanishes, since
the limit $\varepsilon \rightarrow 0$ of (\ref{B10}) is finite and
generates the logarithm that we know to exist in $d=3$,
$p_2^\mathrm{(pole)} \sim \ln r/r$. If we do not take the limit
$\varepsilon \rightarrow 0$, we must keep the structure (\ref{B10})
and see what it generates at the next, cubic, order of iteration. In
addition, we must also add the complementary solution $q_2$ needed
to satisfy the harmonicity condition,
$\partial_\nu(p_2^{\mu\nu}+q_2^{\mu\nu}) =0$. As the calculation of
$q_2$ could be done in $d=3$ without encountering poles (see
Ref.~\cite{BD86}), it will clearly not create problems in $d\neq 3$
apart from the fact that $q_2$, being a homogeneous solution $\Box
q_2 = 0$, will behave near $r=0$ essentially like $h_1$,
\textit{i.e.}, $q_2\sim \sum r^{-p-\varepsilon}$ (which differs from
most of the terms of $p_2$ which were $\sum r^{-p-2\varepsilon}$).

Summarizing so far, the second-order MPM iteration $h_2$ has a
structure, near $r=0$, of the symbolic form
\begin{equation}\label{B11}
h_2 \sim \sum c_1(\varepsilon) \,r^{-p-\varepsilon} +
c_2(\varepsilon) \,r^{-p-2\varepsilon}
+\frac{c_3(\varepsilon)}{\varepsilon} \left[r^{-p-\varepsilon} -
r^{-p-2\varepsilon} \right],
\end{equation}
where the $c_i(\varepsilon)$'s are analytic at $\varepsilon = 0$,
and where we explicitly separated the semi-singular structure in
$\varepsilon$. When inserting the structure (\ref{B11}) into
$N_3(h_1,h_2)$, one finds that the singular behavior of $N_3$ near
$r=0$ can generate several types of singularities in $B$ and
$\varepsilon$. There are simple poles $\propto (B-\varepsilon)^{-1}$
and simple poles $\propto (B-2\varepsilon)^{-1}$, which are natural
generalizations of the structures that generated simple poles in
$F_2(B)$. When looking at the effect of the more complicated
structure given by the third term on the RHS of (\ref{B11}), one
finds that it is best described as generating some ``quasi-double
poles'', namely terms $\propto (B-\varepsilon)^{-1}
(B-2\varepsilon)^{-1}$. The point is that if one were to expand this
term in simple poles with respect to $B$, namely
\begin{equation}\label{B12}
\frac{1}{(B-\varepsilon)(B-2\varepsilon)} =
\frac{1}{\varepsilon(B-2\varepsilon)} -
\frac{1}{\varepsilon(B-\varepsilon)}\,,
\end{equation}
it would seem to involve poles in $1/\varepsilon$. However, all such
poles are ``spurious'' because the source of the trouble which is the
last term in (\ref{B11}) had a finite limit as $\varepsilon
\rightarrow 0$, and because one can easily see that, in our
above-defined MPM algorithm, source terms having a finite limit as
$\varepsilon \rightarrow 0$ generate solutions having also a finite
limit as $\varepsilon \rightarrow 0$.

Finally we find, by induction, that at each iteration order $n$
one has the structure
\begin{equation}\label{B11prime}
h_n \sim \sum d_1(\varepsilon) \,r^{-p-\varepsilon} +
d_2(\varepsilon) \,r^{-p-2\varepsilon} + \cdots +
d_n(\varepsilon)\,r^{-p-n\varepsilon}\,,
\end{equation}
where the coefficients $d_i(\varepsilon)$ might individually have
(simple or multiple) poles in $\varepsilon$, \textit{e.g.}
$d_i(\varepsilon)=c_i(\varepsilon)/\varepsilon^j$, but which always
compensate each other in the complete sum $h_n$. Then we obtain that
the integral
\begin{equation}\label{B13}
F_n^{(d)}(B) \equiv \Box_\mathrm{Ret}^{-1}\Bigl[ \widetilde{r}^B
N_n\left(h_1,\cdots, h_{n-1}\right)\Bigr]
\end{equation}
will have an expansion, near $B=0$, of the generic form\footnote{Our
notation is a little bit over simplified, since the coefficients
$C_{-k}^{(d)}$ depend in fact on a set of integers
$\{q_1,\cdots,q_k\}$. Also we do not indicate the obvious dependence
of the coefficients on the iteration order $n$.}
\begin{equation}\label{B14}
F_n^{(d)}(B) = \sum
\frac{C_{-k}^{(d)}(\mathbf{x},t)}{(B-q_1\varepsilon)
(B-q_2\varepsilon)\cdots(B-q_k\varepsilon)} +
C_{0}^{(d)}(\mathbf{x},t) + C_{1}^{(d)}(\mathbf{x},t) B
+\mathcal{O}\left(B^2\right).
\end{equation}
The ``quasi-multiple poles'' which constitute the first term on the
RHS have $k\leq n-1$ and $1\leq q_i\leq n-1$. As we have seen in
(\ref{B11}) and (\ref{B11prime}), the poles in $1/\varepsilon$ are
in fact spurious, as they have a residue which is always zero.
[We assume here that the seed multipole moments are regular as
$\varepsilon \rightarrow 0$.] So, when writing the result in the
form of (\ref{B14}), we note that the coefficients $C_{-k}^{(d)}$,
$C_0^{(d)}$, \textit{etc.}, are all regular when $d\rightarrow 3$.
One then defines the $d$-dimensional generalization of the finite
part of $F_n^{(d)}$ as being the coefficient of $B^0$ in the
expansion (\ref{B14}):
\begin{equation}\label{B15}
p_n = \mathrm{FP}\,F_n^{(d)}\equiv C_{0}^{(d)}(\mathbf{x},t)\,.
\end{equation}
This coefficient is regular when $\varepsilon \equiv d-3\rightarrow
0$, though it contains apparently singular terms of the type of the
last term on the RHS of (\ref{B11}). Moreover, using the same
reasoning as above, one finds that it satisfies the needed result:
$\Box p_n = N_n$. Note finally that, when $\varepsilon\rightarrow
0$, the quasi-multiple poles in (\ref{B14}) merge together to form
the multiple poles $\propto B^{-k}$, with $k\leq n-1$, that were
found to exist in $d=3$ \cite{BD86}. On the other hand, when
$\varepsilon \neq 0$, the poles form a ``line'' of \textit{simple}
poles located at $B=\varepsilon$, $B=2\varepsilon$, $\cdots$, $B =
(n-1)\varepsilon$. However, it is better not to decompose the
product of simple poles entering (\ref{B14}) in sum of separate
simple poles, because this decomposition would, as in
Eq.~(\ref{B12}), introduce spurious singularities $\propto
\varepsilon^{-j}$.

The main practical outcome of the present subsection is the modified
definition of the operation $\mathrm{FP}\,\Box^{-1}_\mathrm{Ret}$
when working in $d\neq 3$: namely as the coefficient of $B^0$ in an
expansion of the type (\ref{B14}) where, after separating the
shifted poles at $B-\varepsilon$, $\cdots$, $B-(n-1)\varepsilon$,
one expands the remainder in a Taylor series in powers of $B$. Note
that a simple consequence of this definition is that, for instance,
a term of the form $B/(B-q\varepsilon)$ in
$\Box^{-1}_\mathrm{Ret}\left[\widetilde{r}^B N_n\right]$ gives rise
to a finite part equal to 1. Indeed,
\begin{equation}\label{B16}
\mathrm{FP}\left[ \frac{B}{B-q\varepsilon}\right] =
\mathrm{FP}\left[
\frac{B-q\varepsilon+q\varepsilon}{B-q\varepsilon}\right] =
\mathrm{FP}\left[ \frac{q\varepsilon}{B-q\varepsilon} +1\right] =
1\,.
\end{equation}
One might have been afraid that a term of this type,
$B/(B-q\varepsilon)$, could have been ambiguous because, ultimately,
we are sending both $B$ and $\varepsilon$ towards zero without
fixing an ordering between the two limits. However, in the present
$d$-dimensional generalization of the MPM formalism, everything is
precisely defined and unambiguous.

\subsection{Multipolar decomposition of the gravitational field in
$d$ dimensions}\label{secIIIB}

We now sketch the $d$-dimensional generalization of the results
concerning the matching between the MPM exterior metric and the
inner field of a general post-Newtonian (PN) matter system. To start
with, we consider the case of a smooth matter distribution, and will
later allow the matter stress-energy tensor to tend to a
distribution localized on some world-lines. The next subsection will
be devoted to the $d$-dimensional definition of the source multipole
moments. The investigations of this and the next subsection are
based on the works \cite{B95,B98mult,PB02} which derived the
expressions of the source multipole moments of a general PN source,
up to any PN order (in $3$ dimensions). Early derivations of the
relativistic moments, valid up to 1PN order, can be found in
Refs.~\cite{BD89,DI91a}.

We look for a solution, in the form of a PN expansion, of the
$d$-dimensional Einstein field equations. As before we choose some
harmonic coordinates, which means that $\partial_\nu h^{\mu\nu}=0$
where the so-called ``gothic'' metric deviation reads $h^{\mu\nu}
\equiv \sqrt{-g}\, g^{\mu\nu}-\eta^{\mu\nu}$, where $g$ is the
determinant and $g^{\mu\nu}$ the inverse of the usual covariant
metric $g_{\rho\sigma}$. Then the Einstein field equations, relaxed
by the harmonic coordinate conditions, take the form of some
``scalar'' wave equations, similar to (\ref{boxphi}), for each of the
components of $h^{\mu\nu}$,
\begin{equation}\label{EE}
\Box h^{\mu\nu}=\frac{16\pi \,G}{c^4}\,\tau^{\mu\nu}\,,
\end{equation}
where $\Box$ denotes the $d$-dimensional flat space-time wave
operator, and $G$ is the $d$-dimensional Newton constant related to
the usual Newton constant $G_N$ in $3$ dimensions by Eq.~(\ref{GN})
below. The main contribution we shall add in the present subsection,
with respect to our investigation of the scalar wave equation in
Sec.~\ref{secII}, is how to deal with the crucial
\textit{non-linear} gravitational source term in the Einstein field
equations, which makes the RHS of Eq.~(\ref{EE}) to have a support
which is spatially \textit{non-compact}. The RHS of (\ref{EE})
involves what can be called the total stress-energy pseudo tensor of
the non-gravitational and gravitational fields, given by
\begin{eqnarray}\label{tau}
\tau^{\mu\nu} = \vert g\vert T^{\mu\nu}+\frac{c^4}{16\pi
G}\Lambda^{\mu\nu}\bigl(h,\partial h,\partial^2 h\bigr),
\end{eqnarray}
where $T^{\mu\nu}$ is the matter stress-energy tensor, and the
second term represents the gravitational stress-energy distribution,
which can be expanded into non-linearities according to\footnote{In
the MPM formalism of Sec.~\ref{secIIIA}, we used $N_2(h_1) =
\Lambda_2(h_1,h_1)$, $N_3(h_1,h_2) = \Lambda_3(h_1,h_1,h_1) +
\Lambda_2(h_1,h_2) + \Lambda_2(h_2,h_1)$, and so on.}
\begin{eqnarray}\label{Lambda}
\Lambda^{\mu\nu} =
\Lambda_2^{\mu\nu}(h,h)+\Lambda_3^{\mu\nu}(h,h,h)+\cdots\,,
\end{eqnarray}
where the quadratic, cubic, \textit{etc.}, pieces admit symbolic
structures such as $\Lambda_2\sim h\,\partial^2h+\partial h\,\partial
h$ and $\Lambda_3\sim h\,\partial h\,\partial h$.

The solution $h^{\mu\nu}$ of the field equations we consider in this
subsection will be smooth and valid everywhere, inside as well as
outside the matter source localized in the domain $r\leq a$. Inside
the source, or more generally inside the source's near zone
($r\ll\lambda$, where $\lambda$ is the wavelength of the emitted
radiation), $h^{\mu\nu}$ will admit a PN expansion, denoted here as
$\overline{h}^{\mu\nu}$. On the other hand, in the exterior of the
source, $r>a$, $h^{\mu\nu}$ will admit a multipolar expansion,
solution of the vacuum field equations outside the source, and
decomposed into ($d$-dimensional) irreducible spherical harmonics.
As usual, the definition of the multipole expansion is extended by
real analytic continuation in $r$ to any value $r>0$. It will be
necessary to introduce the special notation
$\mathcal{M}(h^{\mu\nu})$ to mean the multipole expansion of
$h^{\mu\nu}$. As we already mentioned, the multipole expansion in
the present formalism is given by the MPM metric of
Sec.~\ref{secIIIA}, which is therefore in the form of a formal
infinite post-\textit{Minkowskian} series up to any order $n$,
\begin{equation}\label{multMPM}
\mathcal{M}(h^{\mu\nu})\equiv h_\mathrm{MPM}^{\mu\nu}\,.
\end{equation}
As mentioned above, though the identification (\ref{multMPM})
is only physically meaningful in the exterior domain $r>a$, it
can be mathematically extended down to any $r>0$ by real
analytic continuation in $r$.

In this subsection we shall show how to relate in $d$ dimensions the
multipolar expansion (\ref{multMPM}) to the properties of the matter
source, in the case of a PN source (\textit{i.e.}, one which is
located deep inside its own near zone, $a\ll\lambda$). Actually, the
derivation below will be a simple $d$-dimensional adaptation of the
proof given in the case of $3$ dimensions in Ref.~\cite{B98mult}
(see notably the Appendix A there).

The heart of the method is to show that one can deal with the
presence of non-compact-support source terms on the RHS of the field
equation (\ref{EE}), by considering a certain quantity
$\Delta^{\mu\nu}$ which satisfies a wave equation whose source does
have a compact support, and thus, whose multipolar expansion can be
computed by using the results of Sec.~\ref{secII} (for each
space-time component $\mu\nu$). This quantity is defined by
\begin{equation}\label{Delta}
\Delta^{\mu\nu}\equiv h^{\mu\nu}-\mathop{\mathrm{FP}}_{B}\,
\Box^{-1}_\mathrm{Ret} \Bigl[ \widetilde{r}^B
\mathcal{M}(\Lambda^{\mu\nu})\Bigr].
\end{equation}
The second term in (\ref{Delta}), that we thus \textit{subtract}
{}from $h^{\mu\nu}$ in order to define this quantity, involves the
finite part operation FP in $d$-dimensions which has been defined in
the previous subsection (\ref{secIIIA}). It contains the
regularization factor $\widetilde{r}^B\equiv (r/r_0)^B$. The use of
the operator $\mathrm{FP}\,\Box^{-1}_\mathrm{Ret}$ is consistent
with Sec.~\ref{secIIIA} because it acts on the \textit{multipole
expansion} of the non-linear source term
$\mathcal{M}(\Lambda^{\mu\nu})$, which is in fact identical to the
formal post-Minkowskian (MPM) infinite series
$\Lambda_\mathrm{MPM}^{\mu\nu}$, \textit{cf.} Eq.~(\ref{multMPM}).
The meaning of the last term on the RHS of (\ref{Delta}) is that
$\mathrm{FP}\,\Box^{-1}_\mathrm{Ret}$ is to be applied to each term
of the MPM expansion of $\widetilde{r}^B
\mathcal{M}(\Lambda^{\mu\nu})$, and that we then consider the formal
summation of this MPM series.

Equation (\ref{Delta}) appears to be the difference\footnote{This
``difference'' has of course nothing to do with the difference
between the dimensional and Hadamard regularizations that we
consider in Sec.~\ref{secV}.} between the solution of the field
equation (\ref{EE}) and the contribution coming only from the
non-linear terms in the exterior of the compact-support source (and
then analytically continued down to $r=0$). Since $h^{\mu\nu}$ is
the retarded integral of the pseudo tensor $\tau^{\mu\nu}$, and
since the multipole expansion of the matter tensor is formally zero:
$\mathcal{M}(T^{\mu\nu})=0$ (because $T^{\mu\nu}$ has a compact
support), we can rewrite (\ref{Delta}) as
\begin{equation}\label{Delta2}
\Delta^{\mu\nu} = \frac{16\pi \,G}{c^4}\biggl\{\Box^{-1}_\mathrm{Ret}
\tau^{\mu\nu}-\mathop{\mathrm{FP}}_{B}\,\Box^{-1}_\mathrm{Ret}
\Bigl[\widetilde{r}^B \mathcal{M} (\tau^{\mu\nu}) \Bigr]\biggr\}.
\end{equation}
Next, we remark that the first term in (\ref{Delta2}) is regular
within the source (for $r\leq a$), and that we can therefore add to
it the same FP procedure as in the second term, without changing its
value --- because for regular sources, the operator $\mathrm{FP}\,
\Box^{-1}_\mathrm{Ret}$ simply gives back the usual retarded
integral. Thus,
\begin{equation}\label{Delta3}
\Delta^{\mu\nu}=\frac{16\pi \,G}{c^4}\mathop{\mathrm{FP}}_{B}\,
\Box^{-1}_\mathrm{Ret} \Bigl[ \widetilde r^B
\Bigl(\tau^{\mu\nu}-\mathcal{M}(\tau^{\mu\nu})\Bigr)\Bigr].
\end{equation}
As we said, the multipole moment formalism we are using is defined
for general \textit{smooth} matter distributions [say $T^{\mu\nu}\in
C^{\infty}(\mathbb{R}^d)$] with compact support. Hence,
$\tau^{\mu\nu}$ is regular inside the source, and
$\Box^{-1}_\mathrm{Ret}\tau^{\mu\nu}$ is a perfectly well-defined
object. Only when general formulas for the multipole moments are in
hand, shall we apply them to point particles (in Sec.~\ref{secV}),
and then shall we need a self-field regularization scheme to cure
the divergencies induced by the point-particle model. Of course the
FP procedure used here should be carefully distinguished from the
self-field regularization.

The point is that $\Delta^{\mu\nu}$, in the form given by
Eq.~(\ref{Delta3}), appears now as the retarded integral of a source
with \textit{compact} support (limited to $r\leq a$). This follows
{}from the fact that $\tau^{\mu\nu}$ agrees numerically with its own
multipole expansion $\mathcal{M}(\tau^{\mu\nu})$ in the exterior of
the source, for $r>a$. Hence we are allowed to use the end results
of Sec.~\ref{secII} which applied to compact-support sources (and
those results can evidently be ``uniformly'' applied to all the
components of $\Delta^{\mu\nu}$). From (\ref{phimultG}) and
(\ref{ShatL})--(\ref{Sellcompact}) we obtain
\begin{equation}\label{MDelta}
\mathcal{M}(\Delta^{\mu\nu})=\frac{16\pi \,G}{c^4}
\sum_{\ell=0}^{+\infty}
\frac{(-)^\ell}{\ell!}\,\partial_L\left[
\int_{-\infty}^{+\infty}ds\,\widehat{\mathcal{F}}_L^{\mu\nu}(s)
\,G_\mathrm{Ret}(\mathbf{x},t-s)\right],
\end{equation}
where the multipole-moment functions read
\begin{equation}\label{calF0initial}
\widehat{\mathcal{F}}_L^{\mu\nu}(t)=\mathop{\mathrm{FP}}_{B}\int
d^d\mathbf{y}\,\vert\widetilde{\mathbf{y}}\vert^B\,
\widehat{y}_L\,\int_{-1}^1
dz \,\delta_\ell^{(\varepsilon)} (z)
\,\Bigl[\tau^{\mu\nu}-\mathcal{M}(\tau^{\mu\nu})
\Bigr](\mathbf{y},t+z\vert\mathbf{y}\vert/c)\,.
\end{equation}
A difference with the multipole moments considered in
Sec.~\ref{secII} is the presence of the FP process with
regularization factor
$\vert\widetilde{\mathbf{y}}\vert^B=\vert\mathbf{y}/r_0\vert^B$. We
shall see later the crucial role played by this FP process.

Now we deal with the integrand
$\tau^{\mu\nu}-\mathcal{M}(\tau^{\mu\nu})$ appearing in the
multipole moments (\ref{calF0initial}), following the same argument
as in Ref.~\cite{B98mult}. Such an integrand has a compact support
limited to $r\leq a$, so we see that in the case of a PN source, for
which $a\ll\lambda$, we can replace it with its formal PN expansion,
because precisely the PN source is confined within the source's near
zone. Hence the PN-expanded moments will be generated by the
PN-expanded integrand
$\overline{\tau}^{\mu\nu}-\overline{\mathcal{M}(\tau^{\mu\nu})}$,
where we denote the formal PN expansion with an overline.

Let us now show that the second term,
$\overline{\mathcal{M}(\tau^{\mu\nu})}$, gives no contribution to
the PN moments. We know that the structure of this term (in $d =
3+\varepsilon$ dimensions) reads
\begin{equation}\label{struct}
\overline{\mathcal{M}(\tau^{\mu\nu})}= \sum
\frac{\widehat{n}_K\,F(t)}{r^{p+q\varepsilon}}\,,
\end{equation}
where $p$ and $q$ are relative integers. This follows from
Eq.~(\ref{B11prime}) above. The argument showing the vanishing of
the term involving $\overline{\mathcal{M}(\tau^{\mu\nu})}$ is that
any term of the type (\ref{struct}) in the moment will ultimately
give [after taking the PN expansion like in (\ref{Sellexpl})] a
spatial integral of the type $\int
d^d\mathbf{x}\,\widehat{n}_K\,r^{B-p'-q'\varepsilon}$ say (times
some function of time), which we know to be exactly \textit{zero} by
analytic continuation in $B$. Therefore, following this argument,
which is in fact the same in $d$ dimensions as in $3$ dimensions, we
are led \textit{in fine} to a PN multipole moment which is simply
generated by the \textit{PN expansion} of the (non-compact-support)
pseudo tensor, $\overline{\tau}^{\mu\nu}$. Hence, we write our
result as
\begin{equation}\label{calF0}
\widehat{\mathcal{F}}_L^{\mu\nu}(t)=\mathop{\mathrm{FP}}_{B}\int
d^d\mathbf{y}\,\vert\widetilde{\mathbf{y}}\vert^B\,
\widehat{y}_L\,\overline{\tau}_{[\ell]}^{\mu\nu}(\mathbf{y},t)\,,
\end{equation}
where the $\ell$-dependent integrand takes the form of the following
PN expansion,
\begin{eqnarray}\label{calF1}
\overline{\tau}_{[\ell]}^{\mu\nu}(\mathbf{y},t)&=&\int_{-1}^1 dz
\,\delta_\ell^{(\varepsilon)} (z)
\,\overline{\tau}^{\mu\nu}(\mathbf{y},t+z\vert\mathbf{y}\vert/c)
\nonumber\\
&=&\sum_{k=0}^{+\infty}\alpha_\ell^k
\left(\frac{\vert\mathbf{y}\vert}{c}\frac{\partial}{\partial
t}\right)^{2k}\overline{\tau}^{\mu\nu}(\mathbf{y},t)\,.
\end{eqnarray}
The PN coefficients $\alpha_\ell^k$ have been given in
(\ref{Smult1b}). Note that the final result in (\ref{calF1})
combines two separate PN expansions: (i) a PN expansion of the type
(\ref{Sell}) (already indicated by an overline notation), and (ii)
the usual PN expansion of $\tau^{\mu\nu}$. The presence of these PN
expansions is crucial to the meaning and validity of the final
expression in (\ref{calF1}). Finally, note that our use (in the
proof above) of the vanishing of the spatial integrals $\int
d^d\mathbf{x}\,\widehat{n}_K\,r^{B-p'-q'\varepsilon}$ implies that
we have transformed the role of the factor
$\vert\widetilde{\mathbf{y}}\vert^B$ from that of regularizing
integrals that are singular at $r=0$, into that of regularizing
integrals that are singular at $r= \infty$. Thereby, in the final
result (\ref{calF1}), the FP procedure is used as a regularization
of the boundary \textit{at infinity} of the integral, which would
otherwise be divergent because of the multipolar factor
$\widehat{y}_L\sim \vert\mathbf{y}\vert^\ell$ multiplying the
non-compact-support (and PN-expanded) $\overline{\tau}^{\mu\nu}$.

\subsection{Symmetric-trace-free source multipole moments
in $d$ dimensions}\label{secIIIC}

In Eq.~(\ref{MDelta}) we have represented the quantity
$\Delta^{\mu\nu}$, Eq.~(\ref{Delta}), in the form of an infinite
superposition of \textit{scalar} multipolar waves, say $\partial_L
\widetilde{\mathcal{F}}_L^{\mu\nu}$ where we associate to any
function of time $\widehat{\mathcal{F}}_L^{\mu\nu}(t)$ a
corresponding spherically symmetric retarded wave
denoted\footnote{In $3$ dimensions we recover
$\widetilde{\mathcal{F}}_L^{\mu\nu}(r,t)=
\widehat{\mathcal{F}}_L^{\mu\nu}(t-r/c)/r$. Recall that in any
dimension the Green function $G_\mathrm{Ret}(\mathbf{x},t)$ is in
fact a function of $r=\vert\mathbf{x}\vert$ and $t$.}
\begin{equation}\label{FLtilde}
\widetilde{\mathcal{F}}_L^{\mu\nu}(r,t) \equiv -4\pi
\int_{-\infty}^{+\infty}ds\,\widehat{\mathcal{F}}_L^{\mu\nu}(s)
\,G_\mathrm{Ret}(\mathbf{x},t-s),
\end{equation}
in which the tensor indices $\mu\nu$ play the role of simple
``spectators''. This expansion is not yet a genuine
\textit{irreducible tensorial} multipole expansion. To transform
Eq.~(\ref{MDelta}) in a tensor multipole expansion, we need to
decompose each ``elementary wave''
$\widetilde{\mathcal{F}}_L^{\mu\nu}$ in \textit{irreducible}
representations of the $d$-dimensional rotation group $O(d)$. As
each (undifferentiated) elementary wave
$\widetilde{\mathcal{F}}_L^{\mu\nu}(r,t)$, Eq.~(\ref{FLtilde}), is
spherically symmetric, the problem of decomposing
$\widetilde{\mathcal{F}}_L^{\mu\nu}(r,t)$ in irreducible components
is reduced to the purely algebraic problem of decomposing its
``source'' $\widehat{\mathcal{F}}_L^{\mu\nu}(t)$, whose expression
is given by Eqs.~(\ref{calF0})--(\ref{calF1}), in irreducible
representations of $O(d)$.

Let us consider in turn the various components of
$\widehat{\mathcal{F}}_L^{\mu\nu}(t)$. The time-time component
$\widehat{\mathcal{F}}_L^{00}$ is already put in irreducible form
because it is STF with respect to the multi-index $L$. In the
language of Young tableaux [for $O(d)$], the STF-$\ell$
representation carried by $\widehat{\mathcal{F}}_L^{00}$ is denoted
by $\ell$ horizontal boxes
\raisebox{-0.8mm}{\includegraphics[scale=0.5]{young1.ps}}. The
time-space component $\widehat{\mathcal{F}}_L^{0i}$ is,
algebraically, the product of an irreducible vector representation
$V^i$ and of an irreducible STF-$\ell$ one $T_L$. In Young tableaux
terms, this corresponds to the product
\raisebox{-0.8mm}{\includegraphics[scale=0.5]{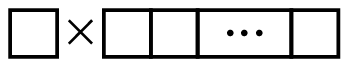}}. In any
dimension $d$, this product gives rise to three irreducible
representations: the STF-$(\ell+1)$ one
\raisebox{-0.8mm}{\includegraphics[scale=0.5]{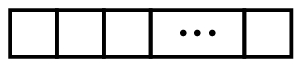}}, the
STF-$(\ell-1)$ one
\raisebox{-0.8mm}{\includegraphics[scale=0.5]{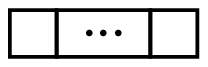}}, and a
\textit{mixed Young tableau} representation
\raisebox{-1.8mm}{\includegraphics[scale=0.5]{young5.ps}}. The first
two representations are easily understood as corresponding to the
STF projection $V_{\langle i} T_{L\rangle}$ of the product $V_i
T_L$, and the contraction $V_a T_{a L-1}$. It is more intricate to
write explicitly the mixed-Young-tableau representation contained in
$V_i T_L$. When $d=3$, it was convenient to use the Levi-Civita
antisymmetric tensor $\varepsilon_{ijk}$ to dualize the two
antisymmetric indices in the mixed Young tableau (\textit{i.e.}, the
two vertical boxes) and replace them by a single vector index.
However, when considering a generic value for the dimension $d$
(initially taken as an integer, and then formally continued to
arbitrary complex values), one is not allowed to use the
specifically 3-dimensional tensor $\varepsilon_{ijk}$. When further
considering the space-space component
$\widehat{\mathcal{F}}_L^{ij}$, one is facing the algebraic problem
of decomposing the product\hspace{-3mm}
\raisebox{-1.4mm}{\includegraphics[scale=0.5]{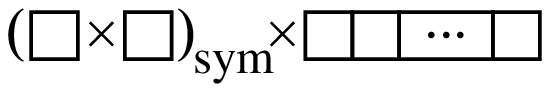}}. Again the
irreducible decomposition of this product contains both relatively
simple \textit{symmetric} representations, such as
\raisebox{-0.8mm}{\includegraphics[scale=0.5]{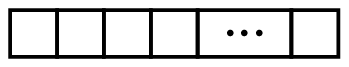}}, and more
involved \textit{mixed}-Young-tableau ones of the ``spin''
\raisebox{-1.8mm}{\includegraphics[scale=0.5]{young5.ps}} or
``Weyl'' \raisebox{-1.8mm}{\includegraphics[scale=0.5]{young8.ps}}
type.\footnote{Note in passing that in the irreducible
decompositions of $\widehat{\mathcal{F}}_L^{0i}$ and
$\widehat{\mathcal{F}}_L^{ij}$ in $3$ dimensions, the Levi-Civita
tensors always appear in pairs, and that the products
$\varepsilon\varepsilon$ which appear can always be entirely
expressed in terms of Kronecker deltas:
$\varepsilon_{abc}\varepsilon^{ijk} \propto \delta_{abc}^{ijk}
\equiv \delta_a^{[i} \delta_b^j \delta_c^{k]}$.} It is quite
possible to write this decomposition in any (integer) dimension $d$,
by using mixed-Young-tableau projectors instead of the
$\varepsilon_{ijk}$ duality operations used in $d=3$. However, for
the present work we can simplify our task and simply ignore all the
mixed tableaux that appear in the irreducible decomposition of
$\widehat{\mathcal{F}}_L^{0i}$ and $\widehat{\mathcal{F}}_L^{ij}$.

Indeed, because of their irreducible character the symmetric Young
tableaux \raisebox{-0.8mm}{\includegraphics[scale=0.5]{young1.ps}}
never mingle with the mixed-symmetry ones when doing linear
operations, as is done when working with a linearized solution of
Einstein's equations. (This is easily checked when looking, for
instance, at the derivation of the multipolar expansion of
linearized gravity given in \cite{DI91b}.) Moreover, when
considering the gauge-invariant content of $h^{\mu\nu}$
(\textit{i.e.}, modulo a linearized gauge transformation), the
symmetric representations will finally give rise to the so-called
mass-type multipole moments, say $I_L$, while the mixed one will
correspond to what are called (in $d=3$) the spin-type or
current-type multipole moments, say $J_L$. The aim of the present
work is to cure the ambiguities, linked to logarithmic divergencies
of integrals calculated in Hadamard's regularization, which appeared
at 3PN order in the calculations of
Refs.~\cite{BIJ02,BFIJ02,BI04mult}. However, the works
\cite{BIJ02,BFIJ02,BI04mult} found that the only multipole moment
which introduce ambiguities (in $d=3$) at the 3PN order is the mass
quadrupole moment $I_{ij}$. Therefore, for our purpose, it is enough
to derive general $d$-dependent formulas for the mass multipole
moments $I_L$ (besides the quadrupole $I_{ij}$ we shall also need to
consider below the mass dipole moment $I_{i}$). We do not need to
consider the definition of the current moments $J_L$ outside of
$d=3$. In view of what we said above, it is therefore enough to
consider only the \textit{symmetric}
(\raisebox{-0.8mm}{\includegraphics[scale=0.5]{young1.ps}}) pieces
in the irreducible decomposition of $\widehat{\mathcal{F}}_L^{0j}$
and $\widehat{\mathcal{F}}_L^{ij}$. We then write
{\allowdisplaybreaks
\begin{subequations}\label{Fdecomp}\begin{eqnarray}
\widehat{\mathcal{F}}_L^{00}&=&R_L\,,\label{Fdecompa}\\
\widehat{\mathcal{F}}_L^{i0}&=&T^{(+)}_{iL}+\delta_{i\langle
i_\ell}T^{(-)}_{L-1\rangle}+ \text{``mixed tableaux''}\,,
\label{Fdecompb}\\
\widehat{\mathcal{F}}_L^{ij}&=&U^{(+2)}_{ijL}+\mathop{\mathrm{STF}}_L
\mathop{\mathrm{STF}}_{ij}\left[\delta_{ii_\ell}U^{(0)}_{jL-1}
+\delta_{ii_\ell} \delta_{ji_{\ell-1}}U^{(-2)}_{L-2}\right]
+\delta_{ij}V_L\nonumber\\
&+& \text{``mixed tableaux''}\,,
\qquad\label{Fdecompc}
\end{eqnarray}\end{subequations}}\noindent
where the angular brackets surrounding indices refer to the STF
projection, and where the tensors $R_L$, $T^{(+)}_{L+1}$,
$T^{(-)}_{L-1}$, $U^{(+2)}_{L+2}$, $U^{(0)}_{L}$, $U^{(-2)}_{L-2}$,
$V_L$ are all STF in their indices (recall our notation for
multi-indices: $L=i_1\cdots i_\ell$, $L+1=i_1\cdots i_{\ell+1}$,
\textit{etc.}). Furthermore, we shall need below the inverse of
Eqs.~(\ref{Fdecomp}), \textit{i.e.}, the expressions of these
tensors in terms of the $\widehat{\mathcal{F}}$'s. These are
{\allowdisplaybreaks
\begin{subequations}\label{inverse}\begin{eqnarray}
R_L &=& \widehat{\mathcal{F}}_L^{00},\\
T^{(+)}_{L+1} &=&
\widehat{\mathcal{F}}_{L\rangle}^{0\langle i_{\ell+1}},\\
T^{(-)}_{L-1}
&=& \frac{\ell(2\ell + d -4)}{(\ell+d-3)(2\ell+d-2)}\,
\widehat{\mathcal{F}}_{aL-1}^{a0}\,,\\
U^{(+2)}_{L+2} &=&
\widehat{\mathcal{F}}_{L\rangle}^{\langle i_{\ell+2}i_{\ell+1}},\\
U^{(0)}_L &=& \frac{2d \,\ell (2\ell + d
-4)}{(d-2)(\ell+d-2)(2\ell+d)} \,\mathop{\mathrm{STF}}_{L}\,
\widehat{\mathcal{F}}_{aL-1}^{\langle ai_\ell\rangle}\,,\\
U^{(-2)}_{L-2} &=& \frac{\ell (\ell-1)(2\ell + d
-6)}{(\ell+d-3)(\ell+d-4)(2\ell +d
-2)}\,\widehat{\mathcal{F}}_{abL-2}^{\langle ab\rangle}\,,\\
V_L &=& \frac{1}{d}\,\widehat{\mathcal{F}}_L^{aa}\,.
\end{eqnarray}\end{subequations}}\noindent

The next step towards the definition of the STF source moments is to
take into account the effect of the harmonicity conditions
($\partial_\nu h^{\mu\nu}=0$) on the multipolar expansion
(\ref{MDelta}), which we henceforth write with the help of the
shorthand notation (\ref{FLtilde}) as
\begin{equation}\label{MDeltashort}
\mathcal{M}(\Delta^{\mu\nu})=-\frac{4\,G}{c^4}\sum_{\ell=0}^{+\infty}
\frac{(-)^\ell}{\ell!}\,
\partial_L\widetilde{\mathcal{F}}_L^{\mu\nu}\,.
\end{equation}
The latter tensor
$\mathcal{M}(\Delta^{\mu\nu})$ is \textit{not} divergence free in the
full non-linear theory. Indeed, by using the same method as the one
employed in $3$ dimensions and which resulted in Eqs.~(4.5)--(4.6) of
\cite{B98mult}, \textit{i.e.}, by using the explicit expressions
(\ref{calF0})--(\ref{calF1}) of the multipole moments, we can derive
the following relation
\begin{equation}\label{rel}
\dot{\widehat{\mathcal{F}}}_L^{\mu 0}-\ell\,
\widehat{\mathcal{F}}_{L-1
\rangle}^{\mu\langle i_\ell}-\frac{1}{2\ell+d}\ddot{\widehat{
\mathcal{F}}}_{aL}^{a\mu} = \widehat{\mathcal{G}}_L^{\mu}\,,
\end{equation}
where the dots mean the time differentiation, and where the new
``multipole-moment'' function $\widehat{\mathcal{G}}_L^{\mu}$ is
given by
\begin{equation}\label{hatcalG0}
\widehat{\mathcal{G}}_L^{\mu}(t)=\mathop{\mathrm{FP}}_{B}\int
d^d\mathbf{y}\,B\,\vert\widetilde{\mathbf{y}}\vert^{B}\,
\widehat{y}_L\,
\frac{y_a}{\vert\mathbf{y}\vert^{2}}\,\overline{\tau}_{[\ell]}^{\mu
a}(\mathbf{y},t)\,.
\end{equation}
Recall that $\overline{\tau}_{[\ell]}^{\mu a}(\mathbf{y},t)$ is
defined by Eq.~(\ref{calF1}) above.\footnote{The result
(\ref{rel})--(\ref{hatcalG0}) is easily checked once we remember
{}from Eq.~(\ref{deltaeps0}) that the kernel function
$\delta_\ell^{(\varepsilon)}(z)$ in $d$ dimensions is equal to
$\delta_{\ell+\frac{\varepsilon}{2}}^{(0)}(z)$. So, the same method
as in $3$ dimensions \cite{B98mult} applies with the simple
replacement $\ell\rightarrow\ell+\frac{\varepsilon}{2}$. In
particular one uses in this derivation
\begin{eqnarray*}
\frac{d}{dz}\left[\delta_{\ell+1}^{(\varepsilon)}(z)\right] &=&
-(2\ell+\varepsilon+3)z\,\delta_{\ell}^{(\varepsilon)}(z)\,,\\
\frac{d^2}{dz^2}\left[\delta_{\ell+1}^{(\varepsilon)}(z)\right] &=&
-(2\ell+\varepsilon+3)(2\ell+\varepsilon+1)
\left[\delta_{\ell}^{(\varepsilon)}(z)
-\delta_{\ell-1}^{(\varepsilon)}(z)\right].
\end{eqnarray*}}
Notice that $\widehat{\mathcal{G}}_L^{\mu}(t)$,
Eq.~(\ref{hatcalG0}), involves an \textit{explicit factor $B$} in its
integrand. Hence, by the properties of the analytic continuation in
$B$, $\widehat{\mathcal{G}}_L^{\mu}(t)$ depends in fact only on the
behavior at the boundary of the integral at infinity,
$\vert\mathbf{y}\vert\rightarrow +\infty$ (hence, for instance, there
is no analog of $\widehat{\mathcal{G}}_L^{\mu}(t)$ in linearized
gravity). The consequence of Eqs.~(\ref{rel})--(\ref{hatcalG0}) is
that the harmonicity conditions can be expressed by saying that the
divergence of the multipole expansion $\mathcal{M}(\Delta^{\mu\nu})$
reads
\begin{subequations}\label{divMDeltashort}\begin{eqnarray}
\partial_\nu\mathcal{M}(\Delta^{\mu\nu})&=&
-\frac{4\,G}{c^4}\sum_{\ell=0}^{+\infty}
\frac{(-)^\ell}{\ell!}\,
\partial_L\widetilde{\mathcal{G}}_L^{\mu}\,,\\
\text{where}\quad\widetilde{\mathcal{G}}_L^{\mu}(r,t) &\equiv& -4\pi
\int_{-\infty}^{+\infty}ds\,\widehat{\mathcal{G}}_L^{\mu}(s)
\,G_\mathrm{Ret}(\mathbf{x},t-s)\,.
\end{eqnarray}\end{subequations}
Next we decompose the components of the function
$\widehat{\mathcal{G}}_L^{\mu}$ in STF guise, which means
\begin{subequations}\label{Gdecomp}\begin{eqnarray}
\widehat{\mathcal{G}}_L^{0}&=&P_L\,,\\
\widehat{\mathcal{G}}_L^{i}&=&Q^{(+)}_{iL}+\delta_{i\langle
i_\ell}Q^{(-)}_{L-1\rangle}+ \text{``mixed tableaux''}\,,
\end{eqnarray}\end{subequations}
where the tensors $P_L$, $Q^{(+)}_{L+1}$ and $Q^{(-)}_{L-1}$ are all
STF, together with the inverse formulas
\begin{subequations}\label{GdecompInv}\begin{eqnarray}
P_L&=&\widehat{\mathcal{G}}_L^{0}\,,\\
Q^{(+)}_{L+1}&=&\widehat{\mathcal{G}}_{L\rangle}^{\langle
i_{\ell+1}}\,,\\
Q^{(-)}_{L-1}&=&\frac{\ell(2\ell+d-4)}{(\ell+d-3)(2\ell+d-2)}
\widehat{\mathcal{G}}_{a L-1}^{a}\,.
\end{eqnarray}\end{subequations}
The relations (\ref{divMDeltashort}) can then be re-stated as
the following constraint equations linking the STF tensors (to
simplify, we set $c=1$ for a while),
{\allowdisplaybreaks\begin{subequations}
\label{harmconstr}
\begin{eqnarray}
\dot{R}_L &=& P_L+\ell \,T^{(+)}_L
+\frac{d+\ell-2}{(\ell+1)(d+2\ell-2)}\,\ddot{T}^{(-)}_L,\\
\dot{T}^{(+)}_L &=& Q^{(+)}_L+(\ell-1) \,
U^{(+2)}_L+\frac{(d-2)(d+\ell-2) (d+2\ell)}{2\ell \,d
(d+2\ell-4)(d+2\ell-2)}\,
\ddot{U}^{(0)}_L+\frac{1}{d+2\ell}\,\ddot{V}_L \,,\\
\dot{T}^{(-)}_L&=& Q^{(-)}_L+\frac{(d-2)(\ell+1)}{2d}\,U^{(0)}_L+
\frac{d+\ell-1}{(\ell+2)(d+2\ell)} \,
\ddot{U}^{(-2)}_L+(\ell+1)\,V_L\,.
\end{eqnarray}\end{subequations}}\noindent

Our aim is now to obtain the linearized multipolar solution, which
is at once solution of the source-free equations and divergenceless,
and which will be exactly equal to the linearized metric
$h^{\mu\nu}_1$ of the MPM formalism. To this end, we introduce the
object $q^{\mu\nu}_1$, given by the following multipole expansion in
$d$ dimensions [recall our notation
(\ref{FLtilde})],{\allowdisplaybreaks
\begin{subequations}\label{q1munu}\begin{eqnarray}
q^{00}_1 &\equiv&
-4\,G\left[-\hbox{$\int$}\widetilde{P}
+\partial_i\left(\hbox{$\int$}
\widetilde{P}_i+\hbox{$\int\!\!\int$}
\widetilde{Q}^{(+)}_i-\frac{3d+1}{2d}
\widetilde{Q}^{(-)}_i\right)\right],\\
q^{i0}_1 &\equiv&
- 4\,G
\left[-\hbox{$\int$}\widetilde{Q}^{(+)}_i
+\frac{3d+1}{2d}\,\dot{\widetilde{Q}}^{(-)}_i-\sum_{\ell\geq
2}\frac{(-)^\ell}{\ell!} \,
\partial_{L-1}\widetilde{P}_{iL-1}\right],\\
q^{ij}_1 &\equiv&
- 4\,G \left[\delta_{ij}\widetilde{Q}^{(-)}
+\sum_{\ell\geq 2}\frac{(-)^\ell}{\ell!}
\,\left(2\,\delta_{ij}
\partial_{L-1}\widetilde{Q}^{(-)}_{L-1}
-6\,\partial_{L-2(i}
\widetilde{Q}^{(-)}_{j)L-2}\right.\right.\nonumber\\
&&\qquad\quad\left.\left. +\partial_{L-2}
\left(\dot{\widetilde{P}}_{ijL-2}+\ell\,\widetilde{Q}^{(+)}_{ijL-2}
-\frac{7\ell+3d-6}{(\ell+1)(2\ell+d-2)}\,
\ddot{\widetilde{Q}}^{(-)}_{ijL-2}
\right)\right)\right].
\end{eqnarray}\end{subequations}}\noindent
Here the integral signs refer to a time \textit{anti-derivative},
\textit{e.g.} $\int\widetilde{P}(r,t)\equiv\int_{-\infty}^t
d\tau\widetilde{P}(r,\tau)$,
$\int\!\!\int\widetilde{Q}(r,t)\equiv\int_{-\infty}^t
d\tau\int_{-\infty}^\tau d\tau' \widetilde{Q}(r,\tau')$. The object
$q^{\mu \nu}_1$, which is given here modulo the mixed tableaux
corresponding notably to the spin-type contributions, exactly
corresponds to the so-called ``harmonicity algorithm'' of
Ref.~\cite{BD86} (in the slightly modified version of it proposed in
Eqs.~(2.12) of \cite{B98quad}; notice that the latter equations are
valid in any $d$). The properties of $q^{\mu\nu}_1$ are that $\Box
q^{\mu\nu}_1=0$ \textit{and}
$\partial_\nu\left[\mathcal{M}(\Delta^{\mu\nu})+q^{\mu\nu}_1\right]=
0$, as one can easily verify by direct
calculation.\footnote{Remember the presence of time
\textit{anti-derivatives} in (\ref{q1munu}). In the present
formalism the metric is past-stationary, and from this one can show
that the functions involved which need to be time integrated are in
fact zero in the past, before the instant $-\mathcal{T}$, so that
there is no problem in defining these anti-derivatives.} One now
introduces the object
\begin{equation}\label{h1munu}
h^{\mu\nu}_{1} \equiv \mathcal{M}(\Delta^{\mu\nu})+q^{\mu\nu}_1\,.
\end{equation}
As in \cite{B98mult} one easily checks that $h^{\mu\nu}_{1}$ defines
a linearized multipolar metric (in harmonic coordinates), which
generates, by MPM iteration, the full metric
$\mathcal{M}(h^{\mu\nu})$. The source multipole moments are then
defined as those which parametrize $h^{\mu\nu}_{1}$. The ``main''
multipole moments will be those which parametrize a specific piece
of the linearized metric sometimes referred to as the ``canonical''
metric and which was introduced long ago in Ref.~\cite{Th80}. The
canonical metric, say $h^{\mu\nu}_{\mathrm{can}\,1}$, is separately
divergenceless, and differs from $h^{\mu\nu}_{1}$ by a linearized
gauge transformation, with gauge vector say $\psi^{\mu}_1$,
\begin{equation}\label{h1h1can}
h^{\mu\nu}_{1} = h^{\mu\nu}_{\mathrm{can}\,1}
+\partial^\mu \psi^\nu_{1}
+\partial^\nu\psi^\mu_{1} -\eta^{\mu\nu}\partial_\rho\psi^\rho_{1}\,.
\end{equation}
It explicitly reads (still consistently omitting the mixed tableaux)
\begin{subequations}\label{h1can}\begin{eqnarray}
h^{00}_{\mathrm{can}\,1}&=&-4\,G \sum_{\ell=0}^{+\infty}
\frac{(-)^\ell}{\ell!} \,\partial_L\widetilde{I}_L\,,\\
h^{i0}_{\mathrm{can}\,1}&=&4\,G \sum_{\ell=1}^{+\infty}
\frac{(-)^\ell}{\ell!}
\,\partial_{L-1}\dot{\widetilde{I}}_{iL-1}\,,\\
h^{ij}_{\mathrm{can}\,1}&=&-4\,G \sum_{\ell=2}^{+\infty}
\frac{(-)^\ell}{\ell!}
\,\partial_{L-2}\ddot{\widetilde{I}}_{ijL-2}\,,
\end{eqnarray}\end{subequations}
where the tilde over the objects $I_L$ means, as in (\ref{FLtilde}),
\begin{equation}\label{ILtilde}
\widetilde{I}_L(r,t) \equiv -4\pi \int_{-\infty}^{+\infty}ds\,I_L(s)
\,G_\mathrm{Ret}(\mathbf{x},t-s)\,.
\end{equation}
Such expressions clearly yield a precise definition of the mass-type
STF multipole moments $I_L(t)$ in $d$ dimensions (and the mixed
tableaux could be used to define some other, ``spin-type'' and
``Weyl-type'', moments). We need now to relate the moments $I_L$
entering (\ref{h1can}) to the STF tensors which were used in the STF
decomposition (\ref{Fdecomp}) of the function
$\widehat{\mathcal{F}}^{\mu\nu}_L(t)$.

To this end it is most convenient to consider the gauge invariant
\textit{linearized curvature} (in $d$ dimensions) associated with
the metric deviation $h_1^{\mu\nu}$, in order to eliminate the
irrelevant linearized gauge transformation in Eq.~(\ref{h1h1can}).
The component $0i0j$ of the curvature, in terms of the gothic metric
deviation, reads
\begin{eqnarray}\label{riemannh}
2R_{0i0j}^\mathrm{lin}[h_1]&=&
\frac{1}{d-1}\left[(d-2)\partial_{i}\partial_{j}h_1^{00}
+\partial_{i}\partial_{j}h_1^{kk}
+\delta_{ij}\partial_{0}^2h_1^{00}
-\delta_{ij}\partial_{0}^2h_1^{kk}\right]\nonumber\\
&+&2\partial_{0} \partial_{(i}h_1^{j)0}+\partial_{0}^2h_1^{ij}\,.
\end{eqnarray}
Since $h_1^{\mu\nu}$ and $h_{\mathrm{can}\,1}^{\mu\nu}$ differ by a
gauge transformation, Eq.~(\ref{h1h1can}), we necessarily have
$R_{0i0j}^\mathrm{lin}[h_1-h_{\mathrm{can}\,1}]=0$, which immediately
gives us (looking at the particular term proportional to the double
gradient $\partial_{i}\partial_{j}$) the looked-for expression of the
moment $I_L$ as being
\begin{equation}\label{ILexpr0}
I_L=R_L+\frac{d}{d-2}V_L
-\frac{2(d-1)}{c (\ell+1)(d-2)}\dot{T}_L^{(-)}
+\frac{d-1}{c^2 (\ell+1)(\ell+2)(d-2)}\ddot{U}_L^{(-2)}
-\frac{2(d-3)}{(\ell+1)(d-2)} Q_L^{(-)},
\end{equation}
where the explicit powers of $c$ have now been restored. We find
that $I_L$ is given in terms of the STF tensors parametrizing our
original multipole-moment function
$\widehat{\mathcal{F}}^{\mu\nu}_L(t)$ and defined by
Eqs.~(\ref{Fdecomp}), and also, in the last term of
Eq.~(\ref{ILexpr0}), of the ``harmonicity'' function
$\widehat{\mathcal{G}}^{\mu}_L(t)$ given by (\ref{hatcalG0}). Note
that the last term of Eq.~(\ref{ILexpr0}) involves a factor $(d-3)$
and therefore is absent in the 3-dimensional formalism of
\cite{B98mult}. Since $\widehat{\mathcal{G}}^{\mu}_L$ involves also
a factor $B$ in its integrand, we see that the contribution induced
in the moments by this term will be proportional to $B(d-3)$; we
shall see that such a contribution is actually zero.

Once we have obtained the moment $I_L(t)$, it is better to express it
back in terms of the original function
$\widehat{\mathcal{F}}^{\mu\nu}_L(t)$, since we know its relation to
the pseudo tensor of the source, given by
Eqs.~(\ref{calF0})--(\ref{calF1}). Using the inverse relations
(\ref{inverse}) and (\ref{GdecompInv}), we get
\begin{eqnarray}\label{ILexpr1}
I_L&=&\frac{d-1}{2(d-2)}\left\{\frac{2}{c^2(d-1)}
\Bigl[(d-2)\widehat{\mathcal{F}}^{00}_L
+\widehat{\mathcal{F}}^{ii}_L\Bigr]\right.\nonumber\\
&&\qquad\qquad~
\left.-\frac{4(d+2\ell-2)}{c^3(d+\ell-2)(d+2\ell)}
\dot{\widehat{\mathcal{F}}}_{iL}^{i0} \right.\nonumber\\
&&\qquad\qquad~\left.+
\frac{2(d+2\ell-2)}{c^4(d+\ell-1)(d+\ell-2)(d+2\ell+2)}
\ddot{\widehat{\mathcal{F}}}_{ijL}^{\langle
ij\rangle}\right.\nonumber\\
&&\qquad\qquad~\left.-
\frac{4(d-3)(d+2\ell-2)}{c^2(d-1)(d+\ell-2)(d+2\ell)}
\widehat{\mathcal{G}}_{iL}^i\right\}.
\end{eqnarray}
If we further introduce the following ``source-rooted'' quantities
\begin{subequations}\label{Sigma}
\begin{eqnarray}
\Sigma&\equiv&\frac{2}{d-1}\,
\frac{(d-2)\tau^{00}+\tau^{ii}}{c^2}\,,\label{Sigmaa}\\
\Sigma_i&\equiv&\frac{\tau^{i0}}{c}\,,\\
\Sigma_{ij}&\equiv&\tau^{ij}\,,
\end{eqnarray}\end{subequations}
we can then write the moment into the following more explicit form
\begin{eqnarray}\label{ILexpr2}
I_L(t)&=&\frac{d-1}{2(d-2)}\mathop{\mathrm{FP}}_{B}\int
d^d\mathbf{y}\,\vert\widetilde{\mathbf{y}}\vert^B
\left\{\widehat{y}_L\,\overline{ \Sigma}_{[\ell]}(\mathbf{y},t)
\right.\nonumber\\
&&\qquad\qquad~~\left.-\frac{4(d+2\ell-2)}
{c^2(d+\ell-2)(d+2\ell)}\,\widehat{y}_{iL}\,
\dot{\overline{\Sigma}}_{i[\ell+1]}(\mathbf{y},t)
\right.\nonumber\\
&&\qquad\qquad~~\left.+\frac{2(d+2\ell-2)}
{c^4(d+\ell-1)(d+\ell-2)(d+2\ell+2)}
\,\widehat{y}_{ijL}\,
\ddot{\overline{\Sigma}}_{ij[\ell+2]}(\mathbf{y},t)
\right.\nonumber\\
&&\qquad\qquad~\left.-
\frac{4(d-3)(d+2\ell-2)}{c^2(d-1)(d+\ell-2)(d+2\ell)}
B \,\widehat{y}_{iL}\frac{y_j}{\vert \mathbf{y}\vert^2}
\overline{\Sigma}_{ij[\ell+1]}(\mathbf{y},t)
\right\},
\end{eqnarray}
in which we denote the relevant infinite PN series of the source
terms [following our earlier notation (\ref{calF1})] by
\begin{eqnarray}\label{series}
\overline{\Sigma}_{[\ell]}(\mathbf{y},t)&=&\int_{-1}^1 dz
\,\delta_\ell^{(\varepsilon)} (z)
\,\overline\Sigma(\mathbf{y},t+z\vert\mathbf{y}\vert/c) \nonumber\\
&=& \sum_{k=0}^{+\infty}\alpha_\ell^k
\left(\frac{\vert\mathbf{y}\vert}{c}\frac{\partial}{\partial
t}\right)^{2k}\overline\Sigma(\mathbf{y},t)\,.
\end{eqnarray}
The numerical coefficients $\alpha_\ell^k$ are given by
Eq.~(\ref{Smult1b}), or more explicitly
\begin{equation}\label{coeffs}
\alpha_\ell^k = \frac{1}{(2k)!!(2\ell+d)
(2\ell+d+2)\cdots(2\ell+d+2k-2)}\,.
\end{equation}
Notice that with our conventions the Newtonian limit, when
$c\rightarrow +\infty$, of the above defined relativistic moment
$I_L$ takes the standard Newtonian expression in any dimension $d$,
\textit{i.e.}, it does not contain any $d$-dependent factors in this
limit:
\begin{equation}\label{newtonian}
I_L = \int d^d\mathbf{y}\,\rho\,\widehat{y}_L +
\mathcal{O}\left(c^{-2}\right),
\end{equation}
where the ``Newtonian'' density of the fluid is $\rho \equiv
T^{00}/c^2$. This is clear from the fact that the factor
$\frac{d-1}{2(d-2)}\equiv f^{-1}$ which appears in front of the
expression of the multipole moment (\ref{ILexpr2}), cancels out
precisely the $d$-dependent factor in the Newtonian approximation for
$\Sigma$, Eq.~(\ref{Sigmaa}), which is given by $\Sigma=f\,\rho +
\mathcal{O}\left(c^{-2}\right)$.

Finally, note that the last term in (\ref{ILexpr1}) or
(\ref{ILexpr2}) is proportional to both $B$ and $\varepsilon = d-3$.
To show that this term does not contribute to $I_L$, we can first
decompose the integral over $d^d\mathbf{y}$ in two parts: (i)~an
integral $\mathcal{I}_1$ over a compact domain $r < \mathcal{R}$
containing the two particles, plus (ii)~an integral $\mathcal{I}_2$
over the outer domain $r > \mathcal{R}$. Even if the integration near
the particles introduces some UV poles $\propto 1/\varepsilon$,
$\mathcal{I}_1$ will be at worst proportional to $\varepsilon
B/\varepsilon = B$, and will [by the definition of the FP process,
Eq.~(\ref{B15})] give a vanishing finite part at $B=0$. Concerning
$\mathcal{I}_2$, we shall prove in Sec.~\ref{secVB} below that, even
if it contains infrared (IR)-type poles, its value is a
\textit{continuous} function of $d$. Now, because of the factor
$(d-3)$, the value in $3$ dimensions is zero, hence this term does
not contribute to the moments and can be ignored in the present work
(this term was neglected in Ref.~\cite{BDEI04}).

\section{Source terms for the 3PN moments in $d$
dimensions}\label{secIV}

In our 3PN calculations of the gravitational wave field, we will
need the expressions of the sources $\Sigma$, $\Sigma_i$ and
$\Sigma_{ij}$, defined in Eqs.~(\ref{Sigma}) above, up to orders
$1/c^6$, $1/c^4$ and $1/c^2$ respectively. The quickest method to
obtain them consists in using the results of our previous work
\cite{BDE04}, in which the 3PN metric $g_{\mu\nu}$ was expanded in
terms of nine retarded potentials, introduced in \cite{BF00} when
$d=3$ and generalized to $d$ dimensions in \cite{BDE04}. Starting
{}from the matter source densities
\begin{equation}
\label{matterSources}
\sigma \equiv \frac{2}{d-1}\,\frac{(d-2)T^{00}+T^{ii}}{c^2}\,,
\qquad
\sigma_i \equiv \frac{T^{0i}}{c},
\qquad
\sigma_{ij}\equiv T^{ij},
\end{equation}
we first defined the ``linear'' potentials
\begin{equation}
\label{PotLin}
V=\Box^{-1}_\mathrm{Ret}\left(-4\pi G \sigma\right),
\qquad
V_i=\Box^{-1}_\mathrm{Ret}\left(-4\pi G \sigma_i\right).
\end{equation}
These linear potentials were then used to construct higher
``non-linear'' potentials, such as
\begin{equation}
\label{Wij}
\hat{W}_{ij}=\Box^{-1}_\mathrm{Ret}\left[-4\pi G
\left(\sigma_{ij}-\delta_{ij}\,\frac{\sigma_{kk}}{d-2}\right)
-\frac{1}{2}\left(\frac{d-1}{d-2}\right)\partial_i
V\partial_j V\right],
\end{equation}
and six other ones (denoted $K$, $\hat R_i$, $\hat X$, $\hat
Z_{ij}$, $\hat Y_i$, $\hat T$) whose field equations are explicitly
given in Eqs.~(2.12) of Ref.~\cite{BDE04}. We computed all of them
for a binary system of point masses, in spatial dimension $d =
3+\varepsilon$ close to 3, at any field point in the case of linear
potentials such as (\ref{PotLin}), and, for the more difficult
non-linear ones like (\ref{Wij}), in the vicinity of the particles as
Laurent-type expansions in powers of the radial distances to them.
The retardations in these potentials were also systematically
expanded to the required PN order.

For our present calculation of the 3PN gravitational wave field,
only the expressions of the first seven potentials ($V$, $V_i$, $K$,
$\hat W_{ij}$, $\hat R_i$, $\hat X$, $\hat Z_{ij}$) will actually be
necessary. From Eqs.~(\ref{EE}) and (\ref{Sigma}) above, the sources
$\Sigma$, $\Sigma_i$ and $\Sigma_{ij}$ may be expressed in terms of
the ``gothic'' metric $\mathfrak{g}^{\mu\nu} \equiv \sqrt{-g}\,
g^{\mu\nu} \equiv \eta^{\mu\nu} + h^{\mu\nu}$ as
\begin{subequations}
\label{Sigh}
\begin{eqnarray}
\Sigma &=& \frac{c^2}{8\pi G}\,\Box\left(
\frac{(d-2) h^{00} + h^{ii}}{d-1}\right),\label{Sigh1}\\
\Sigma_i &=& \frac{c^3}{16\pi G}\,\Box h^{0i},\label{Sigh2}\\
\Sigma_{ij} &=& \frac{c^4}{16\pi G}\,\Box h^{ij},\label{Sigh3}
\end{eqnarray}
\end{subequations}
where $G$ denotes by definition the gravitational constant entering
the $(d+1)$-dimensional Einstein equations (\ref{EE}). As underlined
in Ref.~\cite{BDE04}, it is related to the usual Newton constant (in
3 spatial dimensions) $G_N$ by
\begin{equation}
G = G_N \ell_0^{d-3},
\label{GN}
\end{equation}
where $\ell_0$ is an arbitrary length scale, which will enter our
dimensionally regularized calculation below but will drop out of the
final physical observables.

To identify the sources (\ref{Sigh}), it thus suffices to write
Einstein's equations $R_{\mu\nu} = (8\pi G/c^4)[T_{\mu\nu} -
T^\lambda_\lambda g_{\mu\nu}/(d-1)]$ in harmonic gauge and in terms
of the gothic metric $\mathfrak{g}^{\mu\nu}$. A possible method
would be to use the expression of the Ricci tensor in terms of
$\mathfrak{g}^{\mu\nu}$ that we gave in Eq.~(A9) of
Ref.~\cite{BDE04}, for any dimension $d$. It is however quicker to
use directly the full 3PN form of $g_{\mu\nu}$ that we obtained in
this reference, which can be translated in terms of
$\mathfrak{g}^{\mu\nu}$ thanks to Eqs.~(A3) and (A8) of
\cite{BDE04}. The result not only depends on the nine introduced
potentials ($V$, $V_i$, $K$, $\hat W_{ij}$, $\hat R_i$, $\hat X$,
$\hat Z_{ij}$, $\hat Y_i$, $\hat T$), but the $1/c^8$ order in
$\mathfrak{g}^{00}$ actually depends also on the 4PN ($1/c^8$)
contribution to the spatial metric $g_{ij}$, that was not computed
in \cite{BDE04}. However, the combination entering Eq.~(\ref{Sigh1})
above precisely cancels this uncomputed contribution, and one gets
straightforwardly
{\allowdisplaybreaks
\begin{subequations}
\label{gothmet}
\begin{eqnarray}
\frac{(d-2)h^{00}+h^{ii}}{d-1} &=&
-\frac{2}{c^2} V
-\frac{1}{c^4}\biggl[
2\left(\frac{d-1}{d-2}\right)V^2
- 4 \left(\frac{d-3}{d-2}\right)K
\biggr]
\nonumber\\
&&-\frac{1}{c^6}\biggl[
8\hat X
+ 4 V\hat W
+ \frac{4}{3}\left(\frac{d-1}{d-2}\right)^2 V^3
\nonumber\\
&&- 8 \left(\frac{d-3}{d-1}\right) V_iV_i
- \frac{8(d-1)(d-3)}{(d-2)^2} K V
\biggr]
\nonumber\\
&&-\frac{1}{c^8}\biggl[
32\hat T
+ 16\left(\frac{d-1}{d-2}\right) V\hat X
+ 16 V \hat Z
+ 4\left(\frac{d-1}{d-2}\right) V^2\hat W
\nonumber\\
&&- \frac{8}{d-1} \hat W_{ij}\hat W_{ij}
+ \frac{4}{d-1} \hat W^2
+ \frac{2}{3}\left(\frac{d-1}{d-2}\right)^3 V^4
+ \frac{8(d-1)(d-3)^2}{(d-2)^3} K^2\nonumber\\
&&- 32\left(\frac{d-3}{d-1}\right) \hat R_i V_i
- \frac{8(d-1)^2(d-3)}{(d-2)^3} K V^2
- 8\left(\frac{d-3}{d-2}\right) K\hat W\nonumber\\
&&- 16\left(\frac{d-3}{d-2}\right) V V_i V_i
\biggr]
+\mathcal{O}\left(\frac{1}{c^{10}}\right),
\label{gothmeta}\\
h^{0i} &=&
-\frac{4}{c^3}V_i
-\frac{1}{c^5}\biggl[
8 \hat R_i
+ 4 \left(\frac{d-1}{d-2}\right) V V_i
\biggr]
\nonumber\\
&&
-\frac{1}{c^7}\biggl[
16 \hat Y_i
- 8 \hat W_{ij} V_j
+ 8 \hat W V_i
+ 8\left(\frac{d-1}{d-2}\right) V \hat R_i
\nonumber\\
&&+ 4\left(\frac{d-1}{d-2}\right)^2 V^2 V_i
- \frac{8(d-1)(d-3)}{(d-2)^2} K V_i
\biggr]
+\mathcal{O}\left(\frac{1}{c^9}\right),
\label{gothmetb}\\
h^{ij} &=& -\frac{4}{c^4}\left(\hat W_{ij}
-\frac{1}{2}\delta_{ij}\hat W\right)
-\frac{16}{c^6}\left(\hat Z_{ij}
-\frac{1}{2}\delta_{ij}\hat Z\right)
+\mathcal{O}\left(\frac{1}{c^8}\right),
\label{gothmetc}
\end{eqnarray}
\end{subequations}}\noindent
where $\hat W\equiv \hat W_{ii}$ and $\hat Z\equiv \hat Z_{ii}$
denote the traces of the corresponding potentials. Equations (2.12)
of Ref.~\cite{BDE04} then allow us to compute the d'Alembertian of
these metric coefficients in terms of the first seven potentials,
and one gets the following explicit form for the sources:
{\allowdisplaybreaks
\begin{subequations}
\label{sources}
\begin{eqnarray}
\Sigma &=&
\sigma - \frac{2}{c^2}\left(\frac{d-3}{d-2}\right)\sigma V
- \frac{1}{4\pi G c^2}\left(\frac{d-1}{d-2}\right)
\Delta\left(V^2\right)\nonumber\\
&&+\frac{1}{\pi G c^4}\biggl[
\frac{4 \pi G}{d-2} \sigma_{ii} V
+ 8\pi G\left(\frac{d-3}{d-1}\right)\sigma_i V_i
+ 4\pi G
\left(\frac{d-3}{d-2}\right)^2\sigma\left(\frac{V^2}{2}+K\right)
\nonumber\\
&&-2 V_i \partial_t\partial_i V
-\hat W_{ij}\partial_{ij}V
-\frac{(4-d)(d-1)}{4(d-2)^2}\left(\partial_t V\right)^2
+ 2 \partial_i V_j \partial_j V_i\nonumber\\
&&-\frac{1}{6}\left(\frac{d-1}{d-2}\right)^2\Delta\left(V^3\right)
-\frac{1}{2}\Delta\left(V\hat W\right)
+\frac{d-3}{d-1}\Delta\left(V_i V_i\right)
+\frac{(d-1)(d-3)}{(d-2)^2}\Delta\left(K V\right)\biggr]
\nonumber\\
&&+\frac{1}{\pi G c^6}\biggl[
\frac{16 \pi G}{d-2} \sigma V_i V_i
+ 4\pi G \frac{5-d}{(d-2)^2} \sigma_{ii} V^2
+\frac{8 \pi G}{d-1} \sigma_{ij}\hat W_{ij}
- 8\pi G\left(\frac{d-3}{d-2}\right) \sigma \hat X
\nonumber\\
&&-\frac{4}{3}\pi G\left(\frac{d-3}{d-2}\right)^3\sigma V^3
-8\pi G\left(\frac{d-3}{d-2}\right)^3 \sigma V K
+ 8\pi G \frac{(5-d)(d-3)}{(d-1)(d-2)}\sigma_i V_i V
\nonumber\\
&&+16\pi G\left(\frac{d-3}{d-1}\right)\sigma_i\hat R_i
- 8\pi G \frac{d-3}{(d-2)^2}\sigma_{ii}K
\nonumber\\
&&+\frac{1}{2}\hat W\partial_t^2 V
+\frac{1}{2} V\partial_t^2\hat W
-\frac{1}{2}\frac{(4-d)(d-1)^2}{(d-2)^3} V\left(\partial_t V\right)^2
-\frac{d(d-1)}{(d-2)^2} V_i \partial_t V\partial_i V
\nonumber\\
&&-2\left(\frac{d-1}{d-2}\right) V V_i \partial_t \partial_i V
- 4\left(\frac{d-1}{d-2}\right) V_i \partial_j V_i \partial_j V
+\frac{4}{d-1}\left(\partial_t V_i\right)^2
+\partial_t V\partial_t\hat W
\nonumber\\
&&+4 \partial_i V_j \partial_t \hat W_{ij}
-4 \hat Z_{ij} \partial_{ij} V
- 4 \hat R_i \partial_t\partial_i V
+ 8 \partial_i V_j \partial_j \hat R_i
- 2\left(\frac{d-3}{d-1}\right) V_i \partial_t^2 V_i
\nonumber\\
&&+\frac{(4-d)(d-1)(d-3)}{(d-2)^3} \partial_t V \partial_t K
+ 4 \left(\frac{d-3}{d-2}\right) V_i \partial_t\partial_i K
+ 2 \left(\frac{d-3}{d-2}\right) \hat W_{ij} \partial_{ij} K
\nonumber\\
&&-\frac{1}{12} \left(\frac{d-1}{d-2}\right)^3
\Delta\left(V^4\right)
-\frac{1}{2}\left(\frac{d-1}{d-2}\right)
\Delta\left(V^2\hat W\right)
- \frac{1}{2(d-1)}\Delta\left(\hat W^2\right)
\nonumber\\
&&+\frac{1}{d-1}\Delta\left(\hat W_{ij}\hat W_{ij}\right)
- 2 \left(\frac{d-1}{d-2}\right)
\Delta\left(V\hat X\right)
- 2\Delta\left(V\hat Z\right)
- \frac{(d-1)(d-3)^2}{(d-2)^3} \Delta\left(K^2\right)
\nonumber\\
&&+ 4 \left(\frac{d-3}{d-1}\right) \Delta\left(\hat R_i V_i\right)
+\frac{(d-1)^2(d-3)}{(d-2)^3}\Delta\left(K V^2\right)
+\left(\frac{d-3}{d-2}\right)
\Delta\left(K\hat W\right)
\nonumber\\
&&+ 2 \left(\frac{d-3}{d-2}\right)
\Delta\left(V V_i V_i\right)\biggr]
+ \mathcal{O}\left(\frac{1}{c^8}\right),
\label{sourcesa}\\
\Sigma_i &=& \sigma_i
+\frac{1}{\pi G c^2}\biggl[
\frac{5-d}{d-2} \pi G \sigma_i V
- \frac{d-1}{d-2}\pi G \sigma V_i
\nonumber\\
&&+\frac{1}{2}\left(\frac{d-1}{d-2}\right)\partial_k V\partial_i V_k
+ \frac{d(d-1)}{8(d-2)^2}\partial_t V\partial_i V
-\frac{1}{4}\left(\frac{d-1}{d-2}\right)
\Delta\left(V V_i\right)\biggr]\nonumber\\
&&+\frac{1}{\pi G c^4}\biggl[
\frac{1}{2}\left(\frac{5-d}{d-2}\right)^2 \pi G \sigma_i V^2
- 2\left(\frac{d-1}{d-2}\right)\pi G \sigma \hat R_i
+\frac{2\pi G}{d-2}\sigma_{kk} V_i
\nonumber\\
&&+ 2 \pi G \sigma_k \hat W_{ik}
+2\pi G \sigma_{ik} V_k
+ 2\frac{(d-1)(d-3)}{(d-2)^2}\pi G \sigma V V_i
-2 \frac{(5-d)(d-3)}{(d-2)^2}\pi G \sigma_i K
\nonumber\\
&&+\frac{1}{4}\left(\frac{d-1}{d-2}\right)
V_i \partial_t^2 V
- \frac{1}{4}\left(\frac{d-1}{d-2}\right)
V \partial_t^2 V_i
+ \frac{1}{2}\left(\frac{d-1}{d-2}\right)
\partial_t V \partial_t V_i
-2 V_k \partial_k\partial_t V_i
\nonumber\\
&&+\frac{1}{8}\frac{d(d-1)^2}{(d-2)^3}V\partial_t V\partial_i V
-\frac{1}{4}\left(\frac{d-1}{d-2}\right)^2
V_i \partial_k V \partial_k V
+\frac{1}{4} \frac{d(d-1)}{(d-2)^2}
V_k \partial_i V \partial_k V
\nonumber\\
&&+\left(\frac{d-1}{d-2}\right)
\partial_k V \partial_i \hat R_k
-\hat W_{kl}\partial_{kl} V_i
+\frac{1}{2}\left(\frac{d-1}{d-2}\right)
\partial_t\hat W_{ik}\partial_k V
-\partial_i \hat W_{kl}\partial_k V_l
+\partial_k \hat W_{il}\partial_l V_k
\nonumber\\
&&-\frac{(d-1)(d-3)}{(d-2)^2}
\partial_k K \partial_i V_k
-\frac{d(d-1)(d-3)}{4(d-2)^3}
\left(\partial_t V \partial_i K + \partial_i V \partial_t K\right)
\nonumber\\
&&-\frac{1}{8}\left(\frac{d-1}{d-2}\right)^2
\Delta\left(V^2 V_i\right)
-\frac{1}{2}\left(\frac{d-1}{d-2}\right)
\Delta\left(V \hat R_i\right)
-\frac{1}{2} \Delta\left(\hat W V_i\right)
\nonumber\\
&&+\frac{1}{2} \Delta\left(\hat W_{ij} V_j\right)
+\frac{1}{2} \frac{(d-1)(d-3)}{(d-2)^2}
\Delta\left(V_i K\right)\biggr]
+ \mathcal{O}\left(\frac{1}{c^6}\right),
\label{sourcesb}\\
\Sigma_{ij} &=& \sigma_{ij}
+\frac{1}{8\pi G}\left(\frac{d-1}{d-2}\right)
\left(\partial_i V\partial_j V
-\frac{1}{2}\delta_{ij} \partial_k V \partial_k V\right)
\nonumber\\
&&+\frac{1}{\pi G c^2}
\biggl\{
\frac{4\pi G}{d-2}\sigma_{ij} V
+\left(\frac{d-1}{d-2}\right)
\partial_t V_{(i}\partial_{j)} V
-\partial_i V_k \partial_j V_k
-\partial_k V_i \partial_k V_j
+ 2 \partial_k V_{(i}\partial_{j)}V_k
\nonumber\\
&&-\frac{(d-1)(d-3)}{2(d-2)^2}
\partial_{(i} V \partial_{j)} K
+\delta_{ij}
\biggl[
\frac{1}{2} \partial_k V_m \partial_k V_m
-\frac{1}{2} \partial_k V_m \partial_m V_k
-\frac{d(d-1)}{16(d-2)^2}\left(\partial_t V\right)^2
\nonumber\\
&&-\frac{1}{2} \left(\frac{d-1}{d-2}\right)
\partial_t V_k \partial_k V
+\frac{(d-1)(d-3)}{4(d-2)^2}\partial_k V \partial_k K
\biggr]
\biggr\}
+ \mathcal{O}\left(\frac{1}{c^4}\right).
\label{sourcesc}
\end{eqnarray}
\end{subequations}}\noindent
Note that although we did use the full 3PN expression of the metric
in the intermediate steps of this calculation, the sources $\Sigma$,
$\Sigma_i$ and $\Sigma_{ij}$ actually depend only on the 2PN metric
and on the potential $\hat Z$ (entering the trace of the 3PN spatial
metric $g_{ij}$). The mass-type moment $I_L$ can now be obtained by
inserting the above expressions into
Eqs.~(\ref{ILexpr2})--(\ref{series}), thereby generalizing to $d$
dimensions the 3-dimensional results (3.4)--(3.6) of
Ref.~\cite{BI04mult}.

The above method to derive Eqs.~(\ref{sources}) not only avoids
redoing some of the calculations of Ref.~\cite{BDE04}, but it also
yields the results in a useful form. Indeed, the Laplacians of
product of potentials, say $\Delta(AB)$, are easier to compute than
their expanded form $B \Delta A+A\Delta B+2\partial_k A\partial_k B$
(where $\Delta A$ and $\Delta B$ may be replaced by their
corresponding sources). In particular, when computing the
contributions of such Laplacians to the moment $I_L$,
Eq.~(\ref{ILexpr2}), their lowest-order terms ($k=0$) in
Eq.~(\ref{series}) do not contribute to the difference between the
dimensional and pure-Hadamard-Schwartz (pHS) regularizations; see
Eq.~(4.23) of Ref.~\cite{BI04mult} and Sec.~\ref{secVB} below.
However, the retardation corrections ($k\geq 1$) entering
Eq.~(\ref{series}) do contribute to this difference.

To ease the reading, we classified the various terms of
Eqs.~(\ref{sources}) in different sets, at each successive PN order:
first the compact-support terms (proportional to $\sigma$,
$\sigma_i$ or $\sigma_{ij}$), which do not contribute to the
difference between the dimensional and pHS regularizations below;
second the main non-compact contributions, which are crucial for
this difference; and finally the non-compact terms proportional to
the Laplacian of a product of potentials, which do not contribute to
the difference at lowest order. In each set of terms, we also
gathered at the end those which are proportional to $(d-3)$. These
terms are absent in $d=3$, and notably all those which involve the
potential $K$. Note finally that in expression (\ref{sourcesc}),
none of the terms proportional to $\delta_{ij}$ contributes to our
present calculation, since $\Sigma_{ij}$ is multiplied by the
trace-free tensor $\widehat{y}_{ijL}$ in Eq.~(\ref{ILexpr2}). We
nevertheless quote these terms for completeness, as they may be
useful for future works.

\section{Difference between dimensional and pHS
regularizations}\label{secV}

Let us first recall that the general strategy we are following, in
order to obtain the complete 3PN wave generation from two point
masses, consists of two main steps. They have been devised at the
occasion of the application of dimensional regularization (DR) to
the problem of the 3PN equations of motion \cite{DJSdim,BDE04}, and
are:
\begin{enumerate}
\item[(i)] To obtain the expression of the 3PN mass quadrupole
moment in the case of two point masses, using for the required
self-field regularization the so-called
\textit{pure-Hadamard-Schwartz} (pHS) regularization;
\item[(ii)] To add to the pHS result the \textit{difference} between
DR and the pHS regularization, which, as we shall see, is
exclusively due to the presence of poles in $d$ dimensions
(proportional to $1/\varepsilon$).
\end{enumerate}
Step~(i) has already been achieved in our previous papers devoted to
Hadamard-regularization computations of the multipole moments
\cite{BIJ02,BI04mult}; the present paper deals with step~(ii) of
this general method, and constitutes the central part of our
application of dimensional regularization in the problem. We refer
to \cite{BDE04} for a precise definition of the pHS regularization,
and to \cite{BDEI04} for a summary and discussion of the overall
method. Note that, in order to apply step (ii) we transformed a few
terms in the expression obtained by inserting the effective sources
(\ref{sources}) into the multipole moments $I_L$ so as to exactly
parallel the form used in \cite{BI04mult}. This is notably the case
for terms that will be discussed in Sec.~\ref{secVII} below.

A well known result (see Refs.~\cite{JaraS98,JaraS99,BF00,BFeom}) is
that at the 3PN order, Hadamard's regularization, and in fact any of
its variants like the pHS one, permits the computation of most of
the terms (both in the equations of motion and in the radiation
field at infinity), \textit{except for a few} terms which are
``ambiguous'' in the sense that this particular regularization gives
different results for certain divergent integrals, depending on how
one performs the integration (\textit{e.g.}, by integrating by parts
or not). In fact, the ambiguous integrals are those which exhibit
some \textit{logarithmic} divergencies, corresponding to the
occurrence of poles in $d$ dimensions. As it turns out, the
structure of the ambiguous terms is always of a simple and limited
type, and can therefore be parametrized by means of a few arbitrary
unknown numerical constants called the ``\textit{ambiguity
parameters}''. It was shown in Refs.~\cite{BIJ02,BI04mult} that the
Hadamard regularization of the 3PN mass-quadrupole moment $I_{ij}$
of point particle binaries\footnote{The mass-quadrupole moment is
the only one needed to be computed with full 3PN accuracy, thus it
contains most of the difficult non-linear integrals, and all the
ambiguities associated with Hadamard's regularization.} is complete
up to \textit{three} and only three ambiguity parameters, which were
denoted by $\xi$, $\kappa$ and $\zeta$.

The regularization used in the first work \cite{BIJ02} was a certain
variant of the Hadamard regularization called ``hybrid'', and the
ambiguity parameters $\xi$, $\kappa$ and $\zeta$ were originally
defined with respect to that hybrid regularization. The next
calculation, performed in \cite{BI04mult}, has been based on the pHS
regularization [step~(i)], and therefore we had to perform some
numerical shifts of the values of $\xi$, $\kappa$ and $\zeta$, in
order to take into account the different reference points for their
definition (hybrid regularization in \cite{BIJ02}, pHS one in
\cite{BI04mult}). An important and non trivial check of these
computations has precisely been the very \textit{existence of a
unique numerical shift} for each of the ambiguity parameters, such
that the results of both the computations \cite{BIJ02} and
\cite{BI04mult} are in complete agreement. Indeed, as we said, these
two computations differ in the adopted regularizations, but they
also differ by many details concerning their technical
implementations, like the use of different ``elementary''
potentials. Indeed, in \cite{BIJ02} some instantaneous Poisson-like
versions of the elementary potentials, say $U$, $U_i$, $\cdots$,
were adopted. However, in Ref.~\cite{BI04mult} we preferred to use
the retarded elementary potentials $V$, $V_i$, $\cdots$, which are
the same as in the work on the equations of motion
\cite{BFeom,BDE04}, and also the same as those we employ in the
present paper (see Sec.~\ref{secIV}).

\subsection{Difference for $d$-dimensional spatial
integrals}\label{secVA}

In this section we derive a general formula for the ``difference''
between DR and the pure-Hadamard-Schwartz (pHS) regularization. We
shall not review the meaning and precise definition of the pHS
regularization, and simply refer to Sec.~III of \cite{BDE04} and
Sec.~IV of \cite{BI04mult} for full details. The difference
investigated here concerns the typical (non-compact-support) terms
occurring in the multipole moments, which are in the form of some
spatial integrals over $\mathbb{R}^3$ or $\mathbb{R}^d$. Our
investigation parallels the one of Sec.~IV\,B in \cite{BDE04}, which
dealt with the difference for the case of Poisson and Poisson-like
potentials, appropriate to the equations of motion. However, because
the Poisson potentials depend not only on time $t$ but also on the
field point $\mathbf{x}$, while the integrals we consider here for
the multipole moments are functions of time $t$ only, the derivation
of the end formula will be substantially simpler than in the case of
the equations of motion, so we shall only give the main result.

In $3$ dimensions the generic functions we have to deal with, say
$F(\mathbf{x})$, are smooth on $\mathbb{R}^3$ except at two singular
points $\mathbf{y}_1$ and $\mathbf{y}_2$, around which they admit
Laurent-type expansions in powers (and inverse powers) of
$r_1\equiv\vert\mathbf{x}-\mathbf{y}_1\vert$ and
$r_2\equiv\vert\mathbf{x}-\mathbf{y}_2\vert$.\footnote{The function
$F(\mathbf{x})$ depends also on time $t$, through for instance their
dependence on the velocities $\mathbf{v}_1(t)$ and $\mathbf{v}_2(t)$,
but the (coordinate) $t$ time is purely ``spectator'' in the
regularization process, and thus will not be indicated.} When
$r_1\rightarrow 0$ we have (for any $N\in\mathbb{N}$)
\begin{equation}\label{Fx}
F(\mathbf{x})=\sum_{p_0\leq p\leq N}r_1^p
\mathop{f}_1{}_p(\mathbf{n}_1)+o(r_1^N)\,.
\end{equation}
The Landau symbol $o$ takes its usual meaning; the coefficients
$\mathop{f}_1{}_p(\mathbf{n}_1)$ depend on the unit vector
$\mathbf{n}_1\equiv\vert\mathbf{x}-\mathbf{y}_1\vert/r_1$. Since the
powers $p$ can be positive as well as negative integers, the
expansion (\ref{Fx}) is singular, but there is a maximal order of
divergency, $p_0\in\mathbb{Z}$.

In $d$ dimensions, there is an analogue of the function $F$, which
results from the same detailed PN iteration process as the one
leading to $F$ but performed in $d$ dimensions (see the discussion
in \cite{BDE04}); let us call this $d$-dimensional function
$F^{(d)}(\mathbf{x})$, where $\mathbf{x}\in\mathbb{R}^d$. When
$r_1\rightarrow 0$ this function admits a singular expansion which
is more complicated than in $3$ dimensions, and reads
\begin{equation}\label{Fdx}
F^{(d)}(\mathbf{x})=\sum_{\substack{p_0\leq p\leq N\\
q_0\leq q\leq
q_1}}r_1^{p+q\varepsilon}
\mathop{f}_1{}_{p,q}^{(\varepsilon)}(\mathbf{n}
_1)+o(r_1^N)\,,
\end{equation}
with dimension-dependent coefficients
$\mathop{f}_1{}_{p,q}^{(\varepsilon)}(\mathbf{n}_1)$ (recall that
$\varepsilon\equiv d-3$), and where $p$ and $q$ are relative
integers whose values are limited by some $p_0$, $q_0$ and $q_1$ as
indicated. We will be interested here in integrands
$F^{(d)}(\mathbf{x})$ which have no poles as $\varepsilon
\rightarrow 0$ (the poles in $I_L$ being generated by integrating
these integrands), since this will always be the case at 3PN order.
Therefore, we deduce from the fact that $F^{(d)}(\mathbf{x})$ is
continuous at $d=3$, \textit{i.e.}, $\lim_{d\rightarrow 3}F^{(d)}=F$,
the constraint
\begin{equation}\label{constr}
\sum_{q=q_0}^{q_1}\mathop{f}_1{}_{p,q}^{(\varepsilon=0)}
(\mathbf{n}_1) = \mathop{f}_1{}_{p}(\mathbf{n}_1)\,.
\end{equation}

In the present paper we are interested in spatial integrals $\int
d^d\mathbf{x}\,F^{(d)}(\mathbf{x})$ representing generic terms in
the multipole moments. Here, $F^{(d)}(\mathbf{x})$ is a
\textit{non-compact-support} term in the integrand of the multipole
moments, which follows from Eqs.~(\ref{sources}) in Sec.~\ref{secIV}
above. We do not consider the compact support terms (proportional to
$\sigma$, $\sigma_i$ and $\sigma_{ij}$) since their contribution to
the moments has already been computed in Ref.~\cite{BI04mult} and
they give no contribution in the ``difference'' between the
dimensional and pHS regularizations. Furthermore, we assume in the
definition of the function $F^{(d)}$ that the derivatives of the
elementary potentials therein are taken in the ordinary, non
distributional sense (we further comment below on how the
distributional parts of the derivatives have been taken into account
into the formalism). Furthermore we do not need to consider here the
non-compact-support terms in the multipole moments which have a form
such that their spatial integral depends solely on the boundary at
infinity, $\vert\mathbf{x}\vert\rightarrow +\infty$. These terms
have been discussed in Sec.~IV\,D of \cite{BI04mult}; they provide a
crucial contribution to the multipole moments in $3$ dimensions
computed in \cite{BIJ02,BI04mult}. However, we shall show in
Sec.~\ref{secVB} below that, thanks to the $d$-dimensional
generalization of the finite part process $\mathop{\mathrm{FP}}_{B}$
defined in Sec.~\ref{secIIIA} above, these terms do not contribute
to the difference ``DR - HR'' we are interested in.

Finally, we take for $F^{(d)}$ a generic non-compact-support term,
whose integral cannot be expressed as an integral at infinity,
\textit{i.e.}, not of the form which is discussed in Sec.~IV\,D of
\cite{BI04mult}. The general structure of such $F^{(d)}$ is that of
a multipolar factor $\widehat{x}_L$ times some multilinear
functional, say $\mathcal{P}$, of the elementary potentials (in $d$
dimensions) and their derivatives,
\begin{equation}\label{FcalP}
F^{(d)}(\mathbf{x})=\widehat{x}_L\,\mathcal{P}
[ V, \,V_i, \,\hat{W}_{ij},
\,\cdots, \,\partial_i V, \cdots]\,.
\end{equation}
For the present calculation, the derivatives of potentials in this
definition, $\partial_i V, \cdots$, are ordinary derivatives. Many
terms of Eqs.~(\ref{sources}) are made of a spatial integral applied
to some partial time derivative of a function of the type $F^{(d)}$.
For these terms we always put the time derivatives outside the
integral and perform first the spatial integral using the
regularization, and only then apply the (total) time derivative.

Since we shall prove in Sec.~\ref{secVB} that the difference between
the integrals involving $F^{(d)}$ and $F^{(3)}$ does not involve any
contribution coming from divergencies ``at infinity'', we limit
ourselves to spatial integrals which extend over a finite volume in
the $d$-dimensional space, say the spherical ball
$\mathcal{B}(\mathcal{R})$ defined by $\vert\mathbf{x}\vert <
\mathcal{R}$, where $\mathcal{R}$ denotes some arbitrary constant
radius. The results we shall derive below will not depend on
$\mathcal{R}$. In Hadamard's regularization, and particularly in the
pHS variant of it, the $3$-dimensional spatial integral is defined
by the so-called \textit{partie finie} (Pf) prescription, depending
on two arbitrary constants $s_1$ and $s_2$, say
\begin{equation}\label{H}
H = \mathop{\mathrm{Pf}}_{s_1,s_2}\int_{\mathcal{B}(\mathcal{R})}
d^3\mathbf{x}\,F(\mathbf{x})\,.
\end{equation}
Of course $H$ is in fact a function of time but we do not need to
indicate this. By definition, Hadamard's partie finie integral is
given by the following limit when the radius $s$ of two
``regularizing volumes'' surrounding the singularities tends to
zero, say
\begin{eqnarray}\label{Pf}
H &=& \lim_{s\rightarrow
0}\biggl\{\int_{\mathcal{B}(\mathcal{R})\setminus
\mathcal{B}_1(s)\cup \mathcal{B}_2(s)}d^3\mathbf{x}
\,F(\mathbf{x})\nonumber\\
&&\quad +4\pi\sum_{p=p_0}^{-4}
\frac{s^{p+3}}{p+3}\,\bigl<\mathop{f}_1{}_p\bigr>
+4\pi\ln \left( \frac{s}{s_1}\right)
\bigl<\mathop{f}_1{}_{-3}\bigr>+1\leftrightarrow 2\biggr\}.
\end{eqnarray}
The symbol $1\leftrightarrow 2$ means the same terms but with the
singularities' labels 1 and 2 exchanged. The first term represents
an ordinary integral extending over the region obtained from
$\mathcal{B}(\mathcal{R})$ by excising two spherical balls
$\mathcal{B}_1(s)$ and $\mathcal{B}_2(s)$ centered on the two
singularities, each having the same radius $s$ (evidently we can
always assume $s\ll\mathcal{R}$).\footnote{Two balls with different
radii could be used as well, without changing the results.} The
extra terms in (\ref{Pf}), which are such that they cancel out the
singular part of the first term when $s\rightarrow 0$ (so that the
partie finie exists by definition), involve the usual
(two-dimensional) spherical average
\begin{equation}\label{average}
\bigl<f\bigr>\equiv\int
\frac{d\Omega(\mathbf{n}_1)}{4\pi}\,f(\mathbf{n}_1)\,,
\end{equation}
where $d\Omega(\mathbf{n}_1)$ is the solid angle element around
$\mathbf{n}_1$. The length scales $s_1$ and $s_2$ (one for each
particle) are introduced in Eq.~(\ref{Pf}) in order to
adimensionalize the radius $s$ in the logarithmic terms. They play a
key role at 3PN order, since their appearance signals the presence
of logarithmic divergences which correspond to poles $\propto
1/\varepsilon$ in $d$ dimensions. A way to interpret these constants
is to say that they reflect an arbitrariness in the original choice
of the two regularizing volumes $\mathcal{B}_1(s)$ and
$\mathcal{B}_2(s)$.

In dimensional regularization the situation is much simpler, since
the integral will be (so to speak) ``automatically'' regularized by
means of the analytic continuation of the $d$-dimensional volume
element. Thus, we simply have
\begin{equation}\label{Hd}
H^{(d)} = \int_{\mathcal{B}^{(d)}(\mathcal{R})}d^d\mathbf{x}\,
F^{(d)}(\mathbf{x})\,,
\end{equation}
where $\mathcal{B}^{(d)}(\mathcal{R})$ is the $d$-dimensional ball
with radius $\mathcal{R}$. Given the results of the two
regularizations, (\ref{H}) and (\ref{Hd}), we consider what we call
the difference, which is what we shall have to add to the pHS result
in order to obtain the DR result, namely
\begin{equation}\label{DH}
\mathcal{D}H \equiv H^{(d)} - H\,.
\end{equation}
We shall compute $\mathcal{D}H$ in the limit where
$\varepsilon\rightarrow 0$, keeping the pole part $\propto
\varepsilon^{-1}$ (at 3PN order only simple poles will occur) and
the finite term $\propto \varepsilon^0$, but neglecting
$\mathcal{O}(\varepsilon)$. Using the same method as in
\cite{DJSdim,BDE04}, $\mathcal{D}H$ can be obtained by splitting the
$d$-dimensional integral (\ref{Hd}) into three volumes, two
spherical balls $\mathcal{B}_1^{(d)}(s)$ and
$\mathcal{B}_2^{(d)}(s)$ of radius $s$, which are the
$d$-dimensional analogues of $\mathcal{B}_1(s)$ and
$\mathcal{B}_2(s)$, and the complementary volume in
$\mathcal{B}^{(d)}(\mathcal{R})$, say
$\mathcal{B}^{(d)}(\mathcal{R})\setminus \mathcal{B}_1^{(d)}(s)\cup
\mathcal{B}_2^{(d)}(s)$. It is clear that the integral over the
latter complementary volume reduces, when $\varepsilon\rightarrow 0$
(with fixed $s$), to the integral over
$\mathcal{B}(\mathcal{R})\setminus \mathcal{B}_1(s)\cup
\mathcal{B}_2(s)$, which is the first term in the definition
(\ref{Pf}) of Hadamard's partie finie, and does not contribute to
the difference modulo some negligible terms
$\mathcal{O}(\varepsilon)$. It remains thus the ``local''
contributions of the two volumes $\mathcal{B}_1^{(d)}(s)$ and
$\mathcal{B}_2^{(d)}(s)$, which can be straightforwardly computed
{}from inserting into them the local singular expansions given by
(\ref{Fdx}) and $1\leftrightarrow 2$. We can then connect the result
to the corresponding result in $3$ dimensions by using the
constraint (\ref{constr}). Finally, we obtain for the difference
$\mathcal{D}H$ the following outcome:
\begin{equation}\label{DHres}
\mathcal{D}H =
\frac{\Omega_{2+\varepsilon}}{\varepsilon}\sum_{q=q_0}^{q_1}
\left[\frac{1}{q+1}+\varepsilon \ln s_1\right]
\bigl<\mathop{f}_1{}_{-3,q}^{(\varepsilon)}\bigr>_{2+\varepsilon} +
1\leftrightarrow 2
+ \mathcal{O}(\varepsilon)\,,
\end{equation}
where the spherical average \textit{performed in $d$ dimensions} is
defined by
\begin{equation}\label{averaged}
\bigl< f \bigr>_{d-1} \equiv \int
\frac{d\Omega_{d-1}(\mathbf{n}_1)}{\Omega_{d-1}}\,f(\mathbf{n}_1)\,.
\end{equation}
The volume of the $(d-1)$-dimensional sphere, embedded into
$d$-dimensional space, is given by
$\Omega_{d-1}=2\pi^\frac{d}{2}/\Gamma(\frac{d}{2})$; for instance,
$\Omega_2=4\pi$. Actually, we can see that the $\Omega_{d-1}$'s
cancel out between (\ref{DHres}) and (\ref{averaged}).

Let us now comment on the inclusion in the present formalism of
derivatives in a \textit{distributional} sense. An important feature
of the pHS regularization is the systematic use of distributional
derivatives \textit{\`a la} Schwartz \cite{Schwartz}. It has been
shown both in the contexts of the equations of motion
\cite{JaraS98,BFeom} and of the radiation field
\cite{BIJ02,BI04mult} that the purely distributional parts of
derivatives yield a crucial physical contribution to the results at
the 3PN order. In Hadamard's regularization, various prescriptions
are possible for the distributional derivatives. For instance, some
generalized distributional derivatives, defined in the extended
Hadamard regularization \cite{BFreg}, were used for the 3PN
equations of motion in \cite{BF00,BFeom}. Using different
prescriptions yields different results, which however differ at the
3PN order by some terms having the form of the ambiguous terms, and
therefore which merely change the values of the ambiguity parameters
($\xi$, $\kappa$ and $\zeta$ in the radiation field). Now we showed
\cite{BDE04} that in DR the correct prescription for the derivatives
is the one of the standard distribution theory \cite{Schwartz}. This
is why we have included Schwartz derivatives in the definition of
the pHS regularization, which constitutes in some sense the ``core''
part of DR, by which we mean the part which computes all the
difficult non-linear integrals, but leaves unspecified a few terms
corresponding exclusively to the ``ambiguous'' logarithmic
divergences. Of course, since different variants of Hadamard's
regularization differ precisely in different definitions for the
ambiguity parameters, all of them could be regarded as the ``core''
of DR. However, the point is that the pHS regularization is the only
one for which the final result of DR is to be obtained by adding
exactly the ``difference'' in the way we have computed it in
Eq.~(\ref{DHres}). To summarize, Eq.~(\ref{DHres}) as its stands is
simply to be added to the pHS result, since the latter already
includes the distributional derivatives \textit{\`a la} Schwartz,
whose contributions have been computed in Ref.~\cite{BI04mult}.

\subsection{Proof that the outer-near-zone divergencies do not
contribute to the difference}\label{secVB}

Let us recall the logic that led us to introducing and using the
specific, $d$-dimensionally generalized, finite part (FP) process.
Initially, in the MPM construction of the multipole expansion of the
external metric, when iteratively solving Einstein's equations, we
were faced with some integrands $N_n$ that had a singular behavior
at the origin of the spatial coordinates, \textit{i.e.}, as
$r\rightarrow 0$. One then \textit{defined} the FP of the retarded
integral of $N_n$ by first introducing a factor
$\widetilde{r}^B=(r/r_0)^B$ in the integrand, and then subtracting
the ``quasi-multiple'' shifted poles $C^{(d)}_{-k}(B - q_1
\varepsilon)^{-1} \cdots (B - q_k \varepsilon)^{-1}$ [first term on
the RHS of Eq.~(\ref{B14})], before taking the continuation down to
$B=0$. At this stage, the integrand was, in
principle, defined as a post-Minkowskian expansion, with good
convergence properties at $r \rightarrow \infty$, so that the poles
$\propto (B - q_1 \varepsilon)^{-1} \cdots (B - q_k
\varepsilon)^{-1}$ came only from the region where $r \rightarrow
0$. Later, the external MPM construction was combined with a
straightforward post-Newtonian (PN) iteration of Einstein's
equations, which took into account the interior region containing
the material source $T^{\mu\nu}$. With a generalization of the
argument used in \cite{B98mult}, one could formally relate the
source multipole moments, used in the MPM formalism to parametrize
the source, to integrals over the PN expansion of the effective
stress-energy pseudo tensor $\tau^{\mu\nu} = \vert g \vert \,
T^{\mu\nu} +$ non-linear terms.

This led to what was the starting point of our investigation, namely
to formal expressions for the source multipole moments (of the mass
type) of the symbolic form
\begin{equation}\label{4.1}
I_L = \mathrm{FP} \int d^d \mathbf{x} \,\widetilde{r}^B
\,\widehat{x}_L\,\left\{ \overline{gT} + \overline{\Lambda(h)}
\right\}\,,
\end{equation}
where we recall that the overline denotes a PN (or near zone)
expansion. Note that the presence of this PN expansion process in
Eq.~(\ref{4.1}) is crucial to its validity. Indeed, the argument
used to derive (\ref{4.1}) was based on transforming MPM-expanded
integrands singular when $r \rightarrow 0$ into PN-expanded ones
diverging when $r \rightarrow \infty$. (As discussed in
\cite{BDI04zeta} the formal limit $r \rightarrow \infty$, taken
within a PN-expanded integrand, physically corresponds to the
``outer near zone'' $a \ll r \ll \lambda$, and should not be
confused with a far zone expansion $r \gg \lambda$, in the sense of
spatial infinity $\mathcal{I}^0$.) Technically, the transformation
between the two types of singular integrals was based on the
analytic continuation with respect to $B$ of the integrals, using
the fact that $\int_0^{\infty} dr \, r^{B-p-q\varepsilon} = 0$. In
this reshuffling from the UV ($r \rightarrow 0$) to the IR ($r
\rightarrow \infty$) it was essential to keep the same meaning of
the symbol FP in front of (\ref{4.1}). Indeed, one sees easily, by
separating $\int_0^{\infty} dr \, r^{B-p-q\varepsilon}$ into
$\int_0^{\mathcal R} dr \, r^{B-p-q\varepsilon}$ and $\int_{\mathcal
R}^{\infty} dr \, r^{B-p-q\varepsilon}$ that the MPM-poles $\propto
(B - q \varepsilon)^{-1}$ generated near $r=0$ (when $p=1$) become
transformed in the same (modulo a sign) poles, generated near $r =
\infty$ by the singular behavior of the PN integrand. In addition to
the poles $\propto (B - q_1 \varepsilon)^{-1} \cdots (B - q_k
\varepsilon)^{-1}$ present in the integral (\ref{4.1}), and
generated by the behavior of the PN-expanded non-linear terms
$\overline{\Lambda(h)}$ at $r \rightarrow \infty$, there are also
poles $\propto \varepsilon^{-1}$ associated to the singular behavior
of $\Lambda(h)$ near $\mathbf{x} = \mathbf{y}_1$ and $\mathbf{x} =
\mathbf{y}_2$. But clearly, if we split the integral $\int d^d
\mathbf{x}$ in a part $r < {\mathcal R}$ enclosing the two mass
points, and a complementary part $r > {\mathcal R}$, the latter
integral will have no singularities associated to $\mathbf{x} =
\mathbf{y}_1$ or $\mathbf{x} = \mathbf{y}_2$, and therefore will
have no genuine poles $\varepsilon^{-1}$.

The conclusion is that the restriction of the integral (\ref{4.1}) to
the outer near zone $r > {\mathcal R}$ (corresponding to the IR), say
\begin{equation}\label{4.2}
I_L^\mathrm{IR}(B,\varepsilon) = \int_{r > {\mathcal R}} d^d
\mathbf{x} \,\widetilde{r}^B \,\widehat{x}_L\,\left\{ \overline{gT}
+ \overline{\Lambda(h)} \right\},
\end{equation}
is a meromorphic function of the complex variables $B$ and
$\varepsilon$ which will have, when $B$ and $\varepsilon$ are both
near zero, the same type of quasi-multiple shifted poles structure
as the MPM quantity $F_n^{(d)} (B)$ of Eq.~(\ref{B14}), say
\begin{equation}\label{4.3}
I^{\rm IR} (B,\varepsilon) = \sum \frac{C_{-k}^{(d)}}{(B - q_1
\varepsilon) (B - q_2 \varepsilon) \cdots (B - q_k \varepsilon)} +
C_0^{(d)} + C_1^{(d)} B + \mathcal{O}\left(B^2\right).
\end{equation}
The important point is that, when the expansion is written in the
form (\ref{4.3}), the various coefficients $C_{-k}^{(d)}$,
$C_0^{(d)}$, \textit{etc.}, are regular functions of $d$, which are
continuous at $d=3$.

The structure (\ref{4.3}) proves the result we wanted, namely the
fact that the IR parts of the integrals in $d$ dimensions,
\begin{equation}\label{4.3prime}
I^{\rm IR} (\varepsilon) = \mathop{\mathrm{FP}}_B\left[I^{\rm IR}
(B,\varepsilon)\right],
\end{equation}
admits when $\varepsilon \rightarrow 0$ the same value as the one
given in $3$ dimensions by the original definition of the finite
part FP (in $3$ dimensions). That is to say, the two operations of
taking the FP and the limit $\varepsilon \rightarrow 0$ commute.
Indeed, the definition (\ref{B15}) of the $d$-dimensional FP
operation yields
\begin{equation}\label{4.4}
I^\mathrm{IR} (\varepsilon) = C_0^{(d)} \,,
\end{equation}
so that
\begin{equation}\label{4.5}
\lim_{\varepsilon \rightarrow 0} I^\mathrm{IR}(\varepsilon) =
C_0^{(3)} \,.
\end{equation}
On the other hand, if we interchange the order of the two operations,
we must first consider the limit when $\varepsilon \rightarrow 0$ of
(\ref{4.3}), namely
\begin{equation}\label{4.6}
\lim_{\varepsilon \rightarrow 0} I^\mathrm{IR} (B,\varepsilon) = \sum
\frac{C_{-k}^{(3)}}{B^k} + C_0^{(3)} + C_1^{(3)} B +
\mathcal{O}\left(B^2\right),
\end{equation}
and then, by applying the usual FP operation in $3$ dimensions, we
must discard the poles $B^{-k}$ and evaluate the remainder at $B=0$,
thus
\begin{equation}\label{4.7}
\mathop{\mathrm{FP}}_B \left[\,\lim_{\varepsilon \rightarrow 0}
I^\mathrm{IR}(B,\varepsilon)\right] = C_0^{(3)}\,,
\end{equation}
which is the same result as found in (\ref{4.5}). This shows that
all the IR terms, formally depending on the boundary of the integral
at infinity, notably all those discussed in Sec.~IV\,D of
\cite{BI04mult}, give exactly zero in the difference between DR and
the pHS regularization. Their contribution to the multipole moments
has already been taken into account in Ref.~\cite{BI04mult}. This
shows also that the outer-zone part of the last term in the
expression of the moment, Eq.~(\ref{ILexpr2}), which is proportional
to $B\varepsilon$, is actually zero in the present formalism since
it is zero in $3$ dimensions.

\section{Computation of the ambiguity parameters}\label{secVI}

As we have discussed in Sec.~\ref{secVA}, the end result of the
dimensional regularization (DR) is simply given by the sum of the
pure-Hadamard-Schwartz (pHS) regularization and the
``\textit{difference}'' that we have investigated in the general
analysis of Sec.~\ref{secV}. Now the pHS regularization of the 3PN
mass dipole and quadrupole moments of point particles binaries has
already been computed in our previous work \cite{BI04mult}, in which
the end result of the Hadamard regularization (HR) was obtained as
the sum of the pHS result and of some specific ambiguity part
parametrized by three ambiguity parameters. In the present section
we construct the DR result and impose that it is \textit{physically
equivalent} to the HR one given in \cite{BI04mult}. As we shall
show, this requirement will permit us to uniquely determine the
ambiguity parameters.

\subsection{The 3PN mass-quadrupole moment}\label{secVIA}

Let us first state the end result of \cite{BI04mult} concerning the
3PN mass quadrupole moment as computed with HR. We denote it by
$I_{ij}^{(\mathrm{HR})}$; see Eqs.~(5.9)--(5.10) in \cite{BI04mult}
for its complete expression in the center-of-mass frame. In the
present paper we shall not need the explicit formula for the moment
(which includes many complicated coefficients), but simply its
structure, made of the sum of the pHS moment and some quite simple
ambiguous contribution containing three and only three ambiguity
parameters. The ambiguous part reads
\begin{equation}\label{amb}
\Delta I_{ij}\Bigl[\hat{\xi},\,\hat{\kappa},\,\hat{\zeta}\Bigr] =
\frac{44}{3}\,\frac{G_N^2\,m_1^3}{c^6}\biggl[\left(\hat{\xi} +
\hat{\kappa} \frac{m_1+m_2}{m_1}\right) y_1^{\langle
i}a_1^{j\rangle} + \hat{\zeta} \,v_1^{\langle
i}v_1^{j\rangle}\biggr] + 1\leftrightarrow 2\,,
\end{equation}
where $m_1$ and $m_2$ are the two masses, $y_1^i$, $v_1^i$ and
$a_1^i$ are the position, coordinate velocity and coordinate
acceleration of the particle 1 ($1\leftrightarrow 2$ denotes the
same for the particle 2), and where the angular brackets surrounding
indices mean the STF projection. All the quantities in (\ref{amb})
are defined in $3$ dimensions; $G_N$ is Newton's constant, related
to $G$ in $d$ dimensions by Eq.~(\ref{GN}). Obviously, since
(\ref{amb}) is already of order 3PN (\textit{cf.} the factor
$1/c^6$), the acceleration $a_1^i$ is simply given by the usual
Newtonian value (in $3$ dimensions).

The expression (\ref{amb}) contains three ambiguity parameters
$\{\hat{\xi},\,\hat{\kappa},\,\hat{\zeta}\}$. These are the ones
which would be defined with respect the pHS regularization. However,
the ambiguity parameters were in fact defined earlier in
Ref.~\cite{BIJ02}, which had adopted a different Hadamard-type
regularization, called ``hybrid'', instead of the pHS
one.\footnote{The hybrid regularization mainly differs from the pHS
regularization in the way the ``contact'' (compact-support) terms
are computed. Indeed, the hybrid regularization takes into account
the so-called ``non-distributivity'' of Hadamard's regularization,
which is the fact that $(F\,G)_1\not=(F)_1(G)_1$ in general, where
$(F)_1$ is the partie finie of a singular function $F$ at the point
$\mathbf{y}_1$. (In this respect, the hybrid regularization is like
the extended Hadamard regularization defined in \cite{BFreg}.) This
introduces also some differences in the case of non-compact support
integrals, between the ``case-by-case'' integration followed in
\cite{BIJ02} and the systematic pHS regularization of these
integrals adopted in \cite{BI04mult}.} Accordingly, the pHS
ambiguity parameters $\{\hat{\xi},\,\hat{\kappa},\,\hat{\zeta}\}$
differ from their hybrid counterparts in \cite{BIJ02}, which were
denoted there $\{\xi,\,\kappa,\,\zeta\}$. The result, which
constituted a powerful check of the computations of
\cite{BIJ02,BI04mult}, is that
\begin{subequations}\label{hatamb}\begin{eqnarray}
\hat{\xi} &=& \xi + \frac{1}{22}\,,\\
\hat{\kappa} &=& \kappa\,,\\
\hat{\zeta} &=& \zeta + \frac{9}{110}\,.
\end{eqnarray}\end{subequations}
In the present paper we prefer to stick to the original definition
of the parameters $\{\xi,\,\kappa,\,\zeta\}$, since these have
already been used in the computation of the 3PN binary orbital
phasing \cite{BFIJ02} and in the discussion of the efficiency of the
3PN templates (see \textit{e.g.} \cite{DIJS03}). Hence the final
outcome from HR for the 3PN mass quadrupole moment of the binary
(moving on a general, not necessarily circular, orbit) is written as
\begin{equation}\label{IijH}
I_{ij}^{(\mathrm{HR})}\bigl[r_1',\,r_2',\,r_0;
\,\xi,\,\kappa,\,\zeta\bigr]=
I_{ij}^{(\mathrm{pHS})}\bigl[r_1',\,r_2',\,r_0\bigr]
+\Delta I_{ij}\Bigl[\xi+
\frac{1}{22},\,\kappa,\,\zeta+ \frac{9}{110}\Bigr].
\end{equation}
The pHS part, first term on the RHS, is free of the ambiguities
$\xi$, $\kappa$ and $\zeta$, but depends on the three regularization
scales $r_1'$, $r_2'$ and $r_0$. First, $r_0$ is merely the scale we
have introduced in the general MPM formalism, see (\ref{rtilde}) in
Sec.~\ref{secIIIA}, and which then appears in the definition of the
source multipole moments in Sec.~\ref{secIIIC}. This scale will
disappear when we relate the asymptotic waveform to the local matter
distribution for general extended sources. The other scales $r_1'$
and $r_2'$ are specific to the application to the case of systems of
point particles and come from regularizing self-field effects.
\textit{By definition} of the ambiguity parameters these scales are
taken to be the \textit{same} as the two scales that appear in the
final expression of the 3PN equations of motion in harmonic
coordinates computed in Refs.~\cite{BF00,BFeom}.\footnote{Actually
only the ambiguity parameters $\xi$ and $\kappa$ depend on this
choice; see \cite{BIJ02,BI04mult} for discussions.} They came from
the regularization of Poisson-type integrals in the equations of
motion, where they can be interpreted as some infinitesimal radial
distances used as cut-offs when the field point tends to the
singularities. It should be noted that $r'_1$ and $r'_2$ are
``unphysical'', in the sense that they can be arbitrarily modified
(though they can never be removed) by a coordinate transformation of
the ``bulk'' metric outside the particles \cite{BFeom}, or, more
consistently when we consider the renormalization which follows the
regularization, by suitable shifts of the particles' world-lines
\cite{BDE04}.

To get the DR result we must augment the pHS result
$I_{ij}^{(\mathrm{HR})}\bigl[s_1,s_2,\,r_0;
\,\xi,\,\kappa,\,\zeta\bigr]$ computed for any choice of Hadamard
regularization scales $s_1,s_2$ entering Eq.~(\ref{Pf}), by the
corresponding difference
$\mathcal{D}I_{ij}\bigl[s_1,\,s_2;\,\varepsilon,\,\ell_0\bigr]$,
which is made of the sum of all the contributions $\mathcal{D}H$,
Eq.~(\ref{DHres}), computed for all the individual non-compact
support terms in the 3PN expression of the source quadrupole moment
deduced from the explicit formulas given in Sec.~\ref{secIV}. Hence
this difference reads
\begin{equation}\label{sumdiff}
\mathcal{D}I_{ij}\bigl[s_1,\,s_2;\,\varepsilon,\,\ell_0\bigr] =
\sum_{\substack{\text{non-compact}\\
\text{terms in $I_{ij}$}}}
\mathcal{D}H\bigl[s_1,\,s_2;\,\varepsilon,\,\ell_0\bigr].
\end{equation}
The sum in the RHS runs over all the non-compact support terms
excluding those which are in a form such that they depend only on
the IR behavior of the integral; indeed these terms do not
contribute to the difference (see Sec.~\ref{secVB}). We recall also
that in the calculation of the difference we do not have to take
into account the compact-support terms, nor the distributional parts
of the derivatives since they are already included in the pHS
result. In (\ref{sumdiff}) we indicated that the difference depends
both on the constants $\varepsilon=d-3$ and $\ell_0$ associated with
DR, and on the two scales $s_1$, $s_2$ which were introduced into
the Hadamard partie finie (\ref{Pf}). The DR result is then
\begin{equation}\label{Iijdr}
I_{ij}^{(\mathrm{DR})}\bigl[r_0;\,\varepsilon,\,\ell_0\bigr] =
I_{ij}^{(\mathrm{pHS})}
\bigl[s_1,\,s_2,\,r_0\bigr]+\mathcal{D}I_{ij}\bigl[s_1,\,s_2;\,
\varepsilon,\,\ell_0\bigr].
\end{equation}
The choice of Hadamard regularization length scales $s_1$, $s_2$ in
Eq.~(\ref{Iijdr}) is arbitrary because, as is easily checked $s_1$,
$s_2$ \textit{cancel out} between the two terms in the RHS of
(\ref{Iijdr}), so that, as it should be, $I_{ij}^{(\mathrm{DR})}$
depends only on the DR characteristics $\varepsilon$ and $\ell_0$
(and also on $r_0$ which belongs to our general multipole moment
formalism, and is in fact irrelevant for the present discussion).
Because of this independence on the choice of the scales $s_1$,
$s_2$, we can choose them to be identical to the two specific length
scales $r_1'$, $r_2'$ entering the 3PN equations of motion.
Therefore we can rewrite Eq.~(\ref{Iijdr}) as
\begin{equation}\label{Iijdr'}
I_{ij}^{(\mathrm{DR})}\bigl[r_0;\,\varepsilon,\,\ell_0\bigr] =
I_{ij}^{(\mathrm{pHS})}
\bigl[r'_1,\,r'_2,\,r_0\bigr]+\mathcal{D}I_{ij}\bigl[
r'_1,\,r'_2;\,\varepsilon,\,\ell_0\bigr].
\end{equation}

Let us now impose the \textit{physical equivalence} between the DR
result (\ref{Iijdr'}) and the corresponding final HR result
(\ref{IijH}) containing the ambiguity parameters $\xi$, $\kappa$ and
$\zeta$. In doing this identification, we must remember, from the
work on the 3PN equations of motion \cite{BDE04}, that the ``bare''
particle positions, $\mathbf{y}_1^\mathrm{bare}$ and
$\mathbf{y}_2^\mathrm{bare}$, entering the DR result differ from
their Hadamard counterparts, say $\mathbf{y}_1^\mathrm{ren}$ and
$\mathbf{y}_2^\mathrm{ren}$, entering the equations of motion of
\cite{BF00,BFeom}, by some (purely spatial) \textit{shifts of the
world-lines}, \textit{i.e.},
\begin{subequations}\label{shiftworldline}\begin{eqnarray}
\mathbf{y}_1^\mathrm{bare}(t)&=&
\mathbf{y}_1^\mathrm{ren}(t)+\bm{\eta}_1(t)\,,\\
\mathbf{y}_2^\mathrm{bare}(t)&=&\mathbf{y}_2^\mathrm{ren}(t)
+\bm{\eta}_2(t)\,.
\end{eqnarray}\end{subequations}
These shifts have been uniquely determined in Ref.~\cite{BDE04} and
denoted there by $\bm{\xi}_1$ and $\bm{\xi}_2$ (see Eqs.~(1.13) and
(6.41)--(6.43) in \cite{BDE04}). In the present work, we denote them
by $\bm{\eta}_1$ and $\bm{\eta}_2$ in order to avoid any confusion
with the name of the ambiguity parameter $\xi$. These shifts of the
world-lines are crucial and must be taken into account when
comparing the DR and HR results. Let us insist that the shifts in
Eqs.~(\ref{shiftworldline}) are those which ensured the equivalence
between the DR and HR results for the equations of motion. Having
made contact in \cite{BDE04} between the renormalization scales
entering the two regularization schemes in the context of the 3PN
equations of motion, we must, by consistency, employ them to compare
the DR and HR results for the 3PN multipole moments. The names
$\mathbf{y}_{1,2}^\mathrm{ren}$ come from the fact that the shifts
permit to \textit{renormalize} the DR result for the equations of
motion, in the sense that all the poles $\propto 1/\varepsilon$
appearing in the $d$-dimensional equations of motion were finally
absorbed into the new definition of the world-lines. A non trivial
check of our present calculations will be to verify that the
\textit{same} shifts allow one to get \textit{finite} (when
$\varepsilon\rightarrow 0$) final expressions for \textit{all} the
multipole moments, when expressed in terms of
$\mathbf{y}_{1,2}^\mathrm{ren}$. Note that the definition of the
shifts corresponds to a \textit{non-minimal subtraction}. This
non-minimality was needed to connect the DR result to the
two-parameter \textit{class} of HR results parametrized by arbitrary
values of the scales $r_1'$ and $r_2'$ (see \cite{BDE04} for a
discussion). Hence the shifts $\bm{\eta}_1$ and $\bm{\eta}_2$ depend
on $r_1'$ and $r_2'$ (respectively). We shall comment more on the
renormalization in Sec.~\ref{secVII}. The precise expression of the
shift is
\begin{equation}\label{shift}
\bm{\eta}_1(r_1';\,\varepsilon,\,\ell_0)
=\frac{11}{3}\frac{G_N^2\,m_1^2}{c^6}\left[
\frac{1}{\varepsilon}-\ln\left(
\frac{r_{12}\,{r'}_1^2\,\overline{q}^{3/2}}{\ell_0^3}\right)
+\frac{1983}{1540}\right] \mathbf{a}_1\,,
\end{equation}
together with the shift of the other world-line obtained by
$1\leftrightarrow 2$. Here, $\mathbf{a}_1$ denotes the
\textit{three-dimensional} Newtonian acceleration $\mathbf{a}_1=-G_N
m_2 \,\mathbf{n}_{12}/r_{12}^2$, where
$r_{12}=\vert\mathbf{y}_{1}-\mathbf{y}_{2}\vert$ and
$\mathbf{n}_{12}=(\mathbf{y}_{1}-\mathbf{y}_{2})/r_{12}$
[\textit{i.e.}, $\mathbf{a}_1$ is the same quantity as in
Eq.~(\ref{amb})], and $G_N$ correspondingly denotes the
\textit{three-dimensional} Newtonian constant. The expression
(\ref{shift}) seems to differ from the one given by Eq.~(1.13) in
\cite{BDE04}, but this is because we have used in (\ref{shift}) the
Newtonian acceleration $\mathbf{a}_1$ in $3$ dimensions, while
Eq.~(1.13) in \cite{BDE04} has been written with the help of the
$d$-dimensional analogue $\mathbf{a}_1^{(d)}$. We have
\begin{subequations}\label{a1d}\begin{eqnarray}
\mathbf{a}_1^{(d)}&=&-\frac{2(d-2)^2}{d-1}\,\tilde{k}\,G\,m_2\,
r_{12}^{1-d}\,\mathbf{n}_{12}\,,\\
\mathbf{a}_1^{(3+\varepsilon)}&=&
\left(1+\varepsilon\left[\frac{3}{2}-\ln\left(
\frac{r_{12}\,\overline{q}^{1/2}}{\ell_0}\right)\right]\right)
\mathbf{a}_1^{(d=3)}
+\mathcal{O}\left(\varepsilon^2\right)\,.
\end{eqnarray}\end{subequations}
Here we used $G = G_N \ell_0^{d-3}$, and $\tilde{k} \equiv
\Gamma[(d-2)/2]/\pi^{(d-2)/2}=
1-\frac{1}{2}\varepsilon\ln\overline{q}+\mathcal{O}
\left(\varepsilon^2\right)$, where $\overline{q}\equiv 4\pi e^C$ with
$C=0.577\cdots$ denoting the Euler constant.

Evidently, since the shifts are at 3PN order, the modification of
the mass quadrupole moment brought about by the latter shifts (in the
sense $I_{ij}[\mathbf{y}^\mathrm{bare}] \equiv
I_{ij}[\mathbf{y}^\mathrm{ren}] +
\delta_{\bm{\eta}} I_{ij}$) simply reads
\begin{equation}\label{deltaetaIij}
\delta_{\bm{\eta}}I_{ij}=2\,m_1\,y_1^{\langle
i}\eta_1^{j\rangle}+1\leftrightarrow 2\,,
\end{equation}
where we recall the fact that the Newtonian limit of the quadrupole
in any dimension $d$ takes the standard expression
$I_{ij}=m_1\,y_1^{\langle i}y_1^{j\rangle}+1\leftrightarrow
2+\mathcal{O}\left(c^{-2}\right)$, see Eq.~(\ref{newtonian}).
The physical equivalence between the DR and HR results simply means
that we require that the full DR quadrupole moment, computed for the
``bare'' particle positions entering the DR delta-function source,
$I_{ij}^{(\mathrm{DR})}[r_0;\, \varepsilon,\,\ell_0;\,
\mathbf{y}_1^\mathrm{bare},\, \mathbf{y}_2^\mathrm{bare}]$,
coincides (when $\varepsilon \rightarrow 0$, and for the correct,
looked-for values of $\xi$, $\kappa$, $\zeta$) with the HR result
$I_{ij}^{(\mathrm{HR})}[r'_1,\, r'_2;\, \xi,\, \kappa,\, \zeta;\,
\mathbf{y}_1^\mathrm{HR},\, \mathbf{y}_2^\mathrm{HR}]$. As said
above, the particle positions $\mathbf{y}_a^\mathrm{HR}$ entering
the HR result must be identified with the ``renormalized'' DR
positions $\mathbf{y}_a^\mathrm{ren}$ introduced in
Eqs.~(\ref{shiftworldline}): $\mathbf{y}_a^\mathrm{HR} \equiv
\mathbf{y}_a^\mathrm{ren}$. Re-expressing the DR multipole moment in
terms of the particle arguments $\mathbf{y}_a^\mathrm{HR} =
\mathbf{y}_a^\mathrm{ren}$, this requirement then leads to equating
\begin{equation}\label{equivDRHR}
I_{ij}^{(\mathrm{HR})}\bigl[r_1',\,r_2',\,r_0;\, \xi,\, \kappa,\,
\zeta;\, \mathbf{y}_1^\mathrm{ren},\, \mathbf{y}_2^\mathrm{ren}\bigr]
= \lim_{\varepsilon\rightarrow 0}\left[
I_{ij}^{(\mathrm{DR})}\bigl[r_0;\,\varepsilon,\,\ell_0;\,
\mathbf{y}_1^\mathrm{ren},\, \mathbf{y}_2^\mathrm{ren}\bigr]
+\delta_{\bm{\eta}(r_1',\,r_2';
\,\varepsilon,\,\ell_0)}I_{ij}\right].
\end{equation}
In other words, this equivalence is between HR and the
``renormalized'' result from DR. We find that the poles $\sim
1/\varepsilon$ separately present in the two terms in the brackets of
(\ref{equivDRHR}) cancel, so that the physical, renormalized, DR
quadrupole moment, defined as the RHS of (\ref{equivDRHR}), is
\textit{ finite} when $\varepsilon\rightarrow 0$ and given by the
limit shown.\footnote{Note that the renormalized DR quadrupole moment
is numerically equal to the original, bare quadrupole moment
$I_{ij}^{(\mathrm{DR})}[r_0;\, \varepsilon,\,\ell_0;\,
\mathbf{y}_1^\mathrm{bare},\, \mathbf{y}_2^\mathrm{bare}]$. In
particular, the original, bare quadrupole moment is also
\textit{finite} as $\varepsilon \rightarrow 0$ (when keeping fixed
$\mathbf{y}_a^\mathrm{ren}$ in taking the limit).}

Let us now substitute into Eq.~(\ref{equivDRHR}) the expressions of
the DR and HR quadrupole moments, respectively given by
(\ref{Iijdr'}) and (\ref{IijH}) above. [We henceforth assume that
$I_{ij}^{(\mathrm{DR})}$ on the LHS of Eq.~(\ref{Iijdr'}) is
evaluated for $\mathbf{y}_a = \mathbf{y}_a^\mathrm{ren}$.]
Since, as we have seen, both the HR and the DR results have been
expressed in terms of their core part, given by the pHS
regularization, we see that, when making their comparison in
(\ref{equivDRHR}), we shall be able to remove the pHS part, which is
common to both sides of the equation. In this way, we obtain a
relation for the ambiguity part
$\Delta I_{ij}$ of the HR quadrupole moment in terms of known
quantities, \textit{viz.}
\begin{equation}\label{equamb}
\Delta I_{ij} \Bigl[\xi+\frac{1}{22},\,\kappa,\,
\zeta+\frac{9}{110}\Bigr]
=\lim_{\varepsilon\rightarrow 0}
\Bigl(\mathcal{D}I_{ij}\bigl[r_1',\,r_2';
\,\varepsilon,\,\ell_0\bigr]+\,\delta_{\bm{\eta}(r_1',\,r_2';
\,\varepsilon,\,\ell_0)}I_{ij}\Bigr)\,.
\end{equation}
We must now insert into (\ref{equamb}) the concrete result of the
detailed computation of the difference $\mathcal{D}I_{ij}$, for all
the non-compact-support terms in the explicit expression of the
moment derived in Sec.~\ref{secV}, and following the recipe provided
by Eq.~(\ref{DHres}).

The computation of $\mathcal{D} I_{ij}$ was performed by means of
computer-aided algebraic manipulations, using the
\textit{Mathematica} software. The final result for
$\mathcal{D}I_{ij}$ reads [modulo the neglect of
$\mathcal{O}(\varepsilon)$ terms]
\begin{eqnarray}\label{rawdiff}
\mathcal{D}I_{ij}\bigl[r_1',\,r_2';\,\varepsilon,\,\ell_0\bigr] &=&
\frac{G_N^2\,m_1^3}{c^6}
\biggl[\left(-\frac{22}{2\varepsilon}-\frac{220}{9}
+\frac{22}{3}\ln\left(
\frac{r_{12}\,{r'}_1^2\,\overline{q}^{3/2}}{\ell_0^3}\right)\right)
y_1^{\langle i}a_1^{j\rangle} -\frac{86}{45} \,v_1^{\langle
i}v_1^{j\rangle}\biggr] \nonumber\\
&+& 1\leftrightarrow 2\,,
\end{eqnarray}
with the notation already used in (\ref{shift}) and (\ref{a1d}). We
then modify the result by including the effect of the particular
shift which is given by Eqs.~(\ref{shift}) and (\ref{deltaetaIij}).
Thanks to this shift we see that the dependence of (\ref{rawdiff})
on the constants $r_1'$, $r_2'$, $\varepsilon$ and $\ell_0$ is
cancelled out. More precisely, we find that the RHS of
Eq.~(\ref{equamb}) exactly takes the form of a particular instance
of the general ambiguity term (\ref{amb}), namely
\begin{equation}\label{RHSequamb}
\lim_{\varepsilon\rightarrow 0}
\Bigl(\mathcal{D}I_{ij}\bigl[r_1',\,r_2';
\,\varepsilon,\,\ell_0\bigr]+\,\delta_{\bm{\eta}(r_1',\,r_2';
\,\varepsilon,\,\ell_0)}I_{ij}\Bigr)\, = \Delta
I_{ij}\Bigl[-\frac{9451}{9240},\,0,\,-\frac{43}{330}\Bigr]\,,
\end{equation}
which yields the following constraint (equivalent to three
independent equations) to be satisfied by the three ambiguity
parameters $\xi,\,\kappa,\,\zeta$:
\begin{equation}\label{equamb2}
\Delta I_{ij} \Bigl[\xi+\frac{1}{22},\,\kappa,\,
\zeta+\frac{9}{110}\Bigr] = \Delta
I_{ij}\Bigl[-\frac{9451}{9240},\,0,\,-\frac{43}{330}\Bigr]\,.
\end{equation}
This immediately gives the following values for the ambiguity
parameters,
\begin{subequations}\label{resamb}\begin{eqnarray}
\xi&=&-\frac{9871}{9240}\,,\\
\kappa&=&0\,,\\
\zeta&=&-\frac{7}{33}\,,
\end{eqnarray}\end{subequations}
which finally provide an unambiguous determination of the 3PN
radiation field of compact binaries by DR. As we reviewed in the
Introduction, Eqs.~(\ref{resamb}) represent the end result of DR,
but in fact the results for each of the parameters $\xi$, $\kappa$
and $\zeta$ have also been obtained by means of an independent
calculation. Indeed, $\zeta=-7/33$ has been shown to be a
consequence of the Poincar\'e invariance of the formalism
\cite{BDI04zeta} (we give also an alternative, $d$-dimensional
derivation of this result in Sec.~\ref{secVIII} below), the value
$\xi+\kappa=-9871/9240$ was deduced from the comparison between the
dipole moment and the center-of-mass position within HR
\cite{BI04mult} (the latter test is equivalent to the one we shall
perform below with the mass dipole in DR), and finally we shall be
able to check that $\kappa=0$ in Sec.~\ref{secVII}.\footnote{It is
amusing to notice that our result for $\xi$ happens to be related to
the previous one for the equation-of-motion related ambiguity
parameter $\lambda$ by a simple cyclic permutation of digits:
Compare $$3\,\xi=-\frac{9871}{3080}\qquad\text{with}\qquad\lambda=
-\frac{1987}{3080}\,.$$}

\subsection{The 3PN mass-dipole moment}\label{secVIB}

The mass-dipole moment $M_i$ is quite interesting to consider
because it satisfies a conservation law: $\dot{M_i} - P_i =0$, where
$P_i$ is the total momentum, and, as such, it can be derived
directly from the binary's \textit{equations of motion} (instead of
a \textit{wave generation} formalism), as being linked to the
conserved quantity $K_i = G_i - t P_i$ associated with the boost
symmetry of the Hamiltonian or the Lagrangian of the binary motion.
Indeed, the mass dipole moment is in fact nothing but the
\textit{center-of-mass} vector $G_i\sim \sum m\,y^i$ of the system
of particles.\footnote{Note that the equivalence between the mass
dipole moment $M_i$ and the center-of-mass vector $G_i$ can be
thought of as being a consequence of the equivalence principle
between gravitational and inertial masses,
$m_\mathrm{g}=m_\mathrm{i}$. Indeed, $M_i\sim \sum
m_\mathrm{g}\,y^i$ while $G_i\sim \sum m_\mathrm{i}\,y^i$. (The
equivalence principle is automatically incorporated into the present
formalism, since the motion of the point particles is geodesic, see
\cite{BDE04}.)} Now, the center of mass vector of point particles
binaries is already known at 3PN order. Its explicit expression was
derived both in ADM coordinates \cite{DJSpoinc} and in harmonic
coordinates \cite{ABF01}, Eq.~(4.5) there; see also its implicit
derivation in harmonic coordinates in Ref.~\cite{DJSequiv}. [When
deriving a 3PN conserved quantity we neglect the 2.5PN
radiation-reaction contribution to the equations of motion.] We thus
have the possibility of an excellent verification of our
calculations, since the end result we shall obtain for the 3PN mass
dipole moment $M_i$ in DR \textit{should} perfectly match with the
3PN center-of-mass $G_i$. In our previous paper,
Ref.~\cite{BI04mult}, we have in fact already verified that
$M_i=G_i$ within the HR scheme, in the sense that we
\textit{required} that $M_i=G_i$ holds, and then we deduced from
this requirement the value of a particular combination of ambiguity
parameters, namely $\xi+\kappa=-9871/9240$. In the present section
we shall directly show that $M_i=G_i$ in DR, without any fine tuning
of ambiguity parameters like in HR.

First of all, let us recall from \cite{B98mult} and the discussion
in \cite{BI04mult} that in the present formalism the conserved mass
dipole moment $M_i$ is given by a slightly more complicated
expression than the non-conserved moments $I_L$, with $\ell\geq 2$.
Namely we have $M_i=I_i+\delta I_i$, where $I_i$ is given by the
same expression as for $I_L$ but taken for $\ell=1$, and where
$\delta I_i$ represents a certain correction to it, which is given,
together with the similar corrections present in the mass $M$ and
current dipole $S_i$, in Eqs.~(2.22) of \cite{BI04mult} (in $3$
dimensions). In \cite{BI04mult} we proved that the correction
$\delta I_i$ gives \textit{zero} in the dipole moment \textit{at 3PN
order}, so that $M_i=I_i+\mathcal{O}\left(c^{-7}\right)$. Now,
$\delta I_i$ is in the form of integrals at infinity (\textit{cf.}
the factor $B$ in front of the integrals in Eqs.~(2.22) of
\cite{BI04mult}), and we have proved in Sec.~\ref{secVB} that for
such integrals the results in HR and DR are the same. Hence we
deduce that $\delta I_i$ is also zero when applying DR and that
$M_i=I_i+\mathcal{O}\left(c^{-7}\right)$ is also true in $d$
dimensions, modulo $\mathcal{O}\left(\varepsilon\right)$ terms.
Therefore, we need only discuss here the DR calculation of the main
part of the dipole moment, namely $I_i$.

The 3PN mass dipole moment $I_i$ in HR is ambiguous, but the
structure of the ambiguity part is very simple, as it contains one
and only one ambiguity parameter $\hat{\eta}$, which turned out to
be given by the particular combination
$\hat{\eta}=\hat{\xi}+\hat{\kappa}$ of the parameters $\hat{\xi}$
and $\hat{\kappa}$ which appeared previously in Eq.~(\ref{amb}). See
Sec.~V\,B in \cite{BI04mult} for details. The structure of the
ambiguity in the dipolar case is
\begin{equation}\label{ambdip}
\Delta I_{i}\bigl[\hat{\xi}+\hat{\kappa}\bigr] =
\frac{22}{3}\left(\hat{\xi}+\hat{\kappa}\right)
\frac{G_N^2\,m_1^3}{c^6}\,a_1^{i}
+ 1\leftrightarrow 2\,.
\end{equation}
Using the link we have found in (\ref{hatamb}) we can then write the
HR result for the dipole moment in terms of the combination
$\xi+\kappa$ of the original ambiguity parameters in \cite{BIJ02},
hence
\begin{equation}\label{IiH}
I_{i}^{(\mathrm{HR})}\bigl[r_1',\,r_2';
\,\xi+\kappa\bigr]=I_{i}^{(\mathrm{pHS})}\bigl[r_1',\,r_2'\bigr]
+\Delta I_{i}\Bigl[\xi+\kappa+\frac{1}{22}\Bigr],
\end{equation}
which is the dipolar analogue of Eq.~(\ref{IijH}). However, a minor
difference with (\ref{IijH}) is that the 3PN dipole moment happens
to be independent of the cut-off scale $r_0$. As we said above, the
value of $\xi+\kappa$ could be determined in \cite{BI04mult} by
imposing that the HR result (\ref{IijH}) is in agreement with the
3PN center-of-mass position given in \cite{ABF01}.

Let us now investigate what happens when using DR. Like for the case
of the quadrupole moment, the result for the dipole moment in DR is
given as the sum of the pHS dipole and of the difference
$\mathcal{D}I_{i}$, which is made out of the sum of all the
contributions of the non-compact support terms (excluding as usual
the surface terms at infinity) present in the explicit formulas of
Sec.~\ref{secIV}, say
\begin{equation}\label{sumdiffdip}
\mathcal{D}I_{i}\bigl[s_1,\,s_2;\,\varepsilon,\,\ell_0\bigr] =
\sum_{\substack{\text{non-compact}\\
\text{terms in $I_{i}$}}}
\mathcal{D}H\bigl[s_1,\,s_2;\,\varepsilon,\,\ell_0\bigr],
\end{equation}
where $s_1$, $s_2$ are the two HR scales in the partie-finie integral
(\ref{Pf}), $\varepsilon$ and $\ell_0$ are the DR scales, and each of
the $\mathcal{D}H$'s are computed using Eq.~(\ref{DHres}). Hence,
\begin{subequations}\label{Iidr}\begin{eqnarray}
I_{i}^{(\mathrm{DR})}\bigl[\varepsilon,\,\ell_0\bigr] &=&
I_{i}^{(\mathrm{pHS})}\bigl[s_1,\,s_2\bigr]+\mathcal{D}I_{i}\bigl[
s_1,\,s_2;\,\varepsilon,\,\ell_0\bigr]\label{Iidra}\\
&=& I_{i}^{(\mathrm{pHS})}\bigl[r'_1,\,r'_2\bigr]
+\mathcal{D}I_{i}\bigl[
r'_1,\,r'_2;\,\varepsilon,\,\ell_0\bigr],\label{Iidrb}
\end{eqnarray}\end{subequations}
where, like in the case of the quadrupole moment, we have taken
advantage of the fact that the constants $s_1,\,s_2$ cancel out from
the two terms in the RHS of (\ref{Iidra}), to rewrite the result in
terms of the specific length scales $r'_1,\,r'_2$ which parametrize
the 3PN equations of motion in \cite{BFeom}. The last step is to
renormalize the DR result by absorbing the poles in a spatial shift
of the two particles world-lines. Of course, we must use the same
shift vectors as in Eq.~(\ref{shift}), and these result in the
following modification of the dipole moment,
\begin{equation}\label{deltaetaIi}
\delta_{\bm{\eta}}I_{i} = m_1\,\eta_1^{i}+1\leftrightarrow 2\,,
\end{equation}
which is indeed checked to cancel the poles $\propto 1/\varepsilon$
of the ``bare'' DR dipole moment, so that the following limit when
$\varepsilon\rightarrow 0$ is finite,
\begin{equation}\label{Mi}
M_{i}\bigl[r_1',\,r_2'\bigr] = \lim_{\varepsilon\rightarrow 0}\left[
I_{i}^{(\mathrm{DR})}\bigl[\,\varepsilon,\,\ell_0\bigr]
+\delta_{\bm{\eta}(r_1',\,r_2'; \varepsilon,\ell_0)}I_{i}\right]\,.
\end{equation}
This represents our final ``renormalized'' DR dipole moment. The
final result (\ref{Mi}) for the dipole moment depends on the scales
$r'_1$ and $r'_2$. We recall that this dependence does not
correspond to any physical ambiguity, since $r'_1$ and $r'_2$ have
the character of gauge quantities.

Finally, after having performed the detailed calculation of the
``difference'' (using the same algebraic computer programs as for
the quadrupole), and having added this difference to the result for
the pHS part which was obtained earlier in Ref.~\cite{BI04mult}, we
found that the renormalized DR moment $M_i$ given by Eq.~(\ref{Mi}),
is in complete agreement with the conserved center-of-mass position
$G_i$ associated with the conservative part of the 3PN equations of
motion, namely
\begin{equation}\label{Mi2}
M_{i}\bigl[r_1',\,r_2'\bigr] = G_{i}\bigl[r_1',\,r_2'\bigr],
\end{equation}
where $G_i$ is explicitly given by Eq.~(4.5) in Ref.~\cite{ABF01}. We
view this test as an important verification of our method and our
detailed calculations.

\section{Renormalization and diagrammatic approach}\label{secVII}

We have given above the final results obtained by combining the DR
computation of the singular ($\sim 1/\varepsilon$) contributions to
the multipole moments, coming from the vicinity of the point masses,
with the pHS results of Ref.~\cite{BI04mult}. This way of presenting
our results is in close correspondence with the actual calculations
we did, but it has the defect of somewhat hiding the logical
structure of our DR results. In this section, we shall go back to
basic methodological questions and explain in more details the logic
behind DR. We shall also show how the examination of the structure
of the DR results allows one to perform several checks of these
results.

Let us first recall that Ref.~\cite{D83houches} presented a general
method for dealing with the gravitational interaction of two (non
spinning) \textit{compact} bodies, \textit{i.e.}, bodies whose radii
are of the same order as their gravitational radii. At the time, the
main motivation for considering this situation was the accurate
relativistic description of binary pulsar systems (\textit{i.e.},
binary neutron stars). Today, we have the additional motivation of
accurately describing not only the motion but also the gravitational
radiation from binary black holes (as well as binary neutron stars,
or mixed black-hole neutron-star systems).
Reference~\cite{D83houches} did not assume from the start a formal
``point mass'' representation of the two compact bodies but used
instead a \textit{matching} approach which combined two different
approximation methods: (i) an ``external perturbation scheme'',
\textit{i.e.}, an iterative, weak-field (post-Minkowskian)
approximation scheme valid in a domain outside two world-tubes
containing the two bodies, and (ii) an ``internal perturbation
scheme'' describing the small perturbations of each body by the
far-field of its companion. A useful outcome of this matching
approach was a proof that to a very high approximation, the internal
structures of the compact bodies were \textit{effaced} when seen in
the external scheme. More precisely, \cite{D83houches} (Sec.~5
there) found that the internal structures affected the equations of
motion only starting at the 5PN level, through a term which is of
fractional order $\sim k\,\left[G \, m / (c^2 \, r_{12})\right]^5$.
Here $k$ is a dimensionless Love number describing the quadrupolar
deformation of one of the compact bodies under the influence of the
tidal field generated by its companion. This result can be simply
understood from a well-known Newtonian argument on the influence on
the orbital motion of the Newtonian quadrupole moments induced by
tidal interaction between the two compact objects (see \textit{e.g.}
Sec.~1.2 in \cite{Bliving}). Indeed, the quadrupole moments scale as
$Q \sim k\,m\,a^5/ r_{12}^3$, where $a$ is the typical size of the
objects, hence in the case of compact objects for which $a \sim G \,
m / c^2$ we have in fact $Q \sim (k\,m/ r_{12}^3)(G \, m / c^2)^5$,
which gives rise to the above mentioned correction to the equations
of motion (and orbital phase) at the 5PN order relatively to the
Newtonian acceleration. This \textit{effacement result} is the
rationale for describing, up to 5PN order, two (non spinning)
compact bodies in terms of two point masses. Technically, this means
representing the compact bodies by a ``skeleton'' made of two
massive world-lines, \textit{i.e.}, by a point-particle action
\begin{equation}\label{6.1}
S_\mathrm{pp} = - \sum_a m_a \, c \int \sqrt{-g_{\mu\nu}
(y_a^{\lambda}) \, d y_a^{\mu} \, d y_a^{\nu}} \,.
\end{equation}
Note that the previous reasoning suggests that, starting at the 5PN
level, one will need to augment the effective action (\ref{6.1}) by
further terms, starting with a {\it quadrupole-type} addition to the
\textit{monopole} action (\ref{6.1}). At the 2.5PN level,
Ref.~\cite{D83houches} explicitly showed how to deal with a
point-particle description of the type (\ref{6.1}) by using Riesz
analytical continuation method to (uniquely) regularize the
divergent integrals linked to the use of point particles in
non-linear general relativity. It was also mentioned at the time
\cite{D80} that equivalent (2.5PN) results could be obtained by
using an analytic continuation of the space-time dimension $D$,
instead of a Riesz-type analytic continuation.

The derivation of the equations of motion at the 3PN level turned
out to be technically complicated, but conceptually satisfactory.
Two independent works, \cite{JaraS98,JaraS99} and \cite{BF00,BFeom},
succeeded in computing using Hadamard-type regularizations most of
the complicated non-linear integrals appearing at 3PN order except
for a few of them, which turned out to be ambiguous because of the
appearance of logarithmic divergencies at the 3PN order. Then, two
further independent works, \cite{DJSdim} and \cite{BDE04}, showed
that dimensional regularization gave unique, consistent answers, for
the latter divergent integrals. A satisfactory check of the
consistency of DR was indeed that these two independent calculations
gave perfectly consistent final answers, though they were performed
in \textit{different gauges}, by completely different methods.

In particular, it was found \cite{DJSdim} that, in
Arnowitt-Deser-Misner (ADM) gauge, DR led to \textit{finite}
equations of motion (no poles $\propto 1/\varepsilon$), so that the
full dynamics of the system could be described by an effective
action obtained by adding to the $d$-dimensional ADM-gauge-fixed
gravitational action the usual action for point particles coupled to
gravity, namely Eq.~(\ref{6.1}) above. All the quantities appearing
in this ADM plus point particle action have finite limiting values
when $\varepsilon\rightarrow 0$. By contrast, it was found in
\cite{BDE04} that, in harmonic coordinates, DR led to equations of
motion containing simple poles $\propto 1/\varepsilon$, but that
those poles could be \textit{renormalized away}. There are two ways
of thinking of this renormalization. A first way is to add to the
usual point-particle action (\ref{6.1}) a counter-term describing a
possible (infinitesimal) shift of the world-lines $y_a^{\mu}$ (in
other words a \textit{dipole} term). Then one shows that this
dipolar counter-term is exactly what is needed to absorb the
$1/\varepsilon$ poles and to leave a finite answer for both the
equations of motion and the bulk metric (\textit{i.e.}, the metric
outside the world-lines). A second (technically equivalent) way is
to use only the usual point-particle action (\ref{6.1}) but to
consider that the ``bare'' world-lines $y_a^{\mu}$ entering
(\ref{6.1}) can be decomposed in the way given by
Eq.~(\ref{shiftworldline}), as (choosing a parametrization by the
coordinate time, $c\,t=y_a^0$)
\begin{equation}\label{6.2}
\mathbf{y}_a^\mathrm{bare}(t)
=\mathbf{y}_a^\mathrm{ren}(t)+\bm{\eta}_a(t)\,,
\end{equation}
where $\mathbf{y}_a^\mathrm{ren}$ is finite as $\varepsilon
\rightarrow 0$, but where $\bm{\eta}_a$, though being formally
``small'', namely of 3PN order, contains a pole part $\propto
1/\varepsilon$ which absorbs all the poles appearing in the
harmonic-coordinates calculations.

Summarizing, the explicit 3PN-level calculations of the equations of
motion (and of the pole part of the bulk metric, see Sec.~VI of
\cite{BDE04}) have confirmed the effacement result of
\cite{D83houches}, \textit{i.e.}, technically, the soundness of
describing two compact bodies by the simple effective action
(\ref{6.1}). However, they also showed that, at such a high
non-linearity order, it is crucial to use a fully consistent, and
gauge-invariant regularization method. Dimensional regularization,
which was invented precisely to preserve gauge invariance
\cite{tHooft,Bollini,Breitenlohner}, is the method of choice to use
in this respect.

\subsection{Diagrammatic interpretation of the poles in
DR}\label{secVIIA}

As a start let us explain how one might have described the results
of Sec.~\ref{secVI} for the mass multipole moments in terms of
field-theory diagrams. Classical diagrammatic representations of
non-linear interactions in general relativity have been introduced
and used in several works, notably in \cite{BertottiP60,Dgef96}. In
a previous paper of this series, Ref.~\cite{BDE04}, we have used
diagrams to clarify the structure of the various contributions to
the equations of motion of two point particles. Let us do the same
here for the mass multipole moments given by (\ref{ILexpr2}).

We represent the basic delta-function sources entering $T^{\mu\nu}$
as two world-lines, and each (post-Minkowskian) propagator
$\Box^{-1}$ as a dotted line. The various post-Minkowskian
potentials $V(x)$, $V_i (x)$, $K(x)$, $\hat{X}(x)$, $\hat{W}_{ij}
(x)$, \textit{etc.}, entering the effective sources $\Sigma$,
$\Sigma_i$, $\Sigma_{ij}$ [see Sec.~\ref{secIV}] can then be
represented by drawing some dotted lines which start at the ``bare''
sources $\sigma$, $\sigma_i$, $\sigma_{ij}$, join at some
intermediate vertices, corresponding to the non-linear couplings
entering the definition of the non-linear potentials [such as the
non-compact part of $\hat{W}_{ij}$ given by (\ref{Wij})], and end at
the field point $x$. The simpler ``linear potentials'', such as
$V(x)$ or the ``compact'' part of $\hat{W}_{ij} (x)$ (\textit{i.e.},
the part generated by $\sigma_{ij}$), are just represented by one
dotted line joining a world-line to the field point $x$. A product
of potentials entering the effective sources $\Sigma_{\mu\nu}$, such
as $\partial_i \, V_j (x) \, \partial_j \, V_i (x)$ is represented
by juxtaposing the diagrams of each potential. [In this simplified
diagrammatic representation we do not explicitly indicate the
various derivative operators which enter as ``vertex factors'' at
the common field point $x$. However, we take care of them when they
are important for the convergence properties of the diagram.]
Finally, we can represent the inclusion of the ``multipolar
factors'', such as $\widehat{x}_L$, by adding a circled cross
$\otimes$. It is then understood that one integrates over the
``crossed vertex'', \textit{i.e.}, the field point.

\begin{figure}
\includegraphics[scale=.8]{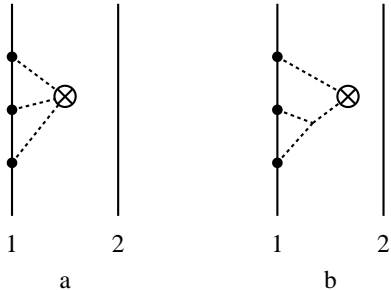}
\caption{Dangerously divergent diagrams contributing to the 3PN
multipole moments. The world-lines of particles 1 and 2 are
represented by vertical solid lines, the propagator $\Box^{-1}$ by
dotted lines, the source points by bullets, and the $\otimes$ symbol
means a multiplication by a multipolar factor, such as
$\widehat{x}_L$, together with a spatial integration $\int d^d
\mathbf{x} \cdots$.}
\label{fig1}
\end{figure}

Using such a representation, the mass multipole moments are given by
the sum of many diagrams. Note first that, when comparing the
diagrams representing the calculation of the 3PN multipole moment to
the diagrams entering the 3PN equations of motion in \cite{BDE04},
one finds that the former have a less complicated structure. Indeed,
Ref.~\cite{BDE04} has shown that the 3PN equations of motion involve
diagrams containing up to \textit{four} independent source points
(located on the world-lines), and up to \textit{five} intermediate
propagators (\textit{i.e.}, five dotted lines): see Figs.~2--4 in
\cite{BDE04}. By contrast, the 3PN multipole moments only involve
(if we treat separately, as was systematically done, the terms that
can be transformed into surface integrals at infinity) diagrams
containing up to \textit{three} source points, and \textit{four}
propagators. Examining the types of singular integrals corresponding
to the possible diagrams, one then finds the same rule of thumb
which was found to hold in \cite{BDE04} for the more complicated
diagrams entering the equations of motion: namely, the only
dangerously\footnote{Here and in the following we focus on the terms
that generate poles $\propto 1/\varepsilon$, and we refer to them as
the ``dangerous'' terms.} diverging diagrams are those containing
(at least) three propagator lines that can simultaneously shrink to
zero size, as a subset of vertices coalesce together on one of the
world-lines. But as there are, in the present problem, at most three
source points, this means that the dangerously divergent diagrams
are only those represented in Fig.~\ref{fig1} below (or their
``mirror'' image obtained by exchanging $1 \leftrightarrow 2$).

These diagrams are also characterized by the fact that they involve,
as post-Minkowskian diagrams (\textit{i.e.}, before explicitly
performing the PN expansion, or the repeated time derivatives, which
can introduce the acceleration of the world-line), either $m_1^3$ or
$m_2^3$ as explicit factor. This reasoning is confirmed by a
scrutiny of the many explicit results reported in \cite{BIJ02} for
separate pieces of the multipole moments. In the presentation of
Ref.~\cite{BIJ02} (which is less systematic than the more recent
re-calculation of \cite{BI04mult}, but more explicit) the
dangerously divergent integrals (in $d=3$) are essentially all the
terms involving the objects $Y_L^{(-3,0)}$, $Y_L^{(-5,0)}$ or
$S_L^{(-5,0)}$, and these terms are all multiplied by $m_1^3$.
Indeed, these objects are integrals of the type $\int d^3 \mathbf{x}
\, r_1^{-3} \varphi (\mathbf{x})$ or $\int d^3 \mathbf{x} \, \Delta
(r_1^{-3}) \, \varphi (\mathbf{x})$, which are logarithmically
divergent in $d=3$, and lead to $1/\varepsilon$ poles in $d = 3 +
\varepsilon$.

Let us exhibit the explicit form of the terms, corresponding to the
diagrams shown in Fig.~\ref{fig1}, which are responsible for the
poles $\propto 1/\varepsilon$ in the final result for the multipole
moments. Let us decompose, as in \cite{BIJ02,BI04mult}, the
expression for $I_L$ in: (i) ``first-order scalar'' part
$\mathrm{SI}_L$ (linear in $\Sigma$), (ii) second-order scalar part
$\mathrm{SII}_L$ (linear in $\partial_t^2 \, \Sigma / c^2$), (iii)
first-order vector part $\mathrm{VI}_L$ (linear in $\partial_t \,
\Sigma_i / c^2$), \textit{etc}. One finds that the ``dangerous''
contributions to $I_L$ are contained only in $\mathrm{SI}_L$,
$\mathrm{SII}_L$ and $\mathrm{VI}_L$. Moreover, one finds that the
velocity-dependent terms that generates poles $\propto
1/\varepsilon$ in intermediate calculations all cancel out in the
final result.\footnote{Such ``cancelled poles'' lead to ambiguities
in the finite part when working in three dimensions. This is taken
care of in our complete results where the calculation is done in $d=
3 + \varepsilon$ before taking the limit $\varepsilon \rightarrow
0$.} We focus here for simplicity on the non-cancelled poles, which
do not depend on velocities. Hence,
\begin{equation}\label{6.3}
I_L^\mathrm{danger} = \mathrm{SI}_L^\mathrm{danger} +
\mathrm{SII}_L^\mathrm{danger} + \mathrm{VI}_L^\mathrm{danger}\,,
\end{equation}
where one checks that among the many contributions generated by
inserting Eq.~(\ref{sources}) into Eq.~(\ref{ILexpr2}) the only
potentially dangerous ones, in the static limit $\mathbf{v}_1
\rightarrow 0$, $\mathbf{v}_2 \rightarrow 0$, come from
{\allowdisplaybreaks
\begin{subequations}\label{6.4-6}\begin{eqnarray}
\mathrm{SI}_L^\mathrm{danger} &=& \frac{1}{\pi \,f\, G \, c^4}
\,\mathrm{FP} \int d^d \mathbf{x} \,\widetilde{r}^B\,\widehat{x}_L
\Biggl\{ - \hat W_{ij} \, \partial_{ij} \, V - \frac{1}{6} \left(
\frac{d-1}{d-2} \right)^2 \, \Delta V^3 - \frac{1}{2} \, \Delta (V
\hat W) \nonumber\\
&&\quad + \frac{1}{c^2} \left[ \frac{1}{2} \, \hat W \, \partial_t^2
\, V + \frac{1}{2} \, V \, \partial_t^2 \, \hat W - 4 \, \hat Z_{ij}
\, \partial_{ij} \, V + 2 \, \frac{d-3}{d-2} \, \hat W_{ij} \,
\partial_{ij} \, K \right] \Biggr\},\label{6.4}\\
\mathrm{SII}_L^\mathrm{danger} &=& \frac{1}{2 (2\ell + d)\,\pi\,
f\,G \, c^6} \,\mathrm{FP} \, \frac{d^2}{dt^2} \int d^d \mathbf{x}
\,\widetilde{r}^B\, \widehat{x}_L \biggl\{ - \vert\mathbf{x}\vert^2
\, \hat W_{ij} \, \partial_{ij} \, V \nonumber \\
&&\quad - (2\ell + d) \,
V \hat W - \frac{1}{3} \, \frac{(2\ell + d)(d-1)^2}{(d-2)^2} \, V^3
\biggr\},\label{6.5}\\
\mathrm{VI}_L^\mathrm{danger} &=& - \frac{4 (2\ell + d-2) }{(\ell +
d - 2)(2\ell + d)\, \pi \, f\,G \, c^6} \,\mathrm{FP} \,
\frac{d}{dt} \int d^d \mathbf{x} \,\widetilde{r}^B\, \widehat x_{iL}
\,\biggl\{ \frac{1}{8} \, \frac{d(d-1)^2}{(d-2)^3} \, V \,
\partial_t \, V \, \partial_i \, V \nonumber \\
&&\quad - \frac{(d-1)^2}{4(d-2)^2} \, V_i \, \partial_k \, V \,
\partial_k \, V + \, \frac{1}{4} \, \frac{d(d-1)}{(d-2)^2} \, V_k \,
\partial_i \, V \, \partial_k \, V + \frac{d-1}{d-2} \, \partial_k
\, V \, \partial_i \, \hat R_k \nonumber \\
&&\quad - \, \hat W_{kl} \, \partial_{kl} \,
V_i + \frac{d-1}{2(d-2)} \,
\partial_t \, \hat W_{ik} \, \partial_k \, V
- \partial_i \, \hat W_{kl} \,
\partial_k \, V_{l} + \, \partial_k
\, \hat W_{il} \, \partial_{l} \, V_k \biggr\}. \label{6.6}
\end{eqnarray}
\end{subequations}}\noindent
where we set $f \equiv \frac{2(d-2)}{d-1}$.

Note that the expression we used for $\mathrm{SII}_L$ in our
calculations has been transformed, from the original form which is
directly arising from the source terms given in Eqs.~(\ref{sources})
above, by operating by parts on the terms proportional to $V \hat W$
and $V^3$. This was done to exactly parallel the calculation of
Ref.~\cite{BI04mult} (see for instance Eq.~(3.4b) there) and thereby
to reduce the problem of evaluating the DR result to a term-by-term
difference between analogous singular integrands. As explained in
Sec.~\ref{secVB}, all the ``gradient terms'' generated when
operating by parts are expressible in terms of surface integrals in
the outer near-zone, and do not contribute to the difference between
DR and pHS. We have therefore suppressed most of these gradient
terms in Eqs.~(\ref{6.4-6}), except in Eq.~(\ref{6.4}) where, as an
example and as a reminder of the presence of such terms, we have
left the terms proportional to the Laplacians of $V^3$ and $V \hat
W$.

Let us explicitly show on the example of $\Delta V^3$ that this
term, though potentially dangerous, does not give rise to any pole.
The linear potential $V$ is naturally decomposed into $V \equiv V_1
+ V_2$ where $V_1 \propto m_1$ is generated by the first particle,
and $V_2 \propto m_2$ by the second. In agreement with
Fig.~\ref{fig1} the dangerous contributions are cubic in $m_1$ or
cubic in $m_2$. In particular, the dangerous pieces in any term
containing $V^3$ are $(V_1)^3$ and $(V_2)^3$. Let us henceforth look
only at the poles generated near the first world-line
(\textit{i.e.}, $\propto m_1^3$). In dimensional regularization, it
is perfectly legitimate to integrate by parts. This transforms the
contribution $\mathrm{FP} \int d^d \mathbf{x} \, \widetilde{r}^B \,
\widehat{x}_L \, \Delta \, V_1^3$ into $\mathrm{FP} \int d^d
\mathbf{x} \, \Delta (\widetilde{r}^B \,\widehat{x}_L) \, V_1^3$.
Using $\Delta (r^B \, \widehat{x}_L) = \Delta (r^{B+\ell} \,
\widehat{n}_L) = B (B + 2\ell + d-2) \, r^{B+\ell-2} \, \widehat
n_L$, we see that the result is proportional to $B$. As we shall see
in detail below the remaining integral $\sim \int d^d \mathbf{x} \,
\varphi (x) \, V_1^3$ generates a pole $\propto 1/\varepsilon$. The
contribution linear in $\Delta V_1^3$ yields therefore a result
proportional to $B/\varepsilon$. But, by the definition of the
$d$-dimensional finite part operation, one has $\mathop{\mathrm{FP}}
(B/\varepsilon) = 0$, so that we have indeed checked the absence of
pole generated by the \textit{a priori} dangerous term $\propto
\Delta V^3$. Similarly for the term $\propto \Delta (V \hat W)$ in
(\ref{6.4}).

Let us consider the various remaining terms in the integrand of
$\mathrm{SI}_L$, Eq.~(\ref{6.4}). We start with the term $\propto
\hat W_{ij}^\mathrm{NC} \, \partial_{ij} \, V$, where $\hat
W_{ij}^\mathrm{NC}$ denotes the so-called ``non compact'' piece of
$\hat W_{ij}$, \textit{i.e.}, the one whose source is $\sim
\partial_i \, V \partial_j \, V$. Again, it is easily seen that the
only dangerous part of the integrand is $\sim \Box^{-1}_\mathrm{Ret}
(\partial_i \, V_1 \, \partial_j \, V_1) \, \partial_{ij} \, V_1$
and its mirror image under the exchange $1 \leftrightarrow 2$. [This
term is an example of the diagram in Fig.~\ref{fig1}b.] We can
compute this term by PN-expanding both $V_1$ and
$\Box^{-1}_\mathrm{Ret}$. This yields a result of the form (in the
static limit)
\begin{equation}\label{6.7}
\mathop{W}_1{}_{\!ij}^\mathrm{\!NC} \,
\partial_{ij} \, V_1 = \alpha_0 \,
\Delta \, U_1^3 + \frac{\beta_0}{c^2} \, a_1^k \, \partial_k \,
U_1^3 + {\mathcal O} (v_1^2) \,,
\end{equation}
where $U_1 = f\,\tilde{k}\, G \, m_1 \, r_1^{-1-\varepsilon}$, with
$\tilde{k} \equiv
\Gamma(\frac{1+\varepsilon}{2})/\pi^{\frac{1+\varepsilon}{2}}$, is
the Newtonian approximation to $V_1$, where $a_1^k \equiv d^2 \,
y_1^k / dt^2$ is the acceleration of the first particle, and where
$\alpha_0$ and $\beta_0$ are numerical coefficients, which depend on
$\varepsilon$. By the same reasoning as used above for the term
$\propto \Delta V_1^3$ one concludes that the term $\alpha_0 \,
\Delta \, U_1^3$ does not generate any pole.

Only the second term on the RHS of (\ref{6.7}) generates a pole
$\propto 1/\varepsilon$ which survives the finite part operation. By
looking at the terms contained in the last bracket on the RHS of
Eq.~(\ref{6.4}) one finds that the only dangerous integrands have
the same form as the second term on the RHS of Eq.~(\ref{6.7}),
namely proportional to $a_1^k \, \partial_k \, U_1^3$. Let us only
give one example of a contribution of this form coming from the last
bracket in (\ref{6.4}). Consider the term (which can be treated to
leading PN order)
\begin{equation}\label{6.9}
\hat{W}^\mathrm{NC} \, \partial_t^2 \, V
= - f^{-1} \,[\Delta^{-1} (\partial_k
\, V \, \partial_k \, V)]^\mathrm{NC} \, \partial_t^2 \, V \, .
\end{equation}
{}From the identity $\partial_k V \,\partial_k V \equiv \Delta
(V^2/2) - V \Delta V$ one has $[\Delta^{-1} (\partial_k \, V \,
\partial_k \, V)]^\mathrm{NC} = V^2/2$, so that the term (\ref{6.9})
is of the type of $V^2 \, \partial_t^2 \, V$. As usual the only
dangerous terms are those proportional to $V_1^2 \, \partial_t^2 \,
V_1$ or $V_2^2 \, \partial_t^2 \, V_2$. Focusing on the first one,
and using the fact that
\begin{equation}\label{6.10}
\partial_t^2 \, V_1 = - \, a_1^k \, \partial_k \, V_1 + {\mathcal O}
(v_1^2) \, ,
\end{equation}
one ends up with an integrand (\ref{6.9}) proportional to $V_1^2 \,
a_1^k \, \partial_k \, V_1$ or, to leading approximation $U_1^2 \,
a_1^k \, \partial_k \, U_1$, which is indeed identical to the second
term in (\ref{6.7}). Finally, we conclude that the
\textit{dangerous} terms in Eq.~(\ref{6.4}) are of the form
\begin{equation}
\label{6.11}
\mathrm{SI}_L^\mathrm{danger} = \mathop{\mathrm{FP}}
\int_\mathrm{loc} d^d \mathbf{x} \, \vert \tilde{\mathbf{x}} \vert^B
\, \widehat{x}_L \left\{ \frac{\beta_\mathrm{SI}}{\pi \, G \, c^6} \,
a_1^k \, \partial_k \, U_1^3 \right\},
\end{equation}
where $\beta_\mathrm{SI}$ is a numerical coefficient which sums
several similar contributions: $\beta_\mathrm{SI} = -\beta_0 +
\cdots$ [we include the factor $f^{-1}=1+\mathcal{O}(\varepsilon)$
into these coefficients], and where the subscript ``loc'' to the
integral reminds us that one can integrate on any local neighborhood
of $\mathbf{x} = \mathbf{y}_1$.

Let us now consider the dangerous terms in $\mathrm{SII}_L$,
Eq.~(\ref{6.5}). In this case one must pay a careful attention to
the dependence of the coefficients on the angular momentum index
$\ell$. Indeed, it is important to note that there was no explicit
dependence on $\ell$ in Eq.~(\ref{6.11}) apart from the factor
$\widehat{x}_L$. By contrast, the coefficients entering (\ref{6.5})
explicitly depend on $\ell$. Since (\ref{6.5}) has an overall factor
$c^{-6}$, it is sufficient to use the leading PN approximations for
$\hat W_{ij}$ and $V$; in view of our previous (1PN-accurate) result
(\ref{6.7}) this means that we can use $\hat W_{ij} \, \partial_{ij}
\, V_1 \simeq \alpha_0 \, \Delta \, U_1^3$ (as usual we focus on the
terms $\propto m_1^3$). The occurrence of an explicit Laplacian
allows us to re-express the first term on the RHS of (\ref{6.5}) by
integrating by parts. This leads to a term proportional to (we keep
only the coefficients depending on $\ell$)
\begin{equation}\label{6.12}
\frac{1}{2\ell + d} \mathrm{FP} \, \frac{d^2}{dt^2} \int d^d
\mathbf{x} \, \Delta \left(\vert \widetilde{\mathbf{x}} \vert^B \,
\vert\mathbf{x}\vert^2 \, \widehat{x}_L\right) \, U_1^3 \, .
\end{equation}
Using $\Delta \left(\vert\mathbf{x}\vert^{B+2} \,
\widehat{x}_L\right) = (B+2) (B+2\ell + d) \,
\vert\mathbf{x}\vert^{B}\, \widehat{x}_L$ we get a contribution of
the form
\begin{equation}\label{6.13}
\frac{1}{\pi \, G \, c^6} \, \mathrm{FP} \, \frac{d^2}{dt^2} \int
d^d \mathbf{x} \, \vert \widetilde{\mathbf{x}} \vert^B \,
\frac{(B+2) (B + 2\ell + d)}{2\ell + d} \, \widehat{x}_L \, U_1^3
\,.
\end{equation}
The pole part $\propto 1/\varepsilon$ of the contribution
(\ref{6.13}) is generated by integrating in the vicinity of the
first world-line. For such a local integral the IR-converging factor
$\vert \widetilde{\mathbf{x}} \vert^B$ has no importance and we can
take the analytic continuation $B \rightarrow 0$ directly in the
(localized) integrand. This leads to the disappearance of the
$\ell$-dependence in the factor appearing in (\ref{6.13}). As for
the last two terms on the RHS of (\ref{6.5}), one sees that the
$\ell$-dependence cancels between the factor $\propto 1/(2\ell + d)$
in front, and the factors $\propto (2\ell + d)$ multiplying the
integrands $\widehat{x}_L \, V \hat W$ and $\widehat{x}_L \, V^3$.
Finally, we conclude that the dangerous terms in $\mathrm{SII}_L$
are of the form
\begin{equation}\label{6.14}
\mathrm{SII}_L^\mathrm{danger} = \frac{d^2}{dt^2} \int_\mathrm{loc}
d^d \mathbf{x} \, \widehat{x}_L \left\{ \frac{\beta_\mathrm{SII}}{\pi
\, G \, c^6} \, U_1^3 \right\} + {\mathcal O} (v_1^2) \,.
\end{equation}
The coefficient $\beta_\mathrm{SII}$ does not depend on $\ell$ (like
was the case for $\beta_\mathrm{SI}$). The repeated time derivative
in (\ref{6.14}) can then be let to act on $U_1^3$ only (modulo ``non
dangerous'' terms) yielding, in view of Eq.~(\ref{6.10}),
$\partial_t^2 \, U_1^3 \simeq - \, a_1^k \, \partial_k \, U_1^3$ so
that
\begin{equation}\label{6.15}
\mathrm{SII}_L^\mathrm{danger} = \int_\mathrm{loc} d^d \mathbf{x} \,
\widehat{x}_L \left\{ - \,
\frac{\beta_\mathrm{SII}}{\pi \, G \, c^6} \,
a_1^k \, \partial_k \, U_1^3 \right\}.
\end{equation}
A similar study of the ``vector'' contribution $\mathrm{VI}_L$,
Eq.~(\ref{6.6}), yields a result of the form
\begin{eqnarray}\label{6.16}
\frac{1}{\pi \, G \, c^6} \,
\frac{(2\ell + d -2)}{(\ell + d - 2)(2\ell +
d)} \, \mathrm{FP} \int_\mathrm{loc} d^d \mathbf{x} \, \vert
\widetilde{\mathbf{x}} \vert^B \, \widehat x_{iL} \Bigl\{
\alpha_\mathrm{VI} \, a_1^i \, \Delta \, U_1^3 + \beta_\mathrm{VI} \,
a_1^k \, \partial_{ik} \, U_1^3 \Bigr\}.
\end{eqnarray}
Integrating by parts the first term, and taking the finite part at
$B=0$ is easily seen to give a vanishing result [because $\Delta
(\widehat x_{iL}) = 0$]. The second term of (\ref{6.16}), with a
coefficient denoted $\beta_\mathrm{VI}$, is \textit{a priori} more
problematic. Integrating by parts does not give a vanishing result
(because $\partial_i \, \widehat x_{iL} \propto \widehat{x}_L$ does
not vanish). If present this term would have a complicated
dependence on $\ell$. However, the overall coefficient
$\beta_\mathrm{VI}$ of this term is the sum of many individual
contributions, and one finds that they \textit{all} cancel out to
yield $\beta_\mathrm{VI} = 0$, so that finally
\begin{equation}\label{6.18}
\mathrm{VI}_L^\mathrm{danger} = 0 \, .
\end{equation}

The result $\beta_\mathrm{VI} = 0$ can be obtained either by
explicit calculations in $d$ dimensions, using notably the explicit
form of $\hat W_{ij}^\mathrm{NC}$, namely (to leading order)
\begin{eqnarray}\label{6.19}
\mathop{\hat{W}}_1{}_{\!ij}^\mathrm{\!NC} = - \, \frac{1}{4} \, (d-1)
(d-2) \, U_1^2 \, \left[ \frac{\widehat{n}_1^{ij}}{(d-1) (d-4)} +
\frac{\delta^{ij}}{d(d-2)^2} \right],
\end{eqnarray}
or by considering the limiting case $d=3$. In this limiting case,
the poles $\propto 1/\varepsilon$ are associated to logarithmically
divergent integrals. Looking at the 3-dimensional results given by
(8.2c), (9.3j) and (9.3k) of Ref.~\cite{BIJ02} for $\mathrm{VI}_L$,
one indeed finds that the terms $a_1^k \, \partial_{ik} \,
Y_{iL}^{(-3,0)}$, corresponding to the $d = 3$ limit of the second
contribution in (\ref{6.16}), do cancel in the final result, though
they appear in intermediate terms: see the first terms on the RHS of
Eqs.~(9.3j)--(9.3k) of \cite{BIJ02} with coefficients $+ 2/63$ and
$- 2/63$ respectively (note a small misprint in (9.3k) of
\cite{BIJ02}: the overall factor $m_1^2$ should be understood as
$m_1^3$). Note also that, in view of the general structure
(\ref{6.16}) derived above, it is enough to check the cancellation
of these terms for the quadrupolar case $(\ell = 2)$ to conclude
that $\beta_\mathrm{VI}=0$.

Summarizing our results so far, we conclude, by adding (\ref{6.11}),
(\ref{6.15}) and (\ref{6.18}) that the pole part $\propto m_1^3$ in
the $\ell$-th mass multipole moment is contained in
\begin{equation}\label{6.20}
I_L^\mathrm{danger} = \int_\mathrm{loc} d^d \mathbf{x} \,
\widehat{x}_L \left\{ \frac{\beta}{\pi \, G \, c^6} \, a_1^k \,
\partial_k \, U_1^3 \right\},
\end{equation}
with a final coefficient $\beta = \beta_\mathrm{SI} -
\beta_\mathrm{SII}$. By summing the various contributions one finds
\begin{equation}\label{6.21}
\beta = - \frac{11}{6} + {\mathcal O} (\varepsilon) \, ,
\end{equation}
where the first term on the RHS is enough to discuss the residue of
the pole $\propto 1/\varepsilon$.

\subsection{Renormalization of poles by shifts of the
world-lines}\label{secVIIB}

The result (\ref{6.20})--(\ref{6.21}) is the explicit expression of
the dangerous part of the two diagrams of Fig.~\ref{fig1}. Let us
now see explicitly why it is nicely renormalized away by using
exactly the same dipole counter-term that was found necessary in
\cite{BDE04}. The pole generated by (\ref{6.20}) can be seen, after
integrating by parts the spatial gradient ($\partial_k$), as coming
{}from an integral of the form
\begin{equation}\label{6.22}
I_\mathrm{loc} = \int_\mathrm{loc} d^d \mathbf{x} \, \varphi
(\mathbf{x}) \, U_1^3\,,
\end{equation}
where $\varphi (\mathbf{x})$ is a smooth function of $\mathbf{x}$
(at least near $\mathbf{x} = \mathbf{y}_1$). Taylor-expanding
$\varphi (\mathbf{x})$ near $\mathbf{x} = \mathbf{y}_1$ one sees
that the pole in (\ref{6.22}) comes from the zero-th term $\varphi
(\mathbf{y}_1)$ which multiplies an integral proportional to
\begin{equation}\label{6.23}
\int_\mathrm{loc} d^{3+\varepsilon}\mathbf{x} \,
r_1^{-3-3\varepsilon} = \Omega_{2+\varepsilon} \int_0^R dr_1 \,
r_1^{-1-2\varepsilon} = \Omega_{2+\varepsilon} \, R^{-2\varepsilon}
/ (-2\varepsilon) \,,
\end{equation}
where we recall that $\Omega_{2+\varepsilon}$ denotes the area of
the $2+\varepsilon$ dimensional sphere. Therefore, in the
limit $\varepsilon \rightarrow 0$, the integral (\ref{6.22}) is
asymptotically equivalent to
\begin{equation}\label{6.24}
I_\mathrm{loc} = - \, \frac{2\pi}{\varepsilon} \, G^3 \, m_1^3 \,
\varphi (\mathbf{y}_1) + \mathcal{O} \left(\varepsilon^0\right).
\end{equation}
This means that, when $\varepsilon \rightarrow 0$ the integrand
$U_1^3$ is asymptotically equivalent (in the formal sense of
distributions in $d$-dimensional space) to
\begin{equation}\label{6.25}
U_1^3 = - \, \frac{2 \, \pi}{\varepsilon} \, G^3 \, m_1^3 \, \delta
(\mathbf{x} - \mathbf{y}_1) + {\mathcal O}
\left(\varepsilon^0\right).
\end{equation}
Inserting this result into (\ref{6.20}) one concludes that the
\textit{pole part} (due to the UV divergencies in the neighborhoods
of $\mathbf{y}_1$ and $\mathbf{y}_2$) of the 3PN-accurate $\ell$-th
mass multipole moment is given by
\begin{equation}\label{6.26}
I_L^\mathrm{pole} = \int d^d \mathbf{x} \, \widehat{x}_L \, a_1^k \,
\partial_k \left\{ - \, \frac{2\beta}{\varepsilon} \, \frac{G^2 \,
m_1^3}{c^6} \, \delta (\mathbf{x} - \mathbf{y}_1) \right\} + 1
\leftrightarrow 2 \,.
\end{equation}
If we compare (\ref{6.26}) with the leading, Newtonian approximation
for $I_L$, namely
\begin{equation}\label{6.27}
I_L^\mathrm{N} = \int d^d \mathbf{x} \, \widehat{x}_L \, \Bigl\{ m_1
\, \delta (\mathbf{x} - \mathbf{y}_1) \Bigr\} + 1 \leftrightarrow 2
\, ,
\end{equation}
we see that the pole part (\ref{6.26}) can be absorbed in a
dipole-like modification [$\sim \partial_k \, \delta (\mathbf{x} -
\mathbf{y}_1)$] of the mass density $m_1\, \delta (\mathbf{x} -
\mathbf{y}_1)$, or equivalently in a shift of the world-line
position $\mathbf{y}_1$. More precisely, if we decompose the full
$\mathbf{y}_1$ (henceforth called the ``bare'' $\mathbf{y}_1$) as in
Eq.~(\ref{6.2}), with $\mathbf{y}_1^\mathrm{ren}$ being finite as
$\varepsilon \rightarrow 0$, but with $\bm{\eta}_1$ designed to
absorb the pole part (\ref{6.26}), one easily checks that one needs
to define
\begin{equation}\label{6.28}
\eta_1^k = - \frac{2\beta}{\varepsilon} \, \frac{G^2 \, m_1^2}{c^6}
\, a_1^k + {\mathcal O} (\varepsilon^0)\,,
\end{equation}
in order to renormalize away this pole. Note that it was crucial to
have no $\ell$-dependence of the coefficients in the dangerous part
(\ref{6.20}) in order to be able to renormalize away the
\textit{infinite sequence} of multipoles by means of the
$\ell$-independent shift $\bm{\eta}_1$ (\ref{6.28}).

In addition, by inserting the numerical value (\ref{6.21}) of the
coefficient $\beta$, one finds that the shift (\ref{6.28}) needed to
absorb the poles in the infinite sequence of multipole moments
coincides with the shift obtained in \cite{BDE04} by the requirement
of renormalizing both the ``bulk metric'' and the equations of
motion. More precisely, Ref.~\cite{BDE04} found that the choice of
the shift recalled above in Eq.~(\ref{shift}) [and which contains
(\ref{6.28}) as its pole part] allowed one not only to get a finite
(pole-less) bulk metric and finite equations of motion, but that the
equations of motion coincide (when, and only when, $\lambda =
-1987/3080$) with the harmonic-gauge equations of motion,
parametrized by $r'_1$ and $r'_2$, and derived using HR in
Refs.~\cite{BF00,BFeom}. We recall that it is necessary to introduce
some length scales $r'_1$ and $r'_2$ associated with the HR of
logarithmically divergent integrals in harmonic gauge.

As we have shown here the dangerous divergencies associated with the
vicinity of the \textit{first} world-line are entirely contained in
the diagrams shown in Fig.~\ref{fig1}, and, therefore, are
proportional to $m_1^3$, without any explicit dependence on the
second mass $m_2$. [There is only an implicit dependence on $m_2$
\textit{via} the fact that the acceleration $\mathbf{a}_1$ is
proportional to $m_2$. But, at the level of the diagrams,
$\mathbf{a}_1$ must be considered as a pure characteristic of the
first world-line.] As a consequence, we see in Eq.~(\ref{6.28}) that
the dipole $m_1 \eta_1^k$ needed to subtract the poles is also
proportional to $m_1^3$. This simple algebraic fact immediately
leads, without calculations, to the result that $\kappa = 0$.
Indeed, the definition of the parameter $\kappa$ in
Ref.~\cite{BIJ02}, was to parametrize a conceivable \textit{a
priori} ambiguity, which is indeed allowed by the weak assumptions
of \cite{BIJ02}, in the renormalization of the logarithmic
divergencies of the type (for the first particle)
\begin{equation}\label{6.29}
m_1^3\, \ln \left(\frac{r'_1}{s_1}\right) = \left(\xi +
\kappa\right) \, m_1^3 + \kappa \, m_1^2 \, m_2\,,
\end{equation}
where $r'_1$ and $s_1$ are two possible choices of regularization
length scales associated to the first particle, and where we have
incorporated the factor $m_1^3$ associated to the divergences linked
to $\mathbf{y}_1$. As (\ref{6.29}) shows, the parameter $\kappa$
corresponds to a mixing between diagrams with three legs on the
first world-line (as in Fig.~\ref{fig1}) and diagrams having two
legs on the first world-line and one on the second. Our diagrammatic
study has shown that the latter diagrams have no dangerous
divergencies, \textit{i.e.}, that they do not introduce any
conceivable ambiguity (even if we were working directly in $d=3$,
using HR). Therefore we conclude that $\kappa = 0$.

The work of this section has shown that the pole in the $\ell$-th
mass moment $I_L$ was given by Eq.~(\ref{6.26}) whose numerical
coefficients contain \textit{no dependence on the value of} $\ell$.
This proves, in particular, that the same shift (\ref{6.28}), or
more precisely (\ref{shift}), yields finite values of both the
quadrupole moment $I_{ij}$ ($\ell = 2$) and the dipole moment $M_i$
($\ell = 1$).\footnote{Recall from Sec.~\ref{secVIB} that the
conserved mass dipole moment $M_i$ reads $M_i=I_i+\delta I_i$, where
$\delta I_i$ represents a certain correction term which, however,
turns out not to contribute at the 3PN order (see \cite{BI04mult}).}
As said above the mass dipole moment $M_i$ coincides with the
Arnowitt-Deser-Misner (ADM) dipole moment or \textit{center-of-mass}
position $G_i$, such that $G_i - P_i \, t$ is conserved, where $P_i$
denotes the total ADM linear momentum.
The comparison between $M_i$ and $G_i$ in \cite{BI04mult} permitted
to fix the value of the combination $\xi + \kappa = -9871/9240$
within the HR scheme, under the assumption that the regularization
scales $s_1$ and $s_2$ represent some unknown but \textit{fixed}
constants, related to $r'_1$ and $r'_2$ by some definite equations,
and in particular take the same values for both the computations of
the quadrupole $I_{ij}$ and the dipole $M_i$. This assumption worked
well in the case of the HR computation of the multipole moments, but
failed to work when it was tried to assume that the same scales
$s_1$ and $s_2$ are \textit{also} those which entered the HR
computation of the equations of motion \cite{BFeom}. Indeed, the
work on the equations of motion used for the relation between $s_1$
and $r'_1$, for the divergences linked to the first particle,
$\ln(r'_1 / s_1) = \mathrm{const} + \lambda \, m_2 / m_1$ where
$\lambda$ was later determined to have a non-zero value, $\lambda =
- 1987/3080$. Such a link is clearly incompatible with (\ref{6.29})
and the value we have found for $\kappa=0$. This means that one is
not \textit{a priori} allowed to assume, when using HR, that the
scales $s_1$ and $s_2$ represent always the same scales, fixed once
and for all, and which can be used in different bodies of
calculations. In this respect the HR is not a fully consistent
regularization scheme. However, it can nevertheless be applied if one
accepts that its incompleteness results in the appearance of some
unknown scales $s_1$ and $s_2$ (generally in front of a few terms
only), which can take different values, depending on the type of
calculation one is doing. By contrast we have proved in
Sec.~\ref{secVIB} above that the same value of $\xi$ is consistent,
in DR, with the renormalized results of both $I_{ij}$ and $M_i =
G_i$. This result constitutes evidently a solid confirmation of the
value $\xi=-9871/9240$.

\subsection{Comments on finite-size effects in the effective
action of compact bodies}\label{secVIIC}

To conclude our discussion of the diagrammatic approach to the
renormalization of the poles which appear in harmonic gauge, let us
briefly comment on the recent claim \cite{Goldberger} that these
poles require the introduction of new terms in the effective action
describing compact (but extended) objects, beyond Eq.~(\ref{6.1})
and the dipole term we found above, linked to the shift (\ref{6.2}).
The modified effective action proposed in Ref.~\cite{Goldberger} has
the form
\begin{equation}\label{6.30}
S'_\mathrm{pp} = S_\mathrm{pp} + S_\mathrm{finite \, size} \, ,
\end{equation}
where $S_\mathrm{pp}$ is the standard point-particle effective action
(\ref{6.1}) and where
\begin{equation}\label{6.31}
S_\mathrm{finite \, size} = \sum_a c_R^{(a)}\!\! \int ds_a \, R
(y_a) + \sum_a c_V^{(a)}\!\! \int ds_a \, R_{\mu\nu} (y_a) \,
u_a^{\mu} \, u_a^{\nu} \,,
\end{equation}
with $u_a^\mu \equiv d y_a^\mu/ds_a$. Several claims were made in
Ref.~\cite{Goldberger}: (i) that the extra terms (\ref{6.31}) are
necessary to ``encapsulate finite size properties of the sources'',
(ii) that they are linked to the same ``dangerous'' diagrams that
were examined in Fig.~6 of \cite{BDE04} and Fig.~\ref{fig1} above,
and (iii) that they entail the presence of genuine ambiguities at
the 3PN level which can only be fixed by a matching calculation. If
these statements were correct, that would mean not only that the
basic ``effacement'' property (modulo 5PN-level ``quadrupole-type''
additional terms to the effective action) is incorrect, but also
that the recent results, \cite{DJSdim,BDE04,BDEI04} and this work,
fixing all 3PN-level ambiguity parameters by DR are flawed.

Let us, however, indicate why we think that the claims (i), (ii) and
(iii) made in Ref.~\cite{Goldberger} are not correct. First, we
mention that the addition of curvature-coupling terms of the type
indicated in (\ref{6.31}) has already been considered in
Ref.~\cite{Nordt94} and in Appendix~A of Ref.~\cite{Dgef98}, which
considered finite-size effects in tensor-scalar gravity. Indeed,
when gravity is partly mediated by a scalar excitation, the internal
characteristics of compact objects are much less effaced than in the
pure spin-2 case. In particular, the coupling to the spherical
inertia moment $I \sim \int d^3 \, \mathbf{x} \, \sigma (\mathbf{x})
\, \mathbf{x}^2$ can introduce extra couplings of the type of the
curvature terms in (\ref{6.31}) (see \cite{Nordt94}) together with
several other scalar-dependent couplings. However, it was shown in
\cite{Dgef98} that the use of suitable field redefinitions can
transform away the curvature couplings (\ref{6.31}) into couplings
explicitly involving the gradient of the scalar field, $\int ds_a \,
N_a (\varphi) \, g^{\mu\nu} \, \partial_{\mu} \, \varphi \,
\partial_{\nu} \, \varphi$. As such a term does not exist in the
pure spin-2 case, one sees that Ref.~\cite{Dgef98} proves that
(\ref{6.31}) can be field-redefined away. Indeed, a simple way to
see it is to recall that the first-order effect of a field
redefinition of the metric $(g'_{\mu\nu} = g_{\mu\nu} + \varepsilon
\, h_{\mu\nu})$ is to modify the effective action by terms
proportional to the Einstein field equations, namely $\delta \,
S_\mathrm{tot} = -(16 \, \pi \, G)^{-1} \int d^D x \, \sqrt{-g} \,
(R^{\mu\nu} - \frac{1}{2} \, R\,g^{\mu\nu} - 8 \, \pi \, G \,
T^{\mu\nu}) \, \varepsilon \, h_{\mu\nu}$ [to simplify, we set the
light velocity $c = 1$ here and below]. Conversely, the (\textit{a
priori} illicit) use of the Einstein field equations \textit{within
an action} is equivalent to a suitably defined field redefinition
$\varepsilon \, h_{\mu\nu}$. Applying this general result to
(\ref{6.30}) we see that the curvature coupling terms (\ref{6.31})
are equivalent to
\begin{eqnarray}\label{6.32}
S'_\mathrm{finite \, size} &=&\sum_a c'^{(a)}_R\!\!\!
\int\!\!\!\!\int\! ds_a \, ds'_a \, \frac{m_a}{\sqrt{-g}} \,
\delta^{(D)} (y_a^{\mu} (s_a) - y_a^{\mu} (s'_a)) \nonumber \\
&+&\sum_a c'^{(a)}_V\!\!\! \int\!\!\!\!\int\! ds_a \, ds'_a \,
\frac{m_a}{\sqrt{-g}} \, u_a^{\mu} (s_a) \, u_a^{\nu} (s_a) \,
u_{a\mu} (s'_a) \, u_{a\nu} (s'_a) \, \delta^{(D)}(y_a^{\mu} (s_a) -
y_a^{\mu} (s'_a))\qquad\quad
\end{eqnarray}
(where $D=d+1$ and $u_a^{\mu} \equiv dy_a^{\mu} / ds_a$) modulo a
field redefinition $g'_{\mu\nu} = g_{\mu\nu} + h_{\mu\nu}$ of the
type
\begin{equation}
\label{6.33}
h_{\mu\nu} (x) = \sum_a \int ds_a \, [c''^{(a)}_R \, g_{\mu\nu} +
c''^{(a)}_V \, u_{\mu} \, u_{\nu}] \, \frac{\delta^{(D)}
(x^{\lambda} - y_a^{\lambda} (s_a))}{\sqrt{-g}} \, .
\end{equation}
Here $c'_R$, $c'_V$, $c''_R$, $c''_V$ are linear combinations of the
coefficients $c_R$, $c_V$ entering (\ref{6.31}), namely $c''_R =
-2c'_R = 16\pi G (c_V-2c_R)/(d-1)$, $c''_V = 2 c'_V = 16 \pi G c_V$.
After using, for instance, the delta function in time, $\delta
(y_a^0 (s_a) - y_a^0 (s'_a))$, to integrate over $s'_a$ (with the
conclusion that $s'_a = s_a$), one easily sees that the result
(\ref{6.32}) is proportional to the $s_a$-integral of the
$d$-dimensional delta function evaluated at a vanishing separation:
$\delta^{(d)} (y_a^i (s_a) - y_a^i (s_a))$. In DR, such a pure
contact term vanishes exactly, so that we have simply
$S'_\mathrm{finite \, size} = 0$. [As Ref.~\cite{Goldberger} uses
also DR, we are entitled in using DR to discuss their claims.]
Therefore we conclude that the proposed curvature-coupling terms
(\ref{6.31}) are equivalent to a field redefinition of the type
(\ref{6.33}). However, (\ref{6.33}) is again a ``contact term'' in
the sense that it vanishes outside of the world-lines, and cannot
therefore affect the external field generated by the world-lines
that we are interested in. In conclusion, the term (\ref{6.31}) can
be essentially completely field-redefined away, and has no physical
import.

We can give another (partial) confirmation of this result by looking
at the form of the pole that Ref.~\cite{Goldberger} claims to be
associated with the diagrams in Fig.~6 of \cite{BDE04}, or
Fig.~\ref{fig1} here (\textit{i.e.}, diagrams (c) and (d) of Fig.~7
of \cite{Goldberger}). Transcribing the Fourier-space result (53) of
\cite{Goldberger} in $x$-space, and considering the combination that
enters the leading term in the multipole moments, one finds that,
according to \cite{Goldberger}, those dangerous diagrams are
equivalent, when $\varepsilon \rightarrow 0$, to an effective
mass-energy distribution of the type
\begin{equation}\label{6.34}
\Sigma_\mathrm{eff}^{\text{Gold.-Roth.}} \equiv \frac{T_{(3)}^{00} +
T_{(3)}^{ii}}{c^2} = \frac{Q}{\varepsilon} \, \frac{G^2 \,
m_1^3}{c^4} \, \Delta \, \delta (\mathbf{x} - \mathbf{y}_1) \, ,
\end{equation}
where $Q$ is a (non zero) numerical constant and $\Delta$ the
Laplacian.

The result (\ref{6.34}) is consistent with part of our analysis
above. Indeed, using Eqs.~(\ref{6.7}), (\ref{6.11}), (\ref{6.16})
and (\ref{6.25}), our analysis has shown that the dangerous terms in
the cubically non-linear ``non compact'' contributions to $\Sigma$,
$\Sigma_i$ and $\Sigma_{ij}$ are equivalent to a term in $\Sigma$ of
the form
\begin{equation}\label{6.35}
\Sigma_\mathrm{eff} = \frac{1}{\varepsilon} \, \frac{G^2 \,
m_1^3}{c^4} \left[ \alpha \, \Delta \, \delta (\mathbf{x} -
\mathbf{y}_1) + \frac{\beta}{c^2} \, a_1^k \, \partial_k \, \delta
(\mathbf{x} - \mathbf{y}_1) \right].
\end{equation}
The equation (\ref{6.34}) is consistent with the \textit{first} term
on the RHS of (\ref{6.35}). But, as we have shown above, this term
has no physical implication; only the \textit{second} term, involving
a \textit{dipole} coupling $a_1^k \, \partial_k \, \delta$, mattered.
This confirms our conclusion that the claims (i), (ii) and (iii) of
\cite{Goldberger} are not correct because the terms they considered
have no physical relevance. Note also that the ``finite size'' effect
(\ref{6.34}) (formally linked to a spherical inertia moment $\int d^3
\, \mathbf{x} \, \sigma (\mathbf{x}) \, \mathbf{x}^2$, as in the
tensor-scalar case of \cite{Dgef98}) is actually a 2PN-level term. If
that term had created physical effects linked to the finite-size of
the source, this would have meant that the 2.5PN equations of motion
\cite{D83houches} had missed some 2PN violation of the effacement
properties. As a final comment let us recall that the ADM-gauge
calculations of \cite{DJSdim} never exhibited any pole. In ADM-gauge
all the 3PN diagrams are finite and the whole discussion of possible
Renormalization-Group dependent quantities evaporates away.

\section{Quadrupole moment of a boosted point
particle}\label{secVIII}

In Sec.~\ref{secVI} we obtained unique values for the three
heretofore unknown parameters $\xi$, $\kappa$ and $\zeta$, by adding
to the HR calculations of the quadrupole moment of an interacting
binary point-mass system the additional contributions
$\mathcal{D}I_{ij}$ coming from a DR treatment of the singularities
near $\mathbf{y}_1$ and $\mathbf{y}_2$. In Sec.~\ref{secVII} we have
shown that a detailed study of the structure of the singular
diagrams represented in Fig.~\ref{fig1} allowed one to check the
values of both $\kappa$ and $\xi$ (using information about the full
computation of the dipole moment in HR to check the latter). Here,
we shall complete our checks by giving an independent calculation of
the third parameter $\zeta$. This calculation will be based on a
full DR evaluation of the quadrupole moment of a moving isolated
particle ($m_1 \neq 0$, $m_2 = 0$). In another paper,
Ref.~\cite{BDI04zeta}, we have already checked the value of $\zeta$
within a purely $3$-dimensional approach, based on the physical
situation of an isolated boosted Schwarzschild (exterior) solution
with mass $m_1$ (and still with $m_2=0$), and without use of any
self-field regularization. Therefore our new, DR-based, computation
of $\zeta$ given here can also be viewed as a further check of the
consistency of DR.

We thus consider the limiting case of a single particle with mass
$m_1$, moving on a straight line. In order to be able to discuss
meaningfully this limiting case, it is important not to use a
center-of-mass frame for the original binary system $m_1$, $m_2$.
Indeed, if we start from a center-of-mass frame before taking the
limit $m_2 \rightarrow 0$, we shall end up with a single particle at
rest and placed at the center of the coordinate frame used to
compute the multipole moments. To simplify the notation, we shall
suppress the index $1$ on the characteristics of the single particle
that we consider. As in \cite{BDI04zeta} we gain also some
simplification by assuming that the origin of the coordinate system
(with respect to which the particle is moving) which is used to
define the multipole moments coincides with the position of the
particle at the time $t=0$. In other words, we consider a single
particle of mass $m$, moving on the world-line $y^i = v^i t$. As was
already used in \cite{BDI04zeta}, the limiting case $m_1 \rightarrow
m$, $m_2 \rightarrow 0$, $y_1^i \rightarrow y^i = v^i t$ of the
mass-type quadrupole moment of a binary system $I_{ij}(m_1,m_2)$,
evaluated by HR in \cite{BIJ02,BI04mult}, takes the form (at 3PN
order)
\begin{equation}\label{7.1}
I_{ij}^\mathrm{HR}(m,0) = m\, y^{\langle i} y^{j\rangle} \left[ 1
+\frac{9}{14}\,\frac{v^2}{c^2} +\frac{83}{168}\,\frac{v^4}{c^4}
+\frac{507}{1232}\,\frac{v^6}{c^6} \right]
+\left(\frac{232}{63}+\frac{44}{3}\,\zeta\right)
\frac{G^2m^3}{c^6}\,v^{\langle i} v^{j\rangle}\,.
\end{equation}
As we see, the $\zeta$-ambiguity enters only in a term $\propto G^2
m^3 v^{\langle i j\rangle}/c^6$. We shall henceforth focus on this
term and show how DR uniquely fixes its coefficient, \textit{i.e.},
the numerical coefficient $\mathcal{C}$ in the expression
\begin{equation}\label{7.2}
I_{ij}^\mathrm{DR}(m,0) = \mathcal{B} \,m \,y^{\langle i}
y^{j\rangle} +\mathcal{C} \,\frac{G^2m^3}{c^6} v^{\langle i}
v^{j\rangle}\,.
\end{equation}
To evaluate the coefficient $\mathcal{C}$ in DR, the first step is
to obtain the $D$-dimensional metric, \textit{in harmonic
coordinates}, generated by a boosted point particle. We shall first
determine the metric generated by a point particle at rest and then
apply Lorentz invariance in $D$ dimensions. There are two ways of
doing this. We can start from the expressions for the harmonically
relaxed Einstein field equations (at 3PN order) explicitly given in
\cite{BDE04} and solve them by iteration, when assuming a source
given by a single delta function. Another method consists in
starting from the well-known $D$-dimensional Schwarzschild solution,
in Schwarzschild-Droste coordinates, and then look for the
\textit{particular} harmonic coordinates selected by the DR
treatment of delta-function sources. We have used both methods and
checked that they fully agree. Let us indicate some details of the
first, more pedestrian, approach.

In the rest frame of a single point particle, the stress-energy
tensor has $T^{00} = m \,c^2 \delta^{(d)}(\mathbf{x})$ as single
non-vanishing component. This yields a ``scalar'' source
$\sigma(\mathbf{x})$, as used in our formalism, see
Eq.~(\ref{matterSources}), of the form
\begin{equation}\label{7.3}
\sigma(\mathbf{x}) = f \,m \,\delta^{(d)}(\mathbf{x})\,,
\end{equation}
with $f\equiv 2(d-2)/(d-1)$, together with $\sigma_i = 0 =
\sigma_{ij}$. The basic scalar potential $V$ generated by $\sigma$,
$\Box V = -4 \pi G \sigma$, is then found to be
\begin{equation}\label{7.4}
V = f \,\tilde{k}\, \frac{G \,m}{r^{d-2}}\,,
\end{equation}
where $\tilde{k} \equiv \Gamma[(d-2)/2]/\pi^{(d-2)/2}$. The other
linear potentials are easily found to vanish, $V_i = 0$, $K=0$.
Going then to the various non-linear potentials, one finds,
successively, $\hat R_i = 0$, $\hat Z_{ij} = 0$, $\hat Y_i = 0$, as
well as $\hat T = 0$. Note that the vanishing of all those
potentials results both from the treatment of contact terms in DR
(namely $r^\alpha \delta^{(d)}(\mathbf{x}) = 0$) and from the
special structure of Einstein's equations (the fact that $\hat
Z_{ij}$ and $\hat T$ vanish is due to the special structure of some
cubic non-linearities in Einstein's equations). Finally, besides
$V$, the only non-vanishing potentials are $\hat W_{ij}$ and $\hat
X$, which are determined by solving
\begin{subequations}\label{7.5-6}
\begin{eqnarray}
\Delta \hat W_{ij} &=&
-\frac{1}{2}\left(\frac{d-1}{d-2}\right)\partial_i V \partial_j V\,,
\label{7.5}\\
\Delta \hat X &=& \hat W_{ij}\partial_{ij} V\,.
\label{7.6}
\end{eqnarray}
\end{subequations}
As in \cite{BDE04}, it is useful to introduce the combination
\begin{equation}\label{7.7}
\mathcal{V} \equiv V -\frac{2}{c^2}\left(\frac{d-3}{d-2}\right) K
+\frac{4}{c^4}\hat X +\frac{16}{c^6}\hat T = V +\frac{4}{c^4}\,\hat
X\,,
\end{equation}
which simplifies the expression of the metric. Indeed, one has
\begin{subequations}\label{7.8-9}
\begin{eqnarray}
g_{00} &=& -\exp\left(-2\frac{\mathcal{V}}{c^2}\right)
+\mathcal{O}\left(\frac{1}{c^{10}}\right),
\label{7.8}\\
g_{ij} &=&
\exp\left(\frac{2}{d-2}\,\frac{\mathcal{V}}{c^2}\right)
\left\{\delta_{ij} + \frac{4}{c^4}\,W_{ij}
\right\}
+\mathcal{O}\left(\frac{1}{c^8}\right),
\label{7.9}
\end{eqnarray}
\end{subequations}
and $g_{0i} = 0$. The ``gothic'' metric $\mathfrak{g}^{\mu\nu} \equiv
\sqrt{-g}\, g^{\mu\nu}$ reads, besides $\mathfrak{g}^{0i} = 0$,
\begin{subequations}
\label{7.10-11}
\begin{eqnarray}
\mathfrak{g}^{00} &=&
-\exp\left(2\,\frac{d-1}{d-2}\,\frac{\mathcal{V}}{c^2}\right)
\left\{1 + \frac{2}{c^4}\,\hat W_{kk}
\right\}
+ \mathcal{O}\left(\frac{1}{c^8}\right),
\label{7.10}\\
\mathfrak{g}^{ij} &=&
\delta_{ij} - \frac{4}{c^4}\,\hat W_{ij}
+\frac{2}{c^4}\,\hat W_{kk}\delta_{ij}
+ \mathcal{O}\left(\frac{1}{c^8}\right).
\label{7.11}
\end{eqnarray}
\end{subequations}
Note that a remarkable simplification occurred in the expression
(\ref{7.11}) of the spatial gothic metric. Indeed, we see form
(\ref{7.4}) that $V/c^2$ is proportional to $Gm/c^2$ and therefore
that $\hat W_{ij}/c^4 \propto (G m/c^2)^2$ while $\hat X/c^6 \propto
(Gm/c^2)^3$. The result (\ref{7.11}) shows that $\mathfrak{g}^{ij} =
\delta^{ij} + \mathcal{O}[(Gm/c^2)^2] +\mathcal{O}[(Gm/c^2)^4]
+\cdots$. The point is that there are no terms $\propto (Gm/c^2)^3$
in the spatial gothic metric. One can even prove, more generally,
that the spatial structure of Einstein's equations is such that
$\mathfrak{g}^{ij}$ (for a particle at rest) contains only
\textit{even} powers of $Gm/c^2$. The only component of the gothic
metric which contains \textit{odd} powers of $Gm/c^2$, and in
particular $(Gm/c^2)^3$, is the time-time component
$\mathfrak{g}^{00}$, Eq.~(\ref{7.10}).

By explicitly solving Eq.~(\ref{7.5}), we find, in $d$ dimensions,
\begin{equation}\label{7.12}
\hat W_{ij} = -\frac{1}{4}(d-1)(d-2)V^2 \left[
\frac{\widehat{n}_{ij}}{(d-1)(d-4)} +\frac{\delta_{ij}}{d(d-2)^2}
\right].
\end{equation}
Inserting this result in the RHS of (\ref{7.6}) then allows one to
solve for $\hat X$, in any dimension $d$, and we find
\begin{equation}\label{7.13}
\hat X = -\frac{1}{24}\left(\frac{d-1}{d-4}\right)V^3.
\end{equation}
Then, from Eq.~(\ref{7.10-11}) we get, still in the rest frame:
\begin{equation}\label{7.14}
\mathfrak{g}^{00} = -A\,, \qquad \mathfrak{g}^{ij} = B\,\delta^{ij}
+ C \,\widehat{n}^{ij},
\end{equation}
where $A$, $B$ and $C$ can be expressed in terms of $V/c^2$ and admit
expansions of the type
\begin{subequations}\label{7.15-17}
\begin{eqnarray}
A &=& 1 + a_1\,\frac{V}{c^2} + a_2\,\frac{V^2}{c^4} +
a_3\,\frac{V^3}{c^6} + a_4\,\frac{V^4}{c^8} + \ldots\,,
\label{7.15}\\
B &=& 1
+ b_2\,\frac{V^2}{c^4}
+ b_4\,\frac{V^4}{c^8}
+ \ldots\,,
\label{7.16}\\
C &=& c_2\,\frac{V^2}{c^4}
+ c_4\,\frac{V^4}{c^8}
+ \ldots\,,
\label{7.17}
\end{eqnarray}
\end{subequations}
where, as said above, $B$ and $C$ contain only \textit{even} powers
of $V/c^2$. The $d$-dependent numerical coefficients $a_1$, $a_2$,
$a_3$, $b_2$ and $c_2$ can be read off the results (\ref{7.4}) and
(\ref{7.10-11})--(\ref{7.13}) above.

It is then easy to ``boost'' the metric (\ref{7.14}) to a moving
frame. It suffices to write it as
\begin{equation}\label{7.18}
\mathfrak{g}^{\mu\nu} = -A \,u^\mu u^\nu + B
\,\left(\eta^{\mu\nu}+u^\mu u^\nu\right) + C \, n^{\langle\mu}
n^{\nu\rangle}\,,
\end{equation}
where $u^\mu$ is the $D$-velocity of the particle, and $n^\mu$ the
unit radial $D$-vector orthogonal to the world-line.\footnote{We
have $u^2\equiv \eta_{\mu\nu} u^\mu u^\nu = -1$, $n^2 \equiv
\eta_{\mu\nu} n^\mu n^\nu = + 1$, $\eta_{\mu\nu} u^\mu n^\nu = 0$,
and $n^{\langle\mu} n^{\nu\rangle} = n^\mu n^\nu
-\frac{1}{D}\eta^{\mu\nu}$.} As the mass $m$ enters only through $V
\propto G\,m$, we see immediately from (\ref{7.18}) that, in the
``laboratory frame'' where the point particle is moving, the only
term in the gothic metric (\ref{7.18}) which is cubic in $G\,m$ is
\begin{equation}\label{7.19}
\left(\mathfrak{g}^{\mu\nu}\right)_{\text{cubic}} = -
a_3\,\frac{V^3}{c^6}\, u^\mu u^\nu\,.
\end{equation}
The explicit value of the coefficient $a_3$ in (\ref{7.19}) is found
to be
\begin{equation}\label{7.20}
a_3 = 8\left(\frac{d-1}{d-2}\right)\left[
\frac{1}{6}\left(\frac{d-1}{d-2}\right)^2
-\frac{1}{8}\left(\frac{d-1}{d-2}\right)
-\frac{1}{24}\left(\frac{d-1}{d-4}\right) \right].
\end{equation}
When $\varepsilon \equiv d-3 \rightarrow 0$, one finds
\begin{equation}\label{7.21}
a_3 = 8 \left[1 - \frac{4}{3}\,\varepsilon+
\mathcal{O}\left(\varepsilon^2\right) \right].
\end{equation}
Finally, to obtain (\ref{7.19}) in the lab-frame, we need to
re-express the rest-frame result (\ref{7.14}) for $V$ in terms of
lab-frame quantities. This is simply done by saying that the
rest-frame radial distance $r$ entering (\ref{7.4}) can be
invariantly characterized as the orthogonal distance $r_{\perp}$
between the world-line and the field point. In any frame,
$r_{\perp}$ is given by
\begin{equation}
\label{7.22}
r_{\perp} = \left(\eta_{\mu\nu}+u_\mu u_\nu\right)
\left(x^\mu-y^\mu\right)\left(x^\nu-y^\nu\right),
\end{equation}
where $y^\mu$ is \textit{any} point on the world-line ($y^\mu$ does
not need to be such that $x^\mu-y^\mu$ be orthogonal to $u^\mu$).
Finally, we get for the part of the gothic metric deviation
$h^{\mu\nu}\equiv \mathfrak{g}^{\mu\nu} - \eta^{\mu\nu}$ which is
cubic in $G \,m$,
\begin{equation}\label{7.23}
h^{\mu\nu}_{\text{cubic}}(\mathbf{x},t) = -\frac{a_3}{c^6}\left(
\frac{f\,\tilde{k}\, G \,m}{r_{\perp}^{1+\varepsilon}}\right)^3 u^\mu
u^\nu\,,
\end{equation}
where $u^0 = 1/\sqrt{1-v^2/c^2}$, $u^i = u^0\,v^i/c \equiv u_i$ and
\begin{equation}\label{7.24}
r_{\perp}^2(t) =
\left(\delta_{ij}+ u_i u_j\right)
\left(x^i - y^i(t)\right)\left(x^j - y^j(t)\right),
\end{equation}
where $y^i(t)$ is the point on the world-line which is
lab-synchronous with the field point (at the same time $t = y^0/c$).

We have focused here on the terms cubic in $G\,m$ because, as
indicated in (\ref{7.1})--(\ref{7.2}), we are only interested in
computing the coefficient $\mathcal{C}$ appearing in front of the
cubic term of (\ref{7.2}). We need now to use the
\textit{definition} of the mass quadrupole moment $I_{ij}$, which is
given by Eq.~(\ref{ILexpr2}), where the RHS is expressed in terms of
the PN expansion of $\tau^{\mu\nu} = \frac{c^4}{16\pi G} \,\Box
h^{\mu\nu}$. Introducing, as in (\ref{Sigma}) above, the notation
$\Sigma\equiv \frac{2}{d-1}[(d-2)\tau^{00}+\tau^{ii}]/c^2$,
$\Sigma^i \equiv \tau^{0i}/c$, $\Sigma^{ij} \equiv \tau^{ij}$,
we finally obtained the following $d$-dimensional expressions for
the cubic terms in these various effective sources:
\begin{subequations}\label{7.25}
\begin{eqnarray}
\Sigma_{\text{cubic}} &=& -\frac{3+\mathcal{O}(\varepsilon)}{\pi}\,
\frac{G^2 m^3}{c^4}\, \frac{1+ \varepsilon+ v^2/c^2}{1-v^2/c^2}\,
\frac{1}{r_{\perp}^{5+3\varepsilon}}\,,
\label{7.25a}\\
\Sigma^i_{\text{cubic}} &=&
-\frac{3+\mathcal{O}(\varepsilon)}{\pi}\,
\frac{G^2 m^3}{c^4}\,
\frac{v^i}{1-v^2/c^2}\,
\frac{1}{r_{\perp}^{5+3\varepsilon}}\,,
\label{7.25b}\\
\Sigma^{ij}_{\text{cubic}} &=&
-\frac{3+\mathcal{O}(\varepsilon)}{\pi}\,
\frac{G^2 m^3}{c^4}\,
\frac{v^i v^j}{1-v^2/c^2}\,
\frac{1}{r_{\perp}^{5+3\varepsilon}}\,,
\label{7.25c}
\end{eqnarray}
\end{subequations}
where $r_{\perp} = r_1 \sqrt{1+(n^i u^i)^2}$ so that we have the
expansion
\begin{equation}\label{7.26}
\frac{1}{r_{\perp}^{5+3\varepsilon}} = \frac{1}{r_1^{5+3\varepsilon}}
\left( 1 - \frac{5+3\varepsilon}{2}\,\frac{(n_1^i v^i)^2}{c^2}
+\cdots\right).
\end{equation}
Here $r_1(t) \equiv \sqrt{\delta_{ij}(x^i-y^i(t))(x^j-y^j(t))}$ is
the usual, lab-instantaneous, distance between the field point
$\mathbf{x}$ and the particle $\mathbf{y}(t)$, and $n_1^i(t) \equiv
(x^i - y^i(t))/r_1$. We have re-installed here the index $1$ to
distinguish the radial distance to the particle, $r_1 =
\vert\mathbf{x}-\mathbf{y}\vert \equiv
\vert\mathbf{x}-\mathbf{y}_1\vert$, from the radial distance to the
origin of the lab-frame coordinate system, everywhere denoted as $r
= \vert\mathbf{x}\vert$.

When inserting the explicit expressions (\ref{7.25}) in the
definition of the quadrupole moment, one ends up with a sum of
$d$-dimensional integrals whose integrands contain several types of
factors: an overall factor $\vert
\widetilde{\mathbf{x}}\vert^B\equiv \vert \mathbf{x}/r_0\vert^B$,
various multipolar factors $\sim \widehat{x}_L$, together with
various spatial derivatives of $r_1^{-5-3\varepsilon}$. We use
$\partial_t f(r_1) = -v^i\partial_i f(r_1)$ to replace time
derivatives acting on the $\Sigma$'s by space derivatives. By
separating the quadrupole moment in several contributions, as is
Eq.~(\ref{ILexpr2}) above, one easily checks that the leading
$\mathcal{O}(c^{-4})$ contribution coming from replacing
$r_{\perp}^{-5-3\varepsilon} \rightarrow r_1^{-5-3\varepsilon}$ is
$\mathrm{SI}_L$ and gives a vanishing contribution (after taking the
$d$-modified finite part). Then it takes more work to check that the
$\mathcal{O}(c^{-6})$ contribution $\mathrm{VI}_L$ coming from the
time derivative of $\Sigma^i$ also gives a vanishing contribution.
One is then left to evaluating an integral of the type
\begin{equation}\label{7.27}
I_{ij} \propto \mathrm{FP}\int d^d \mathbf{x}
\vert\widetilde{\mathbf{x}}\vert^B \left\{
-\frac{5+3\varepsilon}{2}\, \widehat x^{ij} \,\widehat{n}_1^{ab}
\,\widehat v_1^{ab} \,r_1^{-5-3\varepsilon}
+\frac{1}{2(7+\varepsilon)}\, \vert\mathbf{x}\vert^2 \,\widehat
x^{ij} \,\widehat v_1^{ab} \,\partial_{ab} \,r_1^{-5-3\varepsilon}
\right\}.
\end{equation}
The dependence on $\varepsilon$ of the global factor (not displayed
here) does not import for our present calculation. On the contrary,
the relative coefficients $-(5+3\varepsilon)$ and
$1/(7+\varepsilon)$ of the two terms are crucial, as there will
occur below a cancellation between their lowest order contributions.
The trick to compute Eq.~(\ref{7.27}) (for a finite value of $B$) is
to express it, after using some integration by parts, in terms of
parametric derivatives of ``Riesz integrals''. An (Euclidean) Riesz
integral in any dimension $d$ is the integral
\begin{equation}\label{7.28}
R(a,\,b;\,\mathbf{y}_0,\,\mathbf{y}_1) = \int d^d \mathbf{x}\, \vert
\mathbf{x} - \mathbf{y}_0\vert^a \vert \mathbf{x} -
\mathbf{y}_1\vert^b = N_{ab}\, \vert \mathbf{y}_0 -
\mathbf{y}_1\vert^{a+b+d}\,,
\end{equation}
where the numerical coefficient $N_{ab}$ is equal to
\begin{equation}\label{7.28b}
N_{ab} = \pi^{d/2}\, \frac{\Gamma\left(\frac{a+d}{2}\right)
\Gamma\left(\frac{b+d}{2}\right)
\Gamma\left(-\frac{a+b+d}{2}\right)}{\Gamma\left(-\frac{a}{2}\right)
\Gamma\left(-\frac{b}{2}\right)
\Gamma\left(\frac{a+b+2d}{2}\right)}\,.
\end{equation}
We find that we can express (\ref{7.27}) as being proportional to
\begin{equation}
\label{7.29}
\widehat v_1^{ab}\, \frac{\partial^2}{\partial y_0^{\langle i}
\partial y_0^{j\rangle}}\, \frac{\partial^2}{\partial y_1^{\langle
a} \partial y_1^{b\rangle}}
\,R(B+4,\,-3-3\varepsilon;\,\mathbf{y}_0,\,\mathbf{y}_1)\,.
\end{equation}
Here, we have introduced, as extra parameter, the position $y_0^i$
of the origin used to define the multipole moments. Up to now we
have simply taken $y_0^i = 0$, but one could have defined from the
start the multipole moments with factors of the type
$\vert\mathbf{x}- \mathbf{y}_0\vert^B(x-y_0)^{\langle L\rangle}$.
Inserting all needed factors, and explicitly evaluating the
derivatives appearing in (\ref{7.29}), we end up with a final answer
of the type $I_{ij}^\mathrm{cubic} = \mathcal{C} \,G^2 \,m_1^3
\,c^{-6} \,v_1^{\langle ij\rangle}$, \textit{i.e.}, of the form
expected from Eq.~(\ref{7.2}), with a numerical coefficient given,
after appropriate expansion, by
\begin{equation}\label{7.30}
\mathcal{C} = \mathop{\mathrm{FP}}\left[
\frac{B(-14\varepsilon+9B+\cdots)}{7\varepsilon(B-2\varepsilon)}
\right],
\end{equation}
where the ellipsis denote terms of higher order in $\varepsilon$
and/or $B$ that do not contribute.

We explicitly exhibit the near-final form (\ref{7.30}) to emphasize
the subtle nature of the determination of $\mathcal{C}$. The result
is proportional to $B$, which will ultimately be analytically
continued to zero, so that one might \textit{a priori} believe that
$\mathcal{C}$ will vanish when $B\rightarrow 0$. However, this is
not so because $\mathcal{C}$ also contains the shifted pole $\propto
(B-2\varepsilon)^{-1}$. In addition, when $B$ is non zero,
(\ref{7.30}) also exhibits a pole $\propto \varepsilon^{-1}$. As we
explained above, the MPM formalism (and its subsequent PN
re-expansion) imposes a specific finite part operation FP to be
applied to all multipole moments. It consists in first subtracting
the shifted pole terms and then in taking the limit $B\rightarrow 0$
(see Sec.~\ref{secIIIA}). For instance, in the case of a simple pole
of the form $N(B,\varepsilon)/(B-2\varepsilon)$, one must subtract
$N(2\varepsilon,\varepsilon)/(B-2\varepsilon)$
before taking $B\rightarrow 0$, which then leads to the finite part
$[N(0,\varepsilon)-N(2\varepsilon,\varepsilon)]/(-2\varepsilon)$.
Applying this to (\ref{7.30}) yields the final result
\begin{equation}
\label{7.31}
\mathcal{C} = \frac{-2\varepsilon(-14\varepsilon+18\varepsilon)}{7
\varepsilon(-2\varepsilon)} = \frac{4}{7}\,,
\end{equation}
which is exactly the same result as found with an independent
surface-integral evaluation \cite{BDI04zeta}. Comparing this value to
the last term on the RHS of Eq.~(\ref{7.1}), we then conclude that
$\zeta$ is uniquely fixed to the value
\begin{equation}
\label{7.32}
\zeta = -\frac{7}{33}\,,
\end{equation}
in full agreement with our full two-body DR results in (\ref{resamb})
above, and with the regularization-free calculations of
Ref.~\cite{BDI04zeta}.

\acknowledgments

Two of us (L.B. and B.R.I.) would like to thank the Indo-French
collaboration (IFCPAR) under which this work has been carried out.
B.R.I. thanks IHES for hospitality at different stages of the work.
The calculations reported in this paper have been done with the help
of the software \textit{Mathematica}.


\begin{thebibliography}{64}
\expandafter\ifx\csname natexlab\endcsname\relax\def\natexlab#1{#1}\fi
\expandafter\ifx\csname bibnamefont\endcsname\relax
\def\bibnamefont#1{#1}\fi
\expandafter\ifx\csname bibfnamefont\endcsname\relax
\def\bibfnamefont#1{#1}\fi
\expandafter\ifx\csname citenamefont\endcsname\relax
\def\citenamefont#1{#1}\fi
\expandafter\ifx\csname url\endcsname\relax
\def\url#1{\texttt{#1}}\fi
\expandafter\ifx\csname urlprefix\endcsname\relax\def\urlprefix{URL }\fi
\providecommand{\bibinfo}[2]{#2}
\providecommand{\eprint}[2][]{\url{#2}}

\bibitem[{\citenamefont{Cutler et~al.}(1993)\citenamefont{Cutler, Apostolatos,
Bildsten, Finn, Flanagan, Kennefick, Markovic, Ori, Poisson, Sussman
et~al.}}]{3mn}
\bibinfo{author}{\bibfnamefont{C.}~\bibnamefont{Cutler}},
\bibinfo{author}{\bibfnamefont{T.}~\bibnamefont{Apostolatos}},
\bibinfo{author}{\bibfnamefont{L.}~\bibnamefont{Bildsten}},
\bibinfo{author}{\bibfnamefont{L.}~\bibnamefont{Finn}},
\bibinfo{author}{\bibfnamefont{E.}~\bibnamefont{Flanagan}},
\bibinfo{author}{\bibfnamefont{D.}~\bibnamefont{Kennefick}},
\bibinfo{author}{\bibfnamefont{D.}~\bibnamefont{Markovic}},
\bibinfo{author}{\bibfnamefont{A.}~\bibnamefont{Ori}},
\bibinfo{author}{\bibfnamefont{E.}~\bibnamefont{Poisson}},
\bibinfo{author}{\bibfnamefont{G.}~\bibnamefont{Sussman}},
\bibnamefont{et~al.}, \bibinfo{journal}{Phys. Rev. Lett.}
\textbf{\bibinfo{volume}{70}}, \bibinfo{pages}{2984} (\bibinfo{year}{1993}).

\bibitem[{\citenamefont{Cutler and Flanagan}(1994)}]{CF94}
\bibinfo{author}{\bibfnamefont{C.}~\bibnamefont{Cutler}} \bibnamefont{and}
\bibinfo{author}{\bibfnamefont{E.E.}~\bibnamefont{Flanagan}},
\bibinfo{journal}{Phys. Rev. D} \textbf{\bibinfo{volume}{49}},
\bibinfo{pages}{2658} (\bibinfo{year}{1994}).

\bibitem[{\citenamefont{Tagoshi and Nakamura}(1994)}]{TNaka94}
\bibinfo{author}{\bibfnamefont{H.}~\bibnamefont{Tagoshi}} \bibnamefont{and}
\bibinfo{author}{\bibfnamefont{T.}~\bibnamefont{Nakamura}},
\bibinfo{journal}{Phys. Rev. D} \textbf{\bibinfo{volume}{49}},
\bibinfo{pages}{4016} (\bibinfo{year}{1994}).

\bibitem[{\citenamefont{Poisson}(1995)}]{P95}
\bibinfo{author}{\bibfnamefont{E.}~\bibnamefont{Poisson}},
\bibinfo{journal}{Phys. Rev. D} \textbf{\bibinfo{volume}{52}},
\bibinfo{pages}{5719} (\bibinfo{year}{1995}), \bibinfo{note}{erratum Phys.
Rev. D {\bf 55}, 7980, (1997)}.

\bibitem[{\citenamefont{Damour et~al.}(1998)\citenamefont{Damour, Iyer, and
Sathyaprakash}}]{DIS98}
\bibinfo{author}{\bibfnamefont{T.}~\bibnamefont{Damour}},
\bibinfo{author}{\bibfnamefont{B.R.}~\bibnamefont{Iyer}}, \bibnamefont{and}
\bibinfo{author}{\bibfnamefont{B.S.}~\bibnamefont{Sathyaprakash}},
\bibinfo{journal}{Phys. Rev. D} \textbf{\bibinfo{volume}{57}},
\bibinfo{pages}{885} (\bibinfo{year}{1998}).

\bibitem[{\citenamefont{Damour et~al.}(2000{\natexlab{a}})\citenamefont{Damour,
Iyer, and Sathyaprakash}}]{DIS00}
\bibinfo{author}{\bibfnamefont{T.}~\bibnamefont{Damour}},
\bibinfo{author}{\bibfnamefont{B.R.} \bibnamefont{Iyer}}, \bibnamefont{and}
\bibinfo{author}{\bibfnamefont{B.S.} \bibnamefont{Sathyaprakash}},
\bibinfo{journal}{Phys. Rev. D} \textbf{\bibinfo{volume}{62}},
\bibinfo{pages}{084036} (\bibinfo{year}{2000}{\natexlab{a}}),
\eprint{gr-qc/0001023}.

\bibitem[{\citenamefont{Buonanno et~al.}(2003)\citenamefont{Buonanno, Chen, and
Vallisneri}}]{BCV03a}
\bibinfo{author}{\bibfnamefont{A.}~\bibnamefont{Buonanno}},
\bibinfo{author}{\bibfnamefont{Y.}~\bibnamefont{Chen}}, \bibnamefont{and}
\bibinfo{author}{\bibfnamefont{M.}~\bibnamefont{Vallisneri}},
\bibinfo{journal}{Phys. Rev. D} \textbf{\bibinfo{volume}{67}},
\bibinfo{pages}{024016} (\bibinfo{year}{2003}), \eprint{gr-qc/0205122}.

\bibitem[{\citenamefont{Damour et~al.}(2003)\citenamefont{Damour, Iyer,
Jaranowski, and Sathyaprakash}}]{DIJS03}
\bibinfo{author}{\bibfnamefont{T.}~\bibnamefont{Damour}},
\bibinfo{author}{\bibfnamefont{B.R.} \bibnamefont{Iyer}},
\bibinfo{author}{\bibfnamefont{P.}~\bibnamefont{Jaranowski}},
\bibnamefont{and} \bibinfo{author}{\bibfnamefont{B.S.}
\bibnamefont{Sathyaprakash}}, \bibinfo{journal}{Phys. Rev. D}
\textbf{\bibinfo{volume}{67}}, \bibinfo{pages}{064028}
(\bibinfo{year}{2003}), \eprint{gr-qc/0211041}.

\bibitem[{\citenamefont{Ajith et~al.}(2005)\citenamefont{Ajith, Iyer, Robinson,
and Sathyaprakash}}]{AIRS05}
\bibinfo{author}{\bibfnamefont{P.}~\bibnamefont{Ajith}},
\bibinfo{author}{\bibfnamefont{B.}~\bibnamefont{Iyer}},
\bibinfo{author}{\bibfnamefont{C.}~\bibnamefont{Robinson}}, \bibnamefont{and}
\bibinfo{author}{\bibfnamefont{B.}~\bibnamefont{Sathyaprakash}}
(\bibinfo{year}{2005}), \bibinfo{note}{to appear in Phys. Rev. D},
\eprint{gr-qc/0412033}.

\bibitem[{\citenamefont{Arun et~al.}(2005)\citenamefont{Arun, Iyer,
Sathyaprakash, and Sundararajan}}]{AISS05}
\bibinfo{author}{\bibfnamefont{K.}~\bibnamefont{Arun}},
\bibinfo{author}{\bibfnamefont{B.}~\bibnamefont{Iyer}},
\bibinfo{author}{\bibfnamefont{B.}~\bibnamefont{Sathyaprakash}},
\bibnamefont{and}
\bibinfo{author}{\bibfnamefont{P.}~\bibnamefont{Sundararajan}}
(\bibinfo{year}{2005}), \bibinfo{note}{submitted to Phys. Rev. D},
\eprint{gr-qc/0411146}.

\bibitem[{\citenamefont{Blanchet
et~al.}(1995{\natexlab{a}})\citenamefont{Blanchet, Damour, Iyer, Will, and
Wiseman}}]{BDIWW95}
\bibinfo{author}{\bibfnamefont{L.}~\bibnamefont{Blanchet}},
\bibinfo{author}{\bibfnamefont{T.}~\bibnamefont{Damour}},
\bibinfo{author}{\bibfnamefont{B.R.} \bibnamefont{Iyer}},
\bibinfo{author}{\bibfnamefont{C.M.} \bibnamefont{Will}}, \bibnamefont{and}
\bibinfo{author}{\bibfnamefont{A.G.} \bibnamefont{Wiseman}},
\bibinfo{journal}{Phys. Rev. Lett.} \textbf{\bibinfo{volume}{74}},
\bibinfo{pages}{3515} (\bibinfo{year}{1995}{\natexlab{a}}),
\eprint{gr-qc/9501027}.

\bibitem[{\citenamefont{Blanchet
et~al.}(1995{\natexlab{b}})\citenamefont{Blanchet, Damour, and Iyer}}]{BDI95}
\bibinfo{author}{\bibfnamefont{L.}~\bibnamefont{Blanchet}},
\bibinfo{author}{\bibfnamefont{T.}~\bibnamefont{Damour}}, \bibnamefont{and}
\bibinfo{author}{\bibfnamefont{B.R.} \bibnamefont{Iyer}},
\bibinfo{journal}{Phys. Rev. D} \textbf{\bibinfo{volume}{51}},
\bibinfo{pages}{5360} (\bibinfo{year}{1995}{\natexlab{b}}),
\eprint{gr-qc/9501029}.

\bibitem[{\citenamefont{Will and Wiseman}(1996)}]{WWi96}
\bibinfo{author}{\bibfnamefont{C.M.}~\bibnamefont{Will}} \bibnamefont{and}
\bibinfo{author}{\bibfnamefont{A.G.}~\bibnamefont{Wiseman}},
\bibinfo{journal}{Phys. Rev. D} \textbf{\bibinfo{volume}{54}},
\bibinfo{pages}{4813} (\bibinfo{year}{1996}).

\bibitem[{\citenamefont{Blanchet}(1996)}]{B96}
\bibinfo{author}{\bibfnamefont{L.}~\bibnamefont{Blanchet}},
\bibinfo{journal}{Phys. Rev. D} \textbf{\bibinfo{volume}{54}},
\bibinfo{pages}{1417} (\bibinfo{year}{1996}), \eprint{gr-qc/9603048}.

\bibitem[{\citenamefont{Blanchet et~al.}(1996)\citenamefont{Blanchet, Iyer,
Will, and Wiseman}}]{BIWW96}
\bibinfo{author}{\bibfnamefont{L.}~\bibnamefont{Blanchet}},
\bibinfo{author}{\bibfnamefont{B.R.} \bibnamefont{Iyer}},
\bibinfo{author}{\bibfnamefont{C.M.} \bibnamefont{Will}}, \bibnamefont{and}
\bibinfo{author}{\bibfnamefont{A.G.} \bibnamefont{Wiseman}},
\bibinfo{journal}{Class. Quant. Grav.} \textbf{\bibinfo{volume}{13}},
\bibinfo{pages}{575} (\bibinfo{year}{1996}), \eprint{gr-qc/9602024}.

\bibitem[{\citenamefont{Arun et~al.}(2004)\citenamefont{Arun, Blanchet, Iyer,
and Qusailah}}]{ABIQ04}
\bibinfo{author}{\bibfnamefont{W.}~\bibnamefont{Arun}},
\bibinfo{author}{\bibfnamefont{L.}~\bibnamefont{Blanchet}},
\bibinfo{author}{\bibfnamefont{B.R.} \bibnamefont{Iyer}}, \bibnamefont{and}
\bibinfo{author}{\bibfnamefont{M.S.} \bibnamefont{Qusailah}},
\bibinfo{journal}{Class. Quant. Grav.} \textbf{\bibinfo{volume}{21}},
\bibinfo{pages}{3771} (\bibinfo{year}{2004}), \eprint{gr-qc/0404185}.

\bibitem[{\citenamefont{Jaranowski and Sch\"afer}(1998)}]{JaraS98}
\bibinfo{author}{\bibfnamefont{P.}~\bibnamefont{Jaranowski}} \bibnamefont{and}
\bibinfo{author}{\bibfnamefont{G.}~\bibnamefont{Sch\"afer}},
\bibinfo{journal}{Phys. Rev. D} \textbf{\bibinfo{volume}{57}},
\bibinfo{pages}{7274} (\bibinfo{year}{1998}).

\bibitem[{\citenamefont{Jaranowski and Sch\"afer}(1999)}]{JaraS99}
\bibinfo{author}{\bibfnamefont{P.}~\bibnamefont{Jaranowski}} \bibnamefont{and}
\bibinfo{author}{\bibfnamefont{G.}~\bibnamefont{Sch\"afer}},
\bibinfo{journal}{Phys. Rev. D} \textbf{\bibinfo{volume}{60}},
\bibinfo{pages}{124003} (\bibinfo{year}{1999}).

\bibitem[{\citenamefont{Damour et~al.}(2000{\natexlab{b}})\citenamefont{Damour,
Jaranowski, and Sch\"afer}}]{DJSpoinc}
\bibinfo{author}{\bibfnamefont{T.}~\bibnamefont{Damour}},
\bibinfo{author}{\bibfnamefont{P.}~\bibnamefont{Jaranowski}},
\bibnamefont{and}
\bibinfo{author}{\bibfnamefont{G.}~\bibnamefont{Sch\"afer}},
\bibinfo{journal}{Phys. Rev. D} \textbf{\bibinfo{volume}{62}},
\bibinfo{pages}{021501(R)} (\bibinfo{year}{2000}{\natexlab{b}}),
\bibinfo{note}{erratum Phys. Rev. D {\bf 63}, 029903(E) (2000)}.

\bibitem[{\citenamefont{Damour et~al.}(2001{\natexlab{a}})\citenamefont{Damour,
Jaranowski, and Sch\"afer}}]{DJSequiv}
\bibinfo{author}{\bibfnamefont{T.}~\bibnamefont{Damour}},
\bibinfo{author}{\bibfnamefont{P.}~\bibnamefont{Jaranowski}},
\bibnamefont{and}
\bibinfo{author}{\bibfnamefont{G.}~\bibnamefont{Sch\"afer}},
\bibinfo{journal}{Phys. Rev. D} \textbf{\bibinfo{volume}{63}},
\bibinfo{pages}{044021} (\bibinfo{year}{2001}{\natexlab{a}}),
\bibinfo{note}{erratum Phys. Rev. D {\bf 66}, 029901(E) (2002)}.

\bibitem[{\citenamefont{Blanchet and Faye}(2000{\natexlab{a}})}]{BF00}
\bibinfo{author}{\bibfnamefont{L.}~\bibnamefont{Blanchet}} \bibnamefont{and}
\bibinfo{author}{\bibfnamefont{G.}~\bibnamefont{Faye}},
\bibinfo{journal}{Phys. Lett. A} \textbf{\bibinfo{volume}{271}},
\bibinfo{pages}{58} (\bibinfo{year}{2000}{\natexlab{a}}),
\eprint{gr-qc/0004009}.

\bibitem[{\citenamefont{Blanchet and Faye}(2001)}]{BFeom}
\bibinfo{author}{\bibfnamefont{L.}~\bibnamefont{Blanchet}} \bibnamefont{and}
\bibinfo{author}{\bibfnamefont{G.}~\bibnamefont{Faye}},
\bibinfo{journal}{Phys. Rev. D} \textbf{\bibinfo{volume}{63}},
\bibinfo{pages}{062005} (\bibinfo{year}{2001}), \eprint{gr-qc/0007051}.

\bibitem[{\citenamefont{de~Andrade et~al.}(2001)\citenamefont{de~Andrade,
Blanchet, and Faye}}]{ABF01}
\bibinfo{author}{\bibfnamefont{V.}~\bibnamefont{de~Andrade}},
\bibinfo{author}{\bibfnamefont{L.}~\bibnamefont{Blanchet}}, \bibnamefont{and}
\bibinfo{author}{\bibfnamefont{G.}~\bibnamefont{Faye}},
\bibinfo{journal}{Class. Quant. Grav.} \textbf{\bibinfo{volume}{18}},
\bibinfo{pages}{753} (\bibinfo{year}{2001}).

\bibitem[{\citenamefont{Blanchet and Iyer}(2003)}]{BI03CM}
\bibinfo{author}{\bibfnamefont{L.}~\bibnamefont{Blanchet}} \bibnamefont{and}
\bibinfo{author}{\bibfnamefont{B.R.} \bibnamefont{Iyer}},
\bibinfo{journal}{Class. Quant. Grav.} \textbf{\bibinfo{volume}{20}},
\bibinfo{pages}{755} (\bibinfo{year}{2003}), \eprint{gr-qc/0209089}.

\bibitem[{\citenamefont{Blanchet}(1998{\natexlab{a}})}]{B98tail}
\bibinfo{author}{\bibfnamefont{L.}~\bibnamefont{Blanchet}},
\bibinfo{journal}{Class. Quant. Grav.} \textbf{\bibinfo{volume}{15}},
\bibinfo{pages}{113} (\bibinfo{year}{1998}{\natexlab{a}}),
\eprint{gr-qc/9710038}.

\bibitem[{\citenamefont{Blanchet
et~al.}(2002{\natexlab{a}})\citenamefont{Blanchet, Iyer, and Joguet}}]{BIJ02}
\bibinfo{author}{\bibfnamefont{L.}~\bibnamefont{Blanchet}},
\bibinfo{author}{\bibfnamefont{B.R.} \bibnamefont{Iyer}}, \bibnamefont{and}
\bibinfo{author}{\bibfnamefont{B.}~\bibnamefont{Joguet}},
\bibinfo{journal}{Phys. Rev. D} \textbf{\bibinfo{volume}{65}},
\bibinfo{pages}{064005} (\bibinfo{year}{2002}{\natexlab{a}}),
\eprint{gr-qc/0105098}.

\bibitem[{\citenamefont{Blanchet
et~al.}(2002{\natexlab{b}})\citenamefont{Blanchet, Faye, Iyer, and
Joguet}}]{BFIJ02}
\bibinfo{author}{\bibfnamefont{L.}~\bibnamefont{Blanchet}},
\bibinfo{author}{\bibfnamefont{G.}~\bibnamefont{Faye}},
\bibinfo{author}{\bibfnamefont{B.R.} \bibnamefont{Iyer}}, \bibnamefont{and}
\bibinfo{author}{\bibfnamefont{B.}~\bibnamefont{Joguet}},
\bibinfo{journal}{Phys. Rev. D} \textbf{\bibinfo{volume}{65}},
\bibinfo{pages}{061501(R)} (\bibinfo{year}{2002}{\natexlab{b}}),
\eprint{gr-qc/0105099}.

\bibitem[{\citenamefont{Blanchet and Iyer}(2004)}]{BI04mult}
\bibinfo{author}{\bibfnamefont{L.}~\bibnamefont{Blanchet}} \bibnamefont{and}
\bibinfo{author}{\bibfnamefont{B.R.} \bibnamefont{Iyer}},
\bibinfo{journal}{Phys. Rev. D} \textbf{\bibinfo{volume}{71}},
\bibinfo{pages}{024004} (\bibinfo{year}{2004}), \eprint{gr-qc/0409094}.

\bibitem[{\citenamefont{Hadamard}(1932)}]{Hadamard}
\bibinfo{author}{\bibfnamefont{J.}~\bibnamefont{Hadamard}},
\emph{\bibinfo{title}{Le probl\`eme de Cauchy et les \'equations aux
d\'eriv\'ees partielles lin\'eaires hyperboliques}}
(\bibinfo{publisher}{Hermann}, \bibinfo{address}{Paris},
\bibinfo{year}{1932}).

\bibitem[{\citenamefont{Schwartz}(1978)}]{Schwartz}
\bibinfo{author}{\bibfnamefont{L.}~\bibnamefont{Schwartz}},
\emph{\bibinfo{title}{Th\'eorie des distributions}}
(\bibinfo{publisher}{Hermann}, \bibinfo{address}{Paris},
\bibinfo{year}{1978}).

\bibitem[{\citenamefont{Sellier}(1994)}]{Sellier}
\bibinfo{author}{\bibfnamefont{A.}~\bibnamefont{Sellier}},
\bibinfo{journal}{Proc. R. Soc. London, Ser. A}
\textbf{\bibinfo{volume}{445}}, \bibinfo{pages}{69} (\bibinfo{year}{1994}).

\bibitem[{\citenamefont{Blanchet and Faye}(2000{\natexlab{b}})}]{BFreg}
\bibinfo{author}{\bibfnamefont{L.}~\bibnamefont{Blanchet}} \bibnamefont{and}
\bibinfo{author}{\bibfnamefont{G.}~\bibnamefont{Faye}}, \bibinfo{journal}{J.
Math. Phys.} \textbf{\bibinfo{volume}{41}}, \bibinfo{pages}{7675}
(\bibinfo{year}{2000}{\natexlab{b}}), \eprint{gr-qc/0004008}.

\bibitem[{\citenamefont{Pati and Will}(2000)}]{PW00}
\bibinfo{author}{\bibfnamefont{M.E.}~\bibnamefont{Pati}} \bibnamefont{and}
\bibinfo{author}{\bibfnamefont{C.M.}~\bibnamefont{Will}},
\bibinfo{journal}{Phys. Rev. D} \textbf{\bibinfo{volume}{62}},
\bibinfo{pages}{124015} (\bibinfo{year}{2000}).

\bibitem[{\citenamefont{K{\"o}nigsd{\"o}rffer
et~al.}(2003)\citenamefont{K{\"o}nigsd{\"o}rffer, Faye, and
Sch{\"a}fer}}]{KFS03}
\bibinfo{author}{\bibfnamefont{C.}~\bibnamefont{K{\"o}nigsd{\"o}rffer}},
\bibinfo{author}{\bibfnamefont{G.}~\bibnamefont{Faye}}, \bibnamefont{and}
\bibinfo{author}{\bibfnamefont{G.}~\bibnamefont{Sch{\"a}fer}},
\bibinfo{journal}{Phys. Rev. D} \textbf{\bibinfo{volume}{68}},
\bibinfo{pages}{044004} (\bibinfo{year}{2003}).

\bibitem[{\citenamefont{Nissanke and Blanchet}(2005)}]{NB05}
\bibinfo{author}{\bibfnamefont{S.}~\bibnamefont{Nissanke}} \bibnamefont{and}
\bibinfo{author}{\bibfnamefont{L.}~\bibnamefont{Blanchet}},
\bibinfo{journal}{Class. Quant. Grav.} \textbf{\bibinfo{volume}{22}},
\bibinfo{pages}{1007} (\bibinfo{year}{2005}), \eprint{gr-qc/0412018}.

\bibitem[{\citenamefont{'t~Hooft and Veltman}(1972)}]{tHooft}
\bibinfo{author}{\bibfnamefont{G.}~\bibnamefont{'t~Hooft}} \bibnamefont{and}
\bibinfo{author}{\bibfnamefont{M.}~\bibnamefont{Veltman}},
\bibinfo{journal}{Nucl. Phys.} \textbf{\bibinfo{volume}{B44}},
\bibinfo{pages}{139} (\bibinfo{year}{1972}).

\bibitem[{\citenamefont{Bollini and Giambiagi}(1972)}]{Bollini}
\bibinfo{author}{\bibfnamefont{C.G.} \bibnamefont{Bollini}} \bibnamefont{and}
\bibinfo{author}{\bibfnamefont{J.J.} \bibnamefont{Giambiagi}},
\bibinfo{journal}{Phys. Lett. B} \textbf{\bibinfo{volume}{40}},
\bibinfo{pages}{566} (\bibinfo{year}{1972}).

\bibitem[{\citenamefont{Breitenlohner and Maison}(1977)}]{Breitenlohner}
\bibinfo{author}{\bibfnamefont{P.}~\bibnamefont{Breitenlohner}}
\bibnamefont{and} \bibinfo{author}{\bibfnamefont{D.}~\bibnamefont{Maison}},
\bibinfo{journal}{Comm. Math. Phys.} \textbf{\bibinfo{volume}{52}},
\bibinfo{pages}{11} (\bibinfo{year}{1977}).

\bibitem[{\citenamefont{Damour et~al.}(2001{\natexlab{b}})\citenamefont{Damour,
Jaranowski, and Sch\"afer}}]{DJSdim}
\bibinfo{author}{\bibfnamefont{T.}~\bibnamefont{Damour}},
\bibinfo{author}{\bibfnamefont{P.}~\bibnamefont{Jaranowski}},
\bibnamefont{and}
\bibinfo{author}{\bibfnamefont{G.}~\bibnamefont{Sch\"afer}},
\bibinfo{journal}{Phys. Lett. B} \textbf{\bibinfo{volume}{513}},
\bibinfo{pages}{147} (\bibinfo{year}{2001}{\natexlab{b}}).

\bibitem[{\citenamefont{Blanchet
et~al.}(2004{\natexlab{a}})\citenamefont{Blanchet, Damour, and
Esposito-Far{\`e}se}}]{BDE04}
\bibinfo{author}{\bibfnamefont{L.}~\bibnamefont{Blanchet}},
\bibinfo{author}{\bibfnamefont{T.}~\bibnamefont{Damour}}, \bibnamefont{and}
\bibinfo{author}{\bibfnamefont{G.}~\bibnamefont{Esposito-Far{\`e}se}},
\bibinfo{journal}{Phys. Rev. D} \textbf{\bibinfo{volume}{69}},
\bibinfo{pages}{124007} (\bibinfo{year}{2004}{\natexlab{a}}),
\eprint{gr-qc/0311052}.

\bibitem[{\citenamefont{Itoh et~al.}(2001)\citenamefont{Itoh, Futamase, and
Asada}}]{IFA01}
\bibinfo{author}{\bibfnamefont{Y.}~\bibnamefont{Itoh}},
\bibinfo{author}{\bibfnamefont{T.}~\bibnamefont{Futamase}}, \bibnamefont{and}
\bibinfo{author}{\bibfnamefont{H.}~\bibnamefont{Asada}},
\bibinfo{journal}{Phys. Rev. D} \textbf{\bibinfo{volume}{63}},
\bibinfo{pages}{064038} (\bibinfo{year}{2001}).

\bibitem[{\citenamefont{Itoh and Futamase}(2003)}]{itoh1}
\bibinfo{author}{\bibfnamefont{Y.}~\bibnamefont{Itoh}} \bibnamefont{and}
\bibinfo{author}{\bibfnamefont{T.}~\bibnamefont{Futamase}},
\bibinfo{journal}{Phys. Rev. D} \textbf{\bibinfo{volume}{68}},
\bibinfo{pages}{121501(R)} (\bibinfo{year}{2003}).

\bibitem[{\citenamefont{Itoh}(2004)}]{itoh2}
\bibinfo{author}{\bibfnamefont{Y.}~\bibnamefont{Itoh}}, \bibinfo{journal}{Phys.
Rev. D} \textbf{\bibinfo{volume}{69}}, \bibinfo{pages}{064018}
(\bibinfo{year}{2004}).

\bibitem[{\citenamefont{Blanchet
et~al.}(2004{\natexlab{b}})\citenamefont{Blanchet, Damour,
Esposito-Far{\`e}se, and Iyer}}]{BDEI04}
\bibinfo{author}{\bibfnamefont{L.}~\bibnamefont{Blanchet}},
\bibinfo{author}{\bibfnamefont{T.}~\bibnamefont{Damour}},
\bibinfo{author}{\bibfnamefont{G.}~\bibnamefont{Esposito-Far{\`e}se}},
\bibnamefont{and} \bibinfo{author}{\bibfnamefont{B.R.} \bibnamefont{Iyer}},
\bibinfo{journal}{Phys. Rev. Lett.} \textbf{\bibinfo{volume}{93}},
\bibinfo{pages}{091101} (\bibinfo{year}{2004}{\natexlab{b}}),
\eprint{gr-qc/0406012}.

\bibitem[{\citenamefont{Blanchet et~al.}(2005)\citenamefont{Blanchet, Damour,
and Iyer}}]{BDI04zeta}
\bibinfo{author}{\bibfnamefont{L.}~\bibnamefont{Blanchet}},
\bibinfo{author}{\bibfnamefont{T.}~\bibnamefont{Damour}}, \bibnamefont{and}
\bibinfo{author}{\bibfnamefont{B.R.} \bibnamefont{Iyer}},
\bibinfo{journal}{Class. Quant. Grav.} \textbf{\bibinfo{volume}{22}},
\bibinfo{pages}{155} (\bibinfo{year}{2005}), \eprint{gr-qc/0410021}.

\bibitem[{\citenamefont{Damour and Esposito-Far{\`e}se}(1996)}]{Dgef96}
\bibinfo{author}{\bibfnamefont{T.}~\bibnamefont{Damour}} \bibnamefont{and}
\bibinfo{author}{\bibfnamefont{G.}~\bibnamefont{Esposito-Far{\`e}se}},
\bibinfo{journal}{Phys. Rev. D} \textbf{\bibinfo{volume}{53}},
\bibinfo{pages}{5541} (\bibinfo{year}{1996}), \eprint{gr-qc/9506063}.

\bibitem[{\citenamefont{Cardoso et~al.}(2003)\citenamefont{Cardoso, Dias, and
Lemos}}]{Cardoso}
\bibinfo{author}{\bibfnamefont{V.}~\bibnamefont{Cardoso}},
\bibinfo{author}{\bibfnamefont{O.J.C.}~\bibnamefont{Dias}}, \bibnamefont{and}
\bibinfo{author}{\bibfnamefont{J.P.S.}~\bibnamefont{Lemos}},
\bibinfo{journal}{Phys. Rev. D} \textbf{\bibinfo{volume}{67}},
\bibinfo{pages}{064026} (\bibinfo{year}{2003}).

\bibitem[{\citenamefont{Blanchet}(1995)}]{B95}
\bibinfo{author}{\bibfnamefont{L.}~\bibnamefont{Blanchet}},
\bibinfo{journal}{Phys. Rev. D} \textbf{\bibinfo{volume}{51}},
\bibinfo{pages}{2559} (\bibinfo{year}{1995}), \eprint{gr-qc/9501030}.

\bibitem[{\citenamefont{Blanchet}(1998{\natexlab{b}})}]{B98mult}
\bibinfo{author}{\bibfnamefont{L.}~\bibnamefont{Blanchet}},
\bibinfo{journal}{Class. Quant. Grav.} \textbf{\bibinfo{volume}{15}},
\bibinfo{pages}{1971} (\bibinfo{year}{1998}{\natexlab{b}}),
\eprint{gr-qc/9801101}.

\bibitem[{\citenamefont{Poujade and Blanchet}(2002)}]{PB02}
\bibinfo{author}{\bibfnamefont{O.}~\bibnamefont{Poujade}} \bibnamefont{and}
\bibinfo{author}{\bibfnamefont{L.}~\bibnamefont{Blanchet}},
\bibinfo{journal}{Phys. Rev. D} \textbf{\bibinfo{volume}{65}},
\bibinfo{pages}{124020} (\bibinfo{year}{2002}), \eprint{gr-qc/0112057}.

\bibitem[{\citenamefont{Blanchet and Damour}(1989)}]{BD89}
\bibinfo{author}{\bibfnamefont{L.}~\bibnamefont{Blanchet}} \bibnamefont{and}
\bibinfo{author}{\bibfnamefont{T.}~\bibnamefont{Damour}},
\bibinfo{journal}{Annales Inst. H. Poincar\'e Phys. Th\'eor.}
\textbf{\bibinfo{volume}{50}}, \bibinfo{pages}{377} (\bibinfo{year}{1989}).

\bibitem[{\citenamefont{Blanchet and Damour}(1986)}]{BD86}
\bibinfo{author}{\bibfnamefont{L.}~\bibnamefont{Blanchet}} \bibnamefont{and}
\bibinfo{author}{\bibfnamefont{T.}~\bibnamefont{Damour}},
\bibinfo{journal}{Phil. Trans. Roy. Soc. Lond. A}
\textbf{\bibinfo{volume}{320}}, \bibinfo{pages}{379} (\bibinfo{year}{1986}).

\bibitem[{\citenamefont{Blanchet and Damour}(1988)}]{BD88}
\bibinfo{author}{\bibfnamefont{L.}~\bibnamefont{Blanchet}} \bibnamefont{and}
\bibinfo{author}{\bibfnamefont{T.}~\bibnamefont{Damour}},
\bibinfo{journal}{Phys. Rev. D} \textbf{\bibinfo{volume}{37}},
\bibinfo{pages}{1410} (\bibinfo{year}{1988}).

\bibitem[{\citenamefont{Blanchet}(1998{\natexlab{c}})}]{B98quad}
\bibinfo{author}{\bibfnamefont{L.}~\bibnamefont{Blanchet}},
\bibinfo{journal}{Class. Quant. Grav.} \textbf{\bibinfo{volume}{15}},
\bibinfo{pages}{89} (\bibinfo{year}{1998}{\natexlab{c}}),
\eprint{gr-qc/9710037}.

\bibitem[{\citenamefont{Damour and Iyer}(1991{\natexlab{a}})}]{DI91a}
\bibinfo{author}{\bibfnamefont{T.}~\bibnamefont{Damour}} \bibnamefont{and}
\bibinfo{author}{\bibfnamefont{B.R.} \bibnamefont{Iyer}},
\bibinfo{journal}{Annales Inst. H. Poincar\'e, Phys. Th\'eor.}
\textbf{\bibinfo{volume}{54}}, \bibinfo{pages}{115}
(\bibinfo{year}{1991}{\natexlab{a}}).

\bibitem[{\citenamefont{Damour and Iyer}(1991{\natexlab{b}})}]{DI91b}
\bibinfo{author}{\bibfnamefont{T.}~\bibnamefont{Damour}} \bibnamefont{and}
\bibinfo{author}{\bibfnamefont{B.R.} \bibnamefont{Iyer}},
\bibinfo{journal}{Phys. Rev. D} \textbf{\bibinfo{volume}{43}},
\bibinfo{pages}{3259} (\bibinfo{year}{1991}{\natexlab{b}}).

\bibitem[{\citenamefont{Thorne}(1980)}]{Th80}
\bibinfo{author}{\bibfnamefont{K.}~\bibnamefont{Thorne}},
\bibinfo{journal}{Rev. Mod. Phys.} \textbf{\bibinfo{volume}{52}},
\bibinfo{pages}{299} (\bibinfo{year}{1980}).

\bibitem[{\citenamefont{Damour}(1983)}]{D83houches}
\bibinfo{author}{\bibfnamefont{T.}~\bibnamefont{Damour}}, in
\emph{\bibinfo{booktitle}{Gravitational Radiation}}, edited by
\bibinfo{editor}{\bibfnamefont{N.}~\bibnamefont{Deruelle}} \bibnamefont{and}
\bibinfo{editor}{\bibfnamefont{T.}~\bibnamefont{Piran}}
(\bibinfo{publisher}{North-Holland Company}, \bibinfo{address}{Amsterdam},
\bibinfo{year}{1983}), pp. \bibinfo{pages}{59--144}.

\bibitem[{\citenamefont{Blanchet}(2002)}]{Bliving}
\bibinfo{author}{\bibfnamefont{L.}~\bibnamefont{Blanchet}},
\bibinfo{journal}{Living Rev. Rel.} \textbf{\bibinfo{volume}{5}},
\bibinfo{pages}{3} (\bibinfo{year}{2002}), \eprint{gr-qc/0202016}.

\bibitem[{\citenamefont{Damour}(1980)}]{D80}
\bibinfo{author}{\bibfnamefont{T.}~\bibnamefont{Damour}}, \bibinfo{journal}{C.
R. Acad. Sc. Paris} \textbf{\bibinfo{volume}{291}}, \bibinfo{pages}{227}
(\bibinfo{year}{1980}).

\bibitem[{\citenamefont{Bertotti and Plebanski}(1960)}]{BertottiP60}
\bibinfo{author}{\bibfnamefont{B.}~\bibnamefont{Bertotti}} \bibnamefont{and}
\bibinfo{author}{\bibfnamefont{J.}~\bibnamefont{Plebanski}},
\bibinfo{journal}{Ann. Phys. (N. Y.)} \textbf{\bibinfo{volume}{11}},
\bibinfo{pages}{169} (\bibinfo{year}{1960}).

\bibitem[{\citenamefont{Goldberger and Rothstein}(2004)}]{Goldberger}
\bibinfo{author}{\bibfnamefont{W.D.} \bibnamefont{Goldberger}}
\bibnamefont{and} \bibinfo{author}{\bibfnamefont{I.Z.}
\bibnamefont{Rothstein}} (\bibinfo{year}{2004}), \eprint{hep-th/0409156}.

\bibitem[{\citenamefont{Nordtvedt}(1994)}]{Nordt94}
\bibinfo{author}{\bibfnamefont{K.}~\bibnamefont{Nordtvedt}},
\bibinfo{journal}{Phys. Rev. D} \textbf{\bibinfo{volume}{49}},
\bibinfo{pages}{5165} (\bibinfo{year}{1994}).

\bibitem[{\citenamefont{Damour and Esposito-Far{\`e}se}(1998)}]{Dgef98}
\bibinfo{author}{\bibfnamefont{T.}~\bibnamefont{Damour}} \bibnamefont{and}
\bibinfo{author}{\bibfnamefont{G.}~\bibnamefont{Esposito-Far{\`e}se}},
\bibinfo{journal}{Phys. Rev. D} \textbf{\bibinfo{volume}{58}},
\bibinfo{pages}{042001} (\bibinfo{year}{1998}), \eprint{gr-qc/9803031}.

\end{thebibliography}
\end{document}